\documentclass{aa}

\usepackage{graphicx}
\usepackage{txfonts}
\usepackage{amssymb}
\usepackage{listings}
\usepackage{hyperref}
\usepackage{nameref}
\usepackage{csquotes}
\usepackage{rotating}
\usepackage{hyperref}
\usepackage{tabularx}
    \newcolumntype{L}{>{\raggedright\arraybackslash}X}
\usepackage{enumitem}
\usepackage{booktabs}
\usepackage{textcomp}

\usepackage{xcolor}

\begin{document}

    
   \title{Automatic catalog of RR Lyrae from $\sim$14 million VVV light curves:\\How far can we go with traditional machine-learning?}

\titlerunning{Automatic Catalog of RRL from VVV}

   \author{J. B. Cabral\inst{1, 2}
          \and
          F. Ramos\inst{2}
          \and
          S. Gurovich\inst{2,3}
          \and
          P.M. Granitto\inst{1}
          }

   \institute{
        Centro Internacional Franco Argentino de Ciencias de la Información y de Sistemas (CIFASIS, CONICET--UNR),
        Rosario, Santa Fé, Argentina.
        \email{cabral@cifasis-conicet.gov.ar}
        \and
        Instituto De Astronomía Teórica y Experimental – Observatorio Astronómico Córdoba (IATE--OAC--UNC--CONICET), Córdoba, Córdoba, Argentina
        \and
        Western Sydney University (WSU), Locked Bag 1797, Penrith South DC, NSW 2751, Australia
    }

   \date{}


  \abstract
   {The creation of a 3D map of the bulge using RR Lyrae (RRL) is one of the main goals of the VISTA Variables in the Via Lactea Survey (VVV) and VVV(X) surveys. The overwhelming number of sources undergoing analysis undoubtedly requires the use of automatic procedures. In this context, previous studies have introduced the use of machine learning (ML) methods for the task of variable star classification.}
   {Our goal is to develop and test an entirely automatic ML-based procedure for the identification of RRLs in the VVV Survey. This automatic procedure is meant to be used to generate reliable catalogs integrated over several tiles in the survey.}
   {Following the reconstruction of light curves, we extracted a set of period- and intensity-based features, which were already defined in previous works. Also, for the first time, we put a new subset of useful color features to use. We discuss in considerable detail all the appropriate steps needed to define our fully automatic pipeline, namely: the selection of quality measurements; sampling procedures; classifier setup, and model selection.}
   {As a result, we were able to construct an ensemble classifier with an average recall of 0.48 and average precision of 0.86 over 15 tiles. We also made  all our processed datasets available and we published a catalog of candidate RRLs.}
   {Perhaps most interestingly, from a classification perspective based on photometric broad-band data, our results indicate that color is an informative feature type of the RRL objective class that should always be considered in automatic classification methods via ML. We also argue that recall and precision in both tables and curves are high-quality metrics with regard to this highly imbalanced problem. Furthermore, we show for our VVV data-set that to have good estimates, it is important to use the original distribution more abundantly than reduced samples with an artificial balance. Finally, we show that the use of ensemble classifiers helps resolve the crucial model selection step and that most errors in the identification of RRLs are related to low-quality observations of some sources or to the increased difficulty in resolving the RRL-C type given the data.}


   \maketitle
%

\section{Introduction}
\label{section:intro}

The stars of \textit{RR Lyrae} (RRL) were first imaged in the last decade of the 19th century. Although it is uncertain as to who made the first observations, some authors report that it was Williamina Fleming. Fleming led Pickering's Harvard Computer group in early studies of variable stars towards globular clusters  \citep{burnham1978burnham, silbermann1995rr}. Towards the end of the 19th century, J. C. Kapteyn and D. E. Packer independently published results of variable stars which were also likely to be RRL stars \citep{smith2004rr}. In the decade following Fleming's initial work, \citet{bailey1902discussion} classified RRL into three subtypes. Bailey also proposed that RRL be used as ``standard candles'' to measure distances to clusters. Later works by \citet{seares1914color} found that variability in RRL, or so-called ``cluster variables,'' was probably caused by the radial pulsation of RRL stellar atmospheres. Several years later \citet{shapley1918studies}, used RRL to estimate the distance to globular clusters in the Galaxy. Later, \citet{baade1946search} used them to measure the distance to the galactic center.

Now, over a century after Fleming's initial discovery and after a plethora of systematic studies, RRL have come to be known as a class of multi-mode pulsating population II stars with spectral (color) type A or F that reside in the so-called "horizontal branch" of the C-M diagram. They are also considered excellent standard candles with periodic light curves (LC) and characteristic features that are sensitive to the chemical abundance of the primordial fossil gas from which they formed \cite{smith2004rr}. These short-period variables have periods typically ranging between 0.2 and 1.2 days \citep{smith2004rr} and they are old (aged $\sim$ 14 Gyr), with progenitor masses typically about 0.8 Solar $M_{\odot}$ and absolute magnitudes consistent with core helium burning stars. When it comes to how many RRL there are in the Galaxy, \citet{smith2004rr}, estimated some 85 000. We now know, thanks to a recent work by OGLE \citep{2019AcA....69..321S} and Gaia \citep{2018A&A...616A...1G} data releases, that there are over 140000 RRL in The Galaxy. Interestingly from a galaxy formation and evolution perspective, RRL chemical abundances could be used as evidence to piece together the structural formation and evolution of the bulge, disk, and halo components of the Galaxy, as well as in near-by ones. \cite{lee92_aj} presented evidence that bulge RRL stars may  even be older than the Galaxy Halo stars, implying that galaxies may have formed inside-out. So RRL are also important for testing the physics of pulsating stars and the stellar evolution of low mas stars. RRLs have also been used extensively for studies of globular clusters and nearby galaxies. They have been used successfully and systematically as a rung in the extra-galactic distance ladder \citep{de2017toward} which, along with other methods, Cepheid distances, tip of the red giant branch methods, etc. \citep{sakai2001observation}, have provided primary distance estimates to boot-strap distance measurements to external galaxies \citep{clementini2001rr} and reddening determinations within our own.

The list of objectives of the VISTA Variables in the Via Lactea Survey (VVV, \citealt{minniti2010vista}), as designed, includes a $3 (+ 1)$  dimensional mapping of the bulge using RRLs. As shown by \cite{dwek1995morphology}, most of the bulge light is contained within ten degrees of the bulge center, so most of the mass, assuming a constant mass-to-light ratio, should also be contained within this radius.

Our study (and sample) was made after the OGLE-IV release but it precedes the OGLE catalog presented in \citet{2019AcA....69..321S} and Gaia Data release. We use the VISTA data to extract characteristic RRL features based on NIR VVV photometry. A star-by-star approach to the reduction and process analysis of light curves (LC) is too costly, prone to external biases, and largely inefficient due to the overwhelmingly high number of light curves collected by modern surveys like VVV. The automatic methods developed for this study statistically handle subtleties involved in normalizing the observational parameter space to obtain a machine-eye view at the physical parameter space. Our study, we hope, will serve to refine previous attempts at the use of machine learning (ML) and data mining techniques to detect and tag RRL candidates for studies of bulge dynamics of the Galaxy. We also attempt to take into account the inhomogeneity of Galactic extinction and reddening, as well as  varying stellar density at different lines-of-sight, considering these in a computationally statistical manner during the feature extraction and classification steps.

While panchromatic studies from the UV through the near-IR (NIR) and beyond, including spectroscopic studies, will undoubtedly better constrain the observed and physical RRL parameter space, this is not yet possible, so for this study we draw mostly on learning  from visual data taken from the literature and from NIR VVV photometry. Our  objective  class  is  the  set  of known RRL stars from  the literature. These stars are tagged based on \textsc{OGLE-III}/\textsc{OGLE-IV} and other catalogs accessed through VizieR. They are listed in Appendix~\ref{appendix:vizier}.

We wish to note that stellar population synthesis studies based on broadband colors, for example: \cite{2001ApJ...550..212B} and \cite{2010AJ....140..663G} appear to suggest that the evolution of stellar mass in massive disk galaxies is better constrained by NIR photometry than by broadband photometry at shorter effective wavelengths. Although predicted effective temperatures for RRL with given spectral types are more closely matched to bluer filters, we made a compromise and chose to draw our characteristic feature set preferentially on NIR light. Moreover, at NIR wavelengths, Galactic extinction effects (and errors) due to differential reddening are also significantly reduced, so any method of feature engineering for RRL classification using NIR data should also prove to be insightful.

Previous works have attempted to develop ML methods for the identification of variable stars on massive datasets. For example, \citet{richards2011machine} and \citet{richards2012construction} introduced a series of more than 60 features measured over folded light curves and used a dataset with 28 classes of variable stars. \citet{armstrong2015k2} discussed the use of an unsupervised method before a classifier for K2 targets.
\citet{Mackenzie2016ApJ} introduced the use of an automatic method to extract features from a LC based on clustering of patterns detected in the LC. \citet{Shin2009MNRAS,Shin2012AJ} introduced the use of an outlier detection technique for the identification of variable stars and \citet{Moretti2018MNRAS} used another unsupervised technique, PCA, for the same task.
\citet{pashchenko2017MNRAS} considered the task of detecting variable stars over a set of $\sim30000$ OGLE-II sources with 18 features and several ML methods. In the most relevant antecedent to our work, \citet{elorrieta2016machine} described the use of ML methods in the search of RRLs in the VVV. They consider the complete development of the classifier, starting from the collected data (2265 RRLs and 15436 other variables), the extraction of 32 features and the evaluation of several classifiers. As most previous works, they found that methods based on ensembles of trees are the best performers for this task. In this particular case they show an small edge in favor of boosting versions over more traditional random forest (RF) \citep{breiman2001random}.

As always in this line of research, there are several ways in which we can improve upon earlier results, particularly  for the findings of \citet{elorrieta2016machine}. First, they chose not to use color so as to avoid extinction related difficulties and they do not use reddening-free indices. Second, as in most previous works, they use a data sample that is clearly skewed in favor of variable stars. This over-representation of RRLs produces, in general, over-optimistic results, as we  show later in this work. Lastly, they used performance measures (ROC-AUC and F1 at a fixed threshold, see Section~\ref{section:env}) that are not appropriate for highly imbalanced problems and low performance.

Furthermore, modern surveys request for fully automatic methods, starting from the extraction of a subset of stable features through the training and setup of efficient classifiers up to the identification of candidate RRLs. In this work, we attempt to develop a completely automatic method to detect RRLs in the VVV survey. We show the effectiveness of our subset of features (including color features), along with ways to automatically select the complete setup of the classifier and to produce unbiased and useful estimates of its performance.

Although we limit ourselves to classifying RRLs based on VVV aperture magnitudes in this work, we note our method could be adapted for VVV point spread function (PSF) magnitudes or used  to classify other Variables such as Cepheids, Transients, or even employed on other surveys such as the Vera Rubin Observatory’s Legacy Survey of Space and Time (LSST) or Gaia \citep{brough2020vera, ivezic2019lsst}.

This paper is organized as follows. In Section~\ref{section:vvv_data}, we outline the input data based on the VVV survey: type, size, structure, and how we clean it. Section~\ref{section:env} describes the
computational environment  and method used in this work.
Section~\ref{section:features} summarizes the steps taken to build and extract the features from the LC, along with details on how we incorporate our positive class (variable stars) into our dataset, and, finally, we present an evaluation of diverse feature subsets. In Section~\ref{section:model_selection}, we describe the selection of an appropriate ML method and in Section~\ref{section:sampling}, we discuss the best way to sample the data in order to obtain good performance and error estimations. Then we select the final classifier (Section~\ref{section:final}), analyze how some errors are produced (Section~\ref{section:fpfn}), and discuss the data release (Section~\ref{section:carpyncho}). We summarize our work in Section~\ref{section:conclusions}.

\section{The VVV survey and data reduction}
\label{section:vvv_data}

The VVV completed observations in 2015 after about 2000 hours of observations, for which the Milky Way bulge and an adjacent section of the mid-plane (with high star formation rates) were systematically scanned. Its scientific goals were related to the structure, formation, and evolution of the Galaxy
 and galaxies in general.
The data for both VVV and VVV(X) is obtained by VIRCAM, the VISTA infrared camera mounted on ESO’s VISTA survey telescope \citep{sutherland2015visible}, At the time of its commissioning, VIRCAM was the largest NIR camera with 16 non-contiguous 2k x 2k detectors. All 16 detectors are exposed simultaneously capturing a standard ``pawprint'' of a patch of sky. To fill the spaces between the 16 detectors and for more uniform sampling, 6 pawprints are observed consecutively at suitable chosen offsets of a large fraction of a detector. When combined into a "tile," these six pawprints sample a rectangular field of roughly 1.5 x 1.0 degrees, so that each pixel of sky is observed in at least 2 pawprints. There are also two small strips in which there is only one observation. These strips overlap another such strip on the adjacent tile.
Over the course of the survey, the same tile is re-observed for variability studies. In the first year of VVV survey observations more stricter conditions were required and for most tiles five astronomical pass-bands where observed that included in increasing wavelength Z, Y, J, H, and Ks, separated only by a couple hours. However, for a small number of tiles, when the required survey conditions were not met (e.g., seeing), simultaneous multi-band observations for that tile were made later, some, even, in subsequent years. We would like to note that we use the Cambridge Astronomical Survey Unit (CASU) search service to group the multi-epoch data for the construction of our band-merge catalog for CASU version 1.3 release. Finally, during the multi-epoch campaign mostly Ks observations were made, which we use for our variability study for the same release. Typically, for each tile, around 80 observation were recorded for the Ks band and only one, the first tile epoch, for the other four bands. We also note that the input data preprocessed by the CASU VISTA Data Flow System pipeline \citep{emerson_vista_2004}, include  photometric and astrometric corrections, made for each pawprint and tile \citep{gonzalez2017vista} image and catalog pre-processed data produced by CASU as standard FITS files \citep{hanisch2001definition}.

The work in this study is based mainly on the 3rd aperture photometry data (as suggested by CASU for stellar photometry in less crowded fields) from the single pawprint \textsc{CASU FITS} catalogs. The magnitudes are extrapolated or total magnitudes. The FITS catalogs were converted to ascii catalogs using an adapted CASU \texttt{cat\_fit\_list.f} program. Over the course of the survey, given the high data rates ($\sim200$ MB/pawprint), several hundred terabytes of astronomical data were produced. Systematic methods are required to extract, in a homogeneous way, astrophysical information across $\sim1000$ tiles for a wide range of scientific goals. The "first" phase of the pre-reduction process is done by CASU's VISTA Data Flow System (VDFS), which produces, among other results, the pawprint, and tile catalogs that we use as our inputs.

\subsection{Data for this work}
\label{subsection:selected_data}

Based on  the VDFS results, we create two kind of post-processed data sources:

\begin{description}

\item[Tile catalogs:] The $J$, $H$, and $Ks$ consolidated master source list were defined from the first epoch of observations. These data were selected because source color information is available and also because the "first" epoch data have constraints that include more stringent observational and prepossessing data requirements (seeing, photometric conditional limits, etc.) than for any other epoch of observation. The cross-matching proximity algorithm employed was modified for multiple bands. It is based on a KD-tree implementation from the \textit{SciPy} Python library. It reduces the matching complexity of traditional algorithms from $O(N^{2})$ to $O(Log N)$\footnote{The cross-matching assumes that objects A and B  from two different catalogs are the same if object A is the closest to object B, and object B is the closest to  object A}, with a threshold of $1/3 \arcsec$ used to define which sources are the same across different bands. The KD-tree implementation was developed in collaboration with Erik Tollerud for the  \textit{Astropysics} library (the precursor of \textit{Astropy}). Its implementation is also based on the Scipy package \citep{tollerud2012astropysics, robitaille2013astropy, jones_scipy:_2014}.

\item[Pawprint Stacks:] Each pawprint stack is associated to a Tile Catalog and includes the six pawprint observation of the detectors for a single Ks-band epoch.
\end{description}


\section{Environment: software, hardware, methods.}
\label{section:env}

We took \textsc{VDFS} data catalogs of \textsc{CASUVER 1.3} to begin our processing. The processing phase was made within a custom multiprocess environment pipeline called \textit{Carpyncho} \citep{cabral_carpyncho_2016}, developed on top of the \textit{Corral Framework} \citep{cabral2017corral}, the \textit{Python} \footnote{\url{https://www.python.org/}} Scientific Stack \textit{Numpy} \citep{van2011numpy}, \textit{Scipy} \citep{jones_scipy:_2014}, and \textit{Matplotlib} \citep{hunter2007matplotlib}). Additional astronomy routines employed were provided by the \textit{Astropy} \citep{robitaille2013astropy} and \textit{PyAstronomy} \citep{czesla2019pya} libraries. The feature extraction from the light-curve data was handled using \textit{feets} \citep{cabral_fats_2018} packages, which, in turn, had been previously reingineered from an existing package called \textsc{FATS} \citep{nun2015fats} and later incorporated as an affiliated package to \textit{astropy}. The feature extraction was executed in a 50 cores CPU computer provided by the \textsc{IATE} \footnote{\url{http://iate.oac.uncor.edu/}}.
The \textit{Jupyter Notebooks} \citep{ipython_2014} (for interactive analysis), \textit{Scikit-Learn} \citep{pedregosa2011scikit} (for ML), and \textit{Matplotlib} (for the plotting routine) are our tools of choice in the exploratory phase  with the aim of selecting the most useful model and creating the catalog.

We considered the use of four diverse classical ML methods as classifiers in this work. We first selected  the support vector machine (SVM) \citep{vapnik2013nature}, a method that finds a maximum-margin solution in the original feature space (using the so--called linear kernel) or in a transformed feature space for non--linear solutions; we selected the radial-base-function, or RBF kernels, as a second classifier in this case. Third we selected the K-nearest neighbors (KNN) method \citep{mitchell1997machine}, a simple but efficient local solution. Last, we selected RF, an ensemble-based method, which requires almost no tuning and provides competitive performance in most datasets.

In order to estimate performance measures, we used two different strategies. In some cases, we relied on an internal cross-validation procedure, usually known as K-fold CV. In this case, the available data is divided in K separate folds of approximately equal length and we repeat K times the procedure of fitting our methods on (K-1) folds, using the last one as a test set. Finally, the K measurements are averaged. In other cases, we use complete tiles to fit the classifiers and other complete tiles as test sets.

We selected three performance measures that are appropriate for highly imbalanced binary classification problems. The RRLs are considered the positive class and all other sources as the negative class (Section \ref{subsection:Tagging}). RRLs correctly identified by a classifier are called true positives (TP) and those not detected are called false negatives (FN). Other sources wrongly classified as RRLs are called false positives (FP), and those correctly identified are called true negatives (TN). Using the proportion of these four outcomes on a given dataset, we can define two complementary measures. Precision is defined as TP / (TP + FP). It measures the fraction of real RRLs detected over all those retrieved by the classifier. Recall is defined as TP / (TP + FN). It measures the proportion of the total of RRLs that are detected by the classifier. Classifiers that output probabilities can change their decisions simply by changing the threshold at which a case is considered as positive. Using a very low threshold leads to high Recall and low Precision, as more sources are classified as positive. High thresholds, on the other hand, lead to the opposite result, high Precision and low Recall. Precision and Recall should be always evaluated at the same time. One of the best ways to evaluate a classifier in our context (highly imbalanced binary problem) is to consider Precision-Recall curves, in which we plot a set of pairs of values corresponding to different thresholds. In general, curves that approach the top-right corner are considered as better classifiers.

A more traditional measure of performance for binary problems is the accuracy, defined as (TP + TN) / (TP + FP + TN + FN). In relation to accuracy, it is also customary to create ROC-curves and to measure the area under the curve (ROC-AUC) as a global performance measure. We show in Section~\ref{section:model_selection} that these two measures do not provide any information in our case.

\section{Feature extraction: from light-curves to features}
\label{section:features}

To identify and typify a variable-periodic star of any type, it is necessary to precisely measure its magnitude variation in a characteristic time period. To this end, we need to match each source present in a \textit{Tile Catalog} with all the existing observations of the same source in all available Pawprint-Stacks and then to reconstruct its corresponding time series.

All the features that we extracted for this work correspond to sources that share two characteristics:

\begin{itemize}
    \item An average magnitude between $12$ and $16.5$, where the VVV photometry is highly reliable \citep{gran_bulge_2015}.
    \item At
least 30 epochs in its light curve to make the features more reliable, each source needs to have \end{itemize}

\subsection{Proximity \textit{cross-matching}}

We take measurements of positions and magnitudes for every source in each Pawprint Stack catalog with known tile identification number, as well as the mid-time of the observation. Since the input source catalogs do not have star identification numbers, we use cross-matching to determine the correspondence between observed sources in the \textit{Tile Catalog} and the \textit{Pawprint Stack} using our matching algorithm.

\subsection{Date--time correction}

The dates in the \textit{Pawprint Stack} are recorded as the average date across all the "pawprints"\textit{} involved in the "stack" in Modified Julian Days (MJD) in UTC. The recorded MJDs are subject to the light travel time delays caused by the Earth's orbital motion around the Sun. These delays are relevant for short-period variable stars. In consequence, we transform our dates to Heliocentric Julian Days (HJD), which modifies the MJD using differences in the position of the Earth with respect to the Sun and the source \citep{eastman_achieving_2010}.

\subsection{Period}

\begin{figure}
    \centering
    \includegraphics[width=1\columnwidth]{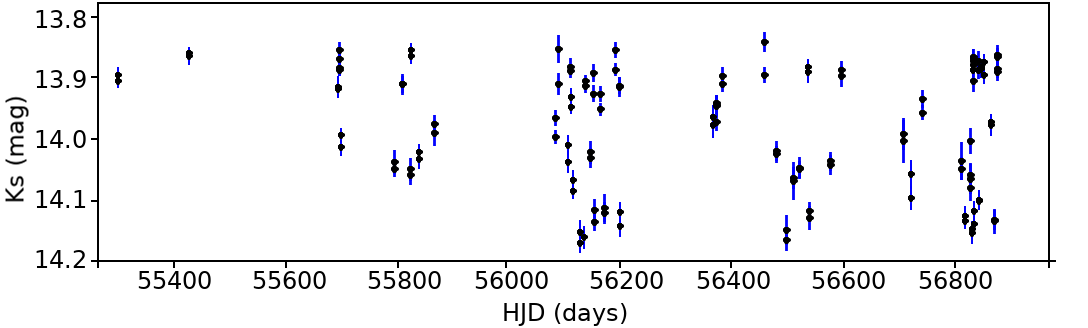}
    \caption{Lightcurve of an RRL AB variable star from  \citet{gran_bulge_2015} identified with the ID \texttt{VVV J2703536.01-412829.4}. The horizontal axis is the measurement date in heliocentric-corrected days (HJD) and the vertical one is the source magnitude.}
    \label{fig:hjd}
    \end{figure}

Once the observation instances for each source have been identified and the HJD calculated, the next step consists of calculating periods for our sources. The VVV is an irregular time-sampled survey, so for sampling consistency across the survey, we make the assumption that the frequency of observational data is random. Figure~\ref{fig:hjd} shows the observed magnitude of an RRL as a function of HDJ time of measurement. No periodic signal is evident  since the cadence sampling is on the order of tens of days, which far exceeds the expected period of RRL stars. We know RRL in the NIR have approximately sinusoidal light curves
with variations between $\sim 0.1$ and $\sim 0.5$ mags as opposed to $\sim 0.2$
and $\sim 2.0$ mags in bluer broad-band filters, and if we assume that therein may lie hidden useful features, to increase sensitivity in the feature analysis that follows, we automatically analyse the Periodogram, extract a period, and proceed by folding the RRL time series by this period. We extract what we expect may be many useful features from the folded LC in the following section.

\begin{figure}
    \centering
    \includegraphics[width=\columnwidth]{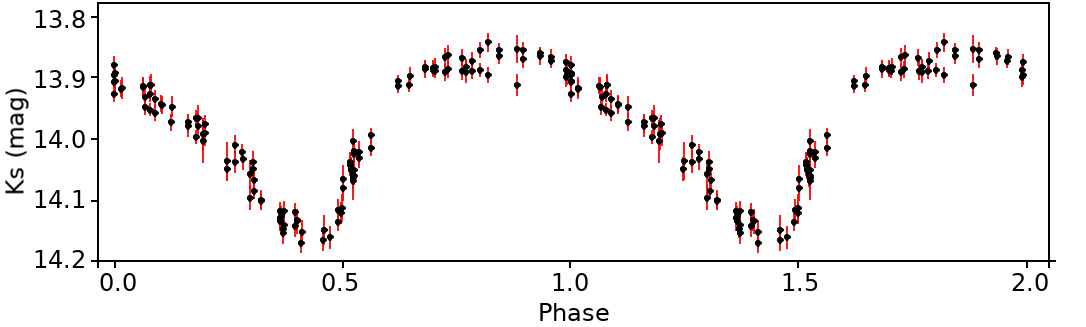}
    \caption{Folded light curve for the same star as Fig.~\ref{fig:hjd}, using a period of $\sim0.757$ days. The horizontal axis is the phase, and the vertical shows the source magnitude. For visual ease, two periods are shown.}
    \label{fig:gls}
    \end{figure}

We note that to recover the period of the RRL source, we use the family of \textit{Fast Lomb-Scargle} methods \citep{lomb_least-squares_1976, scargle_studies_1982, vanderplas2018understanding}.
This method finds the least squares error fit of a sinusoid to the sampled data. Folding the LC by this period puts the LC in-phase, as shown in Fig.~\ref{fig:gls}, allowing for the extraction of interesting features, for example the first Fourier Components that have been found in other work to be sensitive to [Fe/H] \citep[e.g.,][]{kovacs_walk_01}.

\subsection{Tagging: positive and negative samples}
\label{subsection:Tagging}

We selected our positive samples using the following procedure. First, we identified all the variable stars in our Tile Catalogs (Section \ref{subsection:selected_data}) by proximity cross-matching\textit{} (using a threshold of \textit{$1/3$ arc sec}) mainly from the variable-star catalogs of \textit{OGLE-III} \citep{2003AcA....53..291U, 2011AcA....61....1S}, \textit{OGLE-IV} \citep{2015AcA....65....1U, 2014AcA....64..177S} and from other catalogs accessed through VizieR \citep{ochsenbein2000vizier}, listed in Appendix \ref{appendix:vizier} we give all literature RRL with common spatial fields to the VVV. From those variable sources, we selected only the RRLs as our positive class and discarded the rest. Then, any other source present in our Tile Catalogs were considered as members of the negative class.

It is worth mentioning that our positive and negative classes cannot be considered as completely accurate as they were not validated with a manual (visual or spectroscopic) inspection of each source. In this context, it is possible that a reduced number of already known RRLs were missed in our cross-matching process and were, as a consequence, incorrectly labeled as negative. These missed RRLs will most probably appear as FP to our classifiers, increasing the estimated error levels of our methods slightly over the levels that would be estimated with a verified dataset.

\begin{figure*}
\begin{center}
    \includegraphics[width=.6\textwidth]{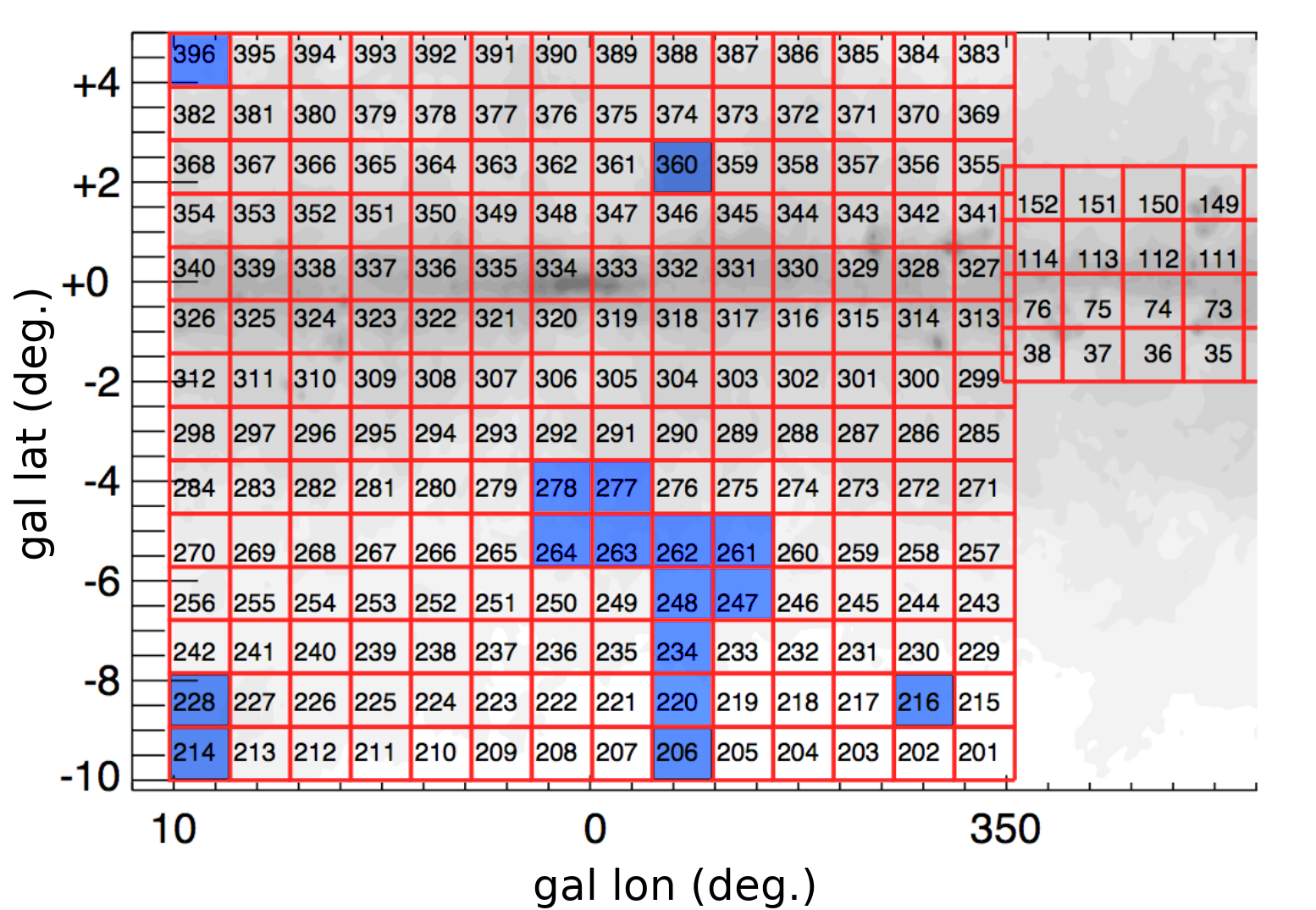}
    \centering \caption{ \label{fig:used_tiles}
    Map of the bulge tiles of the VVV survey. In blue, we mark the tiles used for this paper.}
\end{center}
\end{figure*}

The selected sample of RRLs, Unknown (negative class), and VVV tile-ids are shown in Table~\ref{tab:vars_by_tile} and Fig.~\ref{fig:used_tiles}.

\begin{table}[t!]
\caption[]{Number of epochs in Ks band, RRL, and other stars by tile.}
\begin{center}
\begin{tabular}{l|rrrrl}
\toprule
\textbf{ Tile} & \textbf{Epochs} & \textbf{Size} &  \textbf{RRL} &  \textbf{Unknown }&    \textbf{\%}  \\
\midrule
 b206 &  73 & 157825 &   47 &  157778 &  0.030\% \\
 b214 &  74 & 149557 &   34 &  149523 &  0.023\% \\
 b216 &  73 & 168996 &   43 &  168953 &  0.025\% \\
 b220 &  73 & 209798 &   65 &  209733 &  0.031\% \\
 b228 &  73 & 199853 &   28 &  199825 &  0.014\% \\
 b234 &  73 & 293013 &  126 &  292887 &  0.043\% \\
 b247 &  73 & 406386 &  192 &  406194 &  0.047\% \\
 b248 &  74 & 417839 &  218 &  417621 &  0.052\% \\
 b261 &  74 & 555693 &  252 &  555441 &  0.045\% \\
 b262 &  74 & 573873 &  314 &  573559 &  0.055\% \\
 b263 &  94 & 568110 &  317 &  567793 &  0.056\% \\
 b264 &  94 & 595234 &  307 &  594927 &  0.052\% \\
 b277 &  73 & 718567 &  429 &  718138 &  0.060\% \\
 b278 & 74 & 742153 &  436 &  741717 &  0.059\% \\
 b360 & 74 & 939110 &  669 &  938441 &  0.071\% \\
 b396 & 73 & 486639 &   15 &  486624 &  0.003\% \\
 \hline
 \textbf{Total} &  1216 & 7182646 &  3492 &  7179154 &  0.049\% \\
\bottomrule
\end{tabular}
\end{center}
{\textbf{Note:} Last column shows the proportion of RRLs over other sources, in percentage.}
\label{tab:vars_by_tile}
\end{table}

\subsection{Features extraction}
\label{subsection:features}

The feature extraction process (including the period as previously described) was carried out mostly using \textit{feets} \citep{cabral_fats_2018}.
We obtained a set of sixty-two features, which are summarized in  Table~\ref{tab:cat_features}. A brief description of these features (accompanied by references) can be found in Appendix~\ref{appendixa}. Also, in the following paragraph, we include a more detailed explanation of a particular new set of features considered in this work.

\begin{table*}[tb]
\caption[]{Features selected for the creation of catalogs of \textit{RRL} stars on tiles of \textsc{VVV}}
\centering
\resizebox{.99\textwidth}{!} {
\begin{tabular}{llllllllllll}
\toprule
Features \\
\hline
\midrule
Amplitude & Autocor\_length & Beyond1Std & Con \\
Eta\_e & FluxPercentileRatioMid20 & FluxPercentileRatioMid35 & FluxPercentileRatioMid50 \\
FluxPercentileRatioMid65 & FluxPercentileRatioMid80 & Freq1\_harmonics\_amplitude\_0 & Freq1\_harmonics\_amplitude\_1 \\
Freq1\_harmonics\_amplitude\_2 & Freq1\_harmonics\_amplitude\_3 & Freq1\_harmonics\_rel\_phase\_1 & Freq1\_harmonics\_rel\_phase\_2 \\
Freq1\_harmonics\_rel\_phase\_3 & Freq2\_harmonics\_amplitude\_0 & Freq2\_harmonics\_amplitude\_1 & Freq2\_harmonics\_amplitude\_2 \\
Freq2\_harmonics\_amplitude\_3 & Freq2\_harmonics\_rel\_phase\_1 & Freq2\_harmonics\_rel\_phase\_2 & Freq2\_harmonics\_rel\_phase\_3 \\
Freq3\_harmonics\_amplitude\_0 & Freq3\_harmonics\_amplitude\_1 & Freq3\_harmonics\_amplitude\_2 & Freq3\_harmonics\_amplitude\_3 \\
Freq3\_harmonics\_rel\_phase\_1 & Freq3\_harmonics\_rel\_phase\_2 & Freq3\_harmonics\_rel\_phase\_3 & Gskew \\
LinearTrend & MaxSlope & Mean & MedianAbsDev \\
MedianBRP & PairSlopeTrend & PercentAmplitude & PercentDifferenceFluxPercentile \\
PeriodLS & Period\_fit & Psi\_CS & Psi\_eta \\
Q31 & Rcs & Skew & SmallKurtosis \\
Std & c89\_c3 & c89\_hk\_color & c89\_jh\_color \\
c89\_jk\_color & c89\_m2 & c89\_m4 & n09\_c3 \\
n09\_hk\_color & n09\_jh\_color & n09\_jk\_color \\
n09\_m2 & n09\_m4 & ppmb \\
\bottomrule
\end{tabular}
}
\vspace{6pt}
\label{tab:cat_features}
\end{table*}


Aside from the period, magnitude variation, and LC morphology, variable stars may also be characterized by their temperatures or intrinsic color(s) as a function of time. A color may be calculated as the difference of two standard broad-band magnitudes with one important caveat: colors are sensitive to LOS or external foreground reddening. In fact, its effect increases towards the mid-plane of the Galaxy. At visual wavelengths that are within our range of tiles, this is a serious problem because an absorption of several magnitudes in the $V-band$ is observed near the galactic mid-plane towards the bulge of the Galaxy. Extinction depends on dust grain type, size, spacing, and density, but generally affects shorter wavelengths preferentially, hence causing reddening . We correct the VVV data using different extinction laws and maps obtained from the literature. We also include reddening-free indices that are photometric indices chosen to be relatively insensitive to the effect of extinction. For further information, see \citet{catelan2011vista}.

This work was planned and partly executed prior to the \cite{2014A&A...566A.120S} 3D maps and the incorporation into the updated version of the BEAM-II calculator that provides 3D distances. Thus, to maintain a homogeneous analysis across all fields, we chose to correct for reddening by using the 2D dust-maps from BEAM-I and dust laws employing the work of \cite{gonzalez2011reddening} and \cite{gonzalez2012reddening} for which we estimate the absorption coefficients of our sources with the \textit{BEAM-I - A VVV and 2MASS Bulge Extinction And Metallicity Calculator}  \footnote{\url{http://mill.astro.puc.cl/BEAM/calculator.php}}. Unfortunately the extinction maps of the \textit{VVV(X)} zones were not available when we developed this work, so in this paper, we limit ourselves to the VVV tiles.

The \textit{BEAM} mappings have pixel resolutions of 2 arcmin x 2 arcmin or 6 arcmin x 6 arcmin, depending on the density of the red-clump bulge stars in the field. In some cases, \textit{BEAM} fails to calculate the extinction to a source. In those situations, we replace it with the average of the extinction of the hundred nearest neighbors, weighed by angular proximity.

Thus, for each source we obtain the $A_{v}$ absorption magnitude and use it to calculate the reddening and convert it from the photometry of the \textit{2MASS} survey  \citep{skrutskie2006two} to the photometry of VVV \citep{gonzalez2017vista}, given the two laws of extinction, \cite{cardelli1989relationship} and \cite{nishiyama2009interstellar}. These laws describe the amount of extinction as a function of wavelength.

Finally, the last stumbling block in calculating color is that, unfortunately, the VVV multi-band observations were only made in the first epoch systematically across all fields. Obtaining a good color-index is not possible because, first, the classic calculation of color index (implemented in \textit{feets}) cannot be used, since it requires several observations in both bands to subtract; and second, because the first epoch of two light curves may have been observed in different phases, and the relations between the subtraction of two points of different phases in two bands, even for the same star, is not constant.

Under these conditions, the following strategy was chosen:
The colors were calculated using only the first observation epoch and both extinction laws; in addition, a pseudo phase multi band was calculated, which is related to where in the phase is the first epoch of the source, also including the number of epochs; finally, the pseudo-magnitudes and pseudo-colors (with both extinction laws) proposed in the work of \cite{catelan2011vista} were calculated. These values are reddening free indices.

The following color features were obtained. The reddening free indices are shown with subscript ``m2'', ``m4'', and ``c3'':

\begin{description}
\item[\texttt{c89\_c3} -] $C3$ Pseudo-color using the \cite{cardelli1989relationship} extinction law \citep{catelan2011vista}.

\item[\texttt{c89\_\textit{ab}\_color} -] Extinction-corrected color from the first epoch data between the band $a$ and the band $b$ using the  \cite{cardelli1989relationship} extinction law, where $a$ and $b$ can be the bands $H$, $J,$ and $K_s$.

\item[\texttt{c89\_m2, c89\_m4} -] $m2$ (defined as $H - 1.13 (J - K_s)$) and $m4$ (defined as $K_s - 1.22 (J - H)$) pseudo-magnitudes using the \cite{cardelli1989relationship} extinction law \citep{catelan2011vista}.

\item[\texttt{n09\_c3} -] $C3$ Pseudo-color using the \cite{nishiyama2009interstellar} extinction law o  \citep{catelan2011vista}.

\item[\texttt{n09\_\textit{ab}\_color} -] Extinction corrected color from the first epoch data between the band $a$ and the band $b$ using the  \cite{nishiyama2009interstellar} extinction law, where $a$ and $b$ can be the bands $H$, $J$ and $K_s$.

\item[\texttt{n09\_m2, n09\_m4}] $m2$ and $m4$ pseudo-magnitudes using the \cite{nishiyama2009interstellar} extinction law  \citep{catelan2011vista}.

\item[\texttt{ppmb} -] ``\textit{Pseudo-Phase Multi-Band}''. This index sets the first time in phase with respect to the average time in all bands, using the period calculated by \textit{feets}.

$$
PPMB = frac(\frac{|mean(HJD_H, HJD_J, HJD_{K_s}) - T_0|}{P}),
$$

where $HJD_H$, $HJD_J$ and $HJD_{K_s}$ are the time of observations in the band $H$, $J$ and $K_s$; $T_0$ is the time of observation of maximum magnitude in the $K_s$ band; $mean$ calculates the mean of the three times, $frac$ returns only the decimal part of the number, and $P$ is the extracted period.

\end{description}

\subsection{Evaluation of feature subsets}
\label{subsection:feat_analysis}

We consider three subsets of features based on the root of feature type. We bundle all color-derived features into the "color feature" sub-type. Features that are related to the extracted period are placed into the "period based" sub-type, including, for example,  \texttt{PeriodLS}, \texttt{Period\_fit}, \texttt{Psi\_eta}, \texttt{Psi\_CS}, \texttt{ppmb} and all the Fourier components (See Appendix~\ref{appendixa}). This subset depends on the correct determination of the period as a first step, which makes them less reliable than features that are measured directly from the observations. Then, in the third and final subset, we include all classical features with this last property that we call "magnitude-based" features.

We designed a first experiment and established a handle on the predictive power of our selected feature sets and restricted subsets. We selected four tiles in the VVV Bulge footprint; $b234$, $b261$, $b278,$ and $b360$, as a compromise between RRL number density and coverage. For each tile, we created a reduced dataset with all the RRL stars and 5,000 unknown sources. We use RF as classifier, with the setup explained in the next section (we will also support this decision in that section). For every tile, we train four classifiers: one with the full set of features, another with only the period + magnitude subsets, a third with the color + magnitude subsets and a fourth with the period + color feature subsets. Every classifier was then tested using all other tiles as test sets (obviously using the same feature subsets).

\begin{figure*}
\centering
    \includegraphics[width=.99\textwidth]{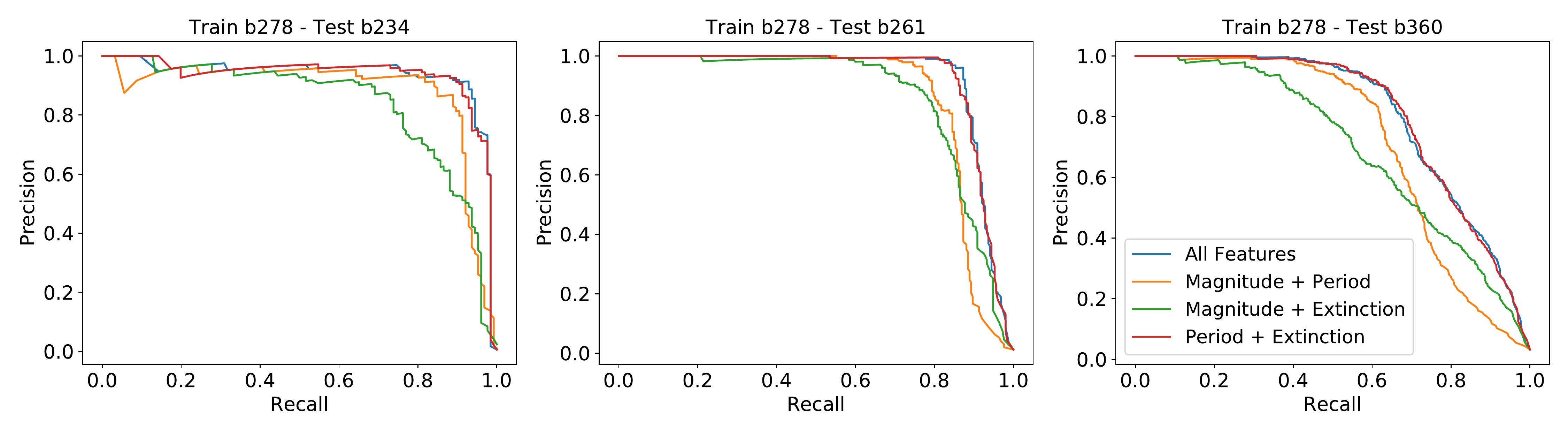}
    \caption{Recall vs. precision curves of trained RF classifiers for different feature subsets. On the left, center, and right panels, the test tiles are $b234$, $b261$, and $b360$, respectively. All models were trained on tile $b278$.
    \label{fig:s_features:curve}}
\end{figure*}

\begin{figure*}
\centering
    \includegraphics[width=.99\textwidth]{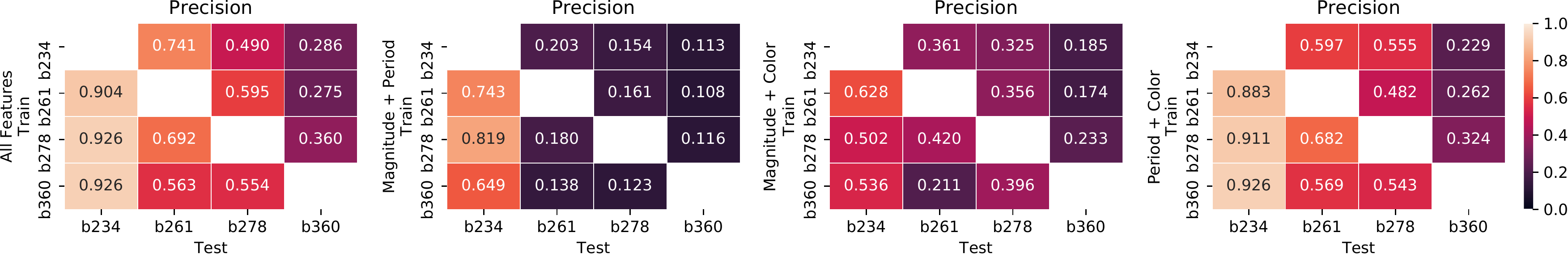}
    \caption{Precision for different subsets of features for the tiles $b234$, $b261$, $b278,$ and $b360$ for a fixed recall ($\sim0.9$). From left to right: the first panel includes the total set of features; in the second, only the magnitude- and period-based features is used; in the third, the magnitude- and extinction-based ones; and, finally, the extinction- and period-based features are used. In all cases, the rows indicate which tile was used for train the RF classifier and the columns indicate the one used for the test.
    \label{fig:s_features:heatmap}}
\end{figure*}

First, we evaluate the results using precision--recall curves.
While all the validation curves can be found in Appendix~\ref{appendix:feat_analysis}, we present the curves trained only with the tile $b278$ in Fig.~\ref {fig:s_features:curve}. We selected $b278$ not only because it is the one with the most RRL, but also for historical reasons; Baade's window is centred therein. It can be seen in Fig.~\ref{fig:s_features:curve} that the Period subset of features contains most of the information since  its removal causes the greatest loss in performance for all test cases (the green curve corresponding to magnitude + color is always the lowest and furthest from the 1-1 upper-right corner point). However the Color subset of features is also significantly relevant since its removal produces considerable loss to performance (but to a smaller degree than does the removal of the Period subset). Finally, the Magnitude subset is the least informative of all subsets. Its removal produces marginal performance loss, which is somewhat expected if we consider RRL as standard candles that are isotropically distributed about the bulge, the typical RRL's apparent magnitude should be weakly correlated to their class type. The full set of curves shows the same behaviour (see  Appendix~\ref{appendix:feat_analysis}).

In order to make a quantitative comparison, we selected a fixed value of Recall ($\sim0.9$) and produced tables with the corresponding precision values for a RF classifier trained with each subset of features. Figure~\ref {fig:s_features:heatmap} shows that in all cases, the results are consistent with the previous qualitative analysis as far as the relative importance of our RRL feature subtype.

\section{Model selection}
\label{section:model_selection}

In a second experiment, we compared the performance of the four classifiers considered in this work: SVM with linear kernel, SVM with RBF kernel, KNN and RF.

Most of the classifiers have hyperparameters that need to be set to optimal values. We carried out a grid search considering all possible combinations of values for each hyperparameter over a fixed list. We used a 10 k-folds setup on tile $b278$, considering the precision for a fixed recall of $\sim 0.9$ as the performance measure.

With this setup, we selected the following hyperparameters values:

\begin{description}
    \item[SVM-Linear:] $C = 50$.
    \item[SVM-RBF:] $C = 10$ and $\gamma = 0.003$.
    \item[KNN:] $K = 5$ with a $manhattan$ metric; also, the importance of the neighbor class wasn't weighted by distance.
    \item[RF:] We created $500$ decision trees with Information-Gain as metric, the maximum number of random selected features for each tree is the $log_2$ of the total number of features, and the minimum number of observations in each leaf is $2$.
\end{description}

\begin{figure*}
\centering
    \includegraphics[width=.99\textwidth]{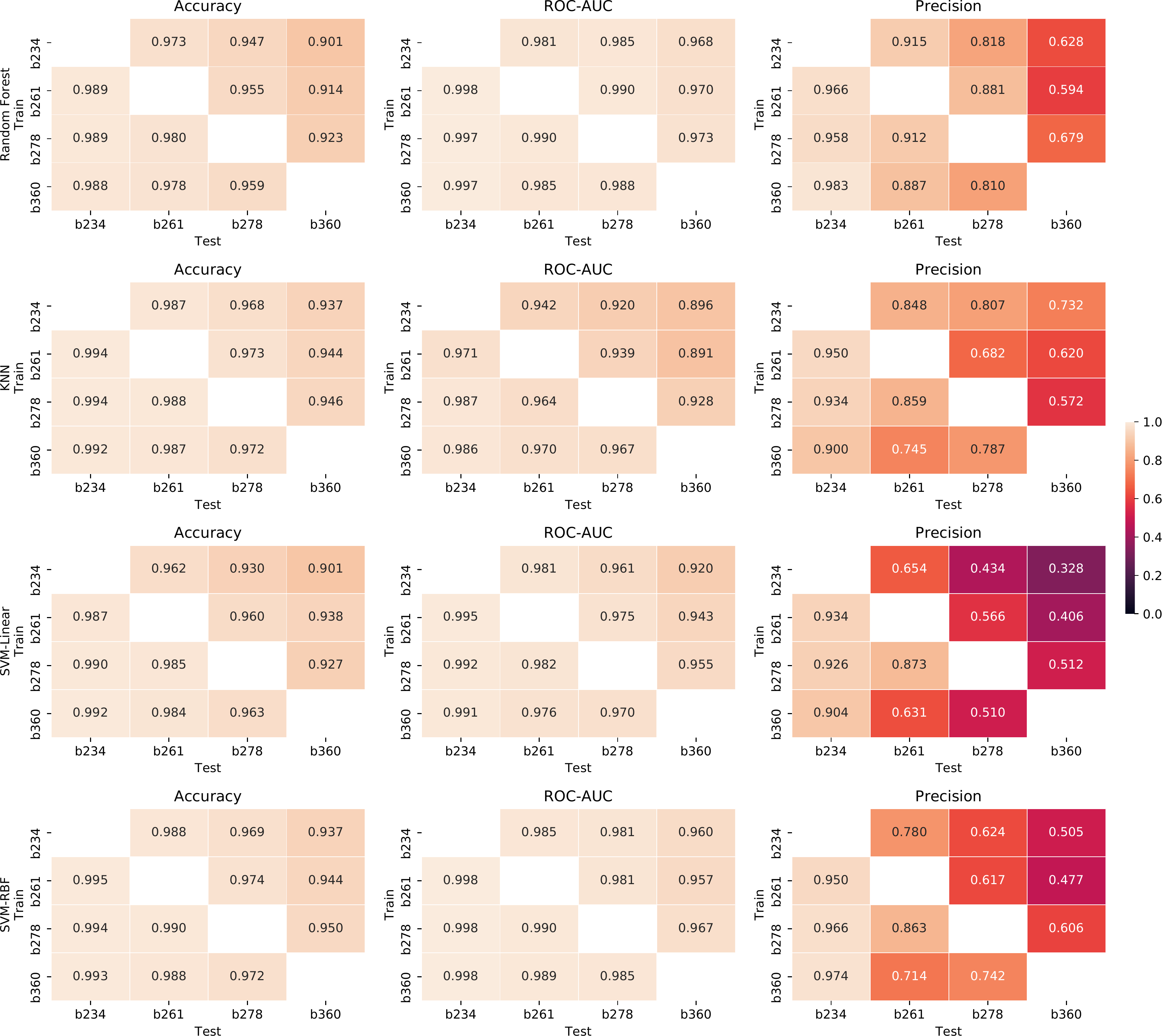}
    \caption{Accuracy, ROC-AUC, and precision for different classifiers. All details are the same as in Fig.~\ref{fig:s_features:heatmap}.}
    \label{fig:s_models:heatmap}
\end{figure*}

We trained the four classifiers using the same datasets and general setup as in the previous experiments. Again, we selected a threshold for each classifier that results in a fixed Recall of $\sim0.9$. Figure~\ref{fig:s_models:heatmap} presents the results of the experiment. In this case, we show (in addition to the Precision), the ROC-AUC and the accuracy (for the same threshold as precision).

\begin{figure*}
\centering
    \includegraphics[width=.99\textwidth]{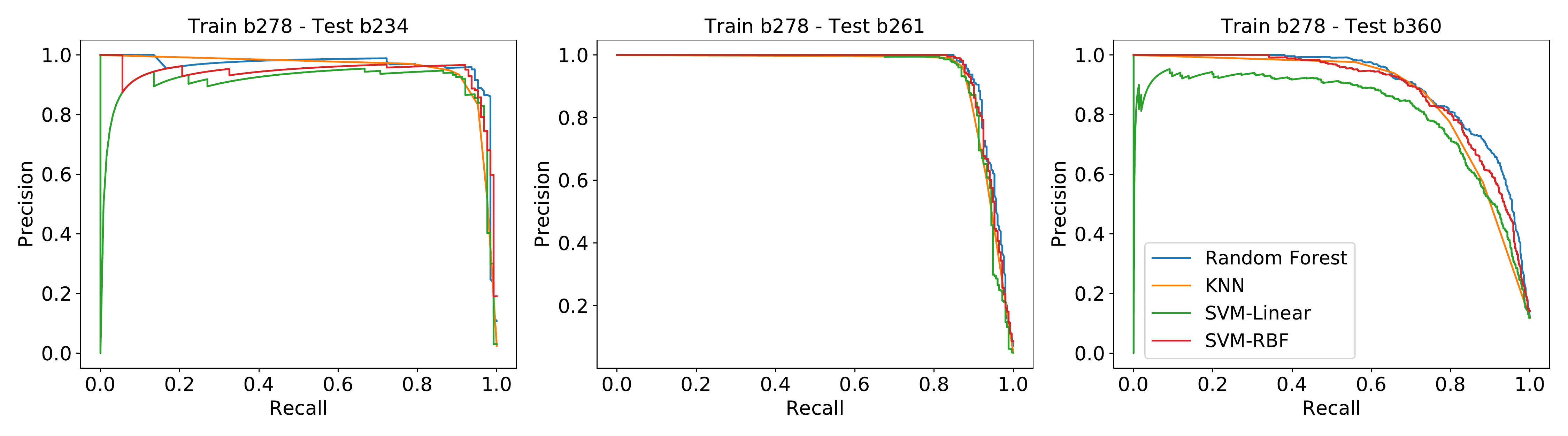}
    \caption{Recall vs. recision curves for different classification methods. In the left, center, and right panel, the test tiles are $b234$, $b261,$ and $b360$, respectively.  All models were trained on tile $b278$.
    \label{fig:s_models:curve}}
\end{figure*}

 In the first place, all classifiers have a similar (and quite high) accuracy and ROC-AUC for all test and train cases. Both measurements give the same weight to positive and negative classes and, as a  consequence, both are dominated by the overwhelming majority of negative class. The comparison with Precision values (with very different values for different cases) shows why both measurements are not informative in the context of this work.
When analysing, then, only the precision results, the best method is clearly RF, followed by KNN, SVM-RBF, and, finally, SVM-Linear.

In Fig.~\ref{fig:s_models:curve}, we show the complete precision vs. recall curves for tile $b278$ (the complete set of curves for all tiles can be found in the Appendix~\ref{appendix:model_selection}). All figures show that in any situation RF is the best method (or comparable to the best with few exceptions). Given these results, we proceed to selecting RF as the choice classifier for the remainder of this work.


\section{Sampling size and class imbalance}
\label{section:sampling}
\begin{figure*}
\centering
    \includegraphics[width=.97\textwidth]{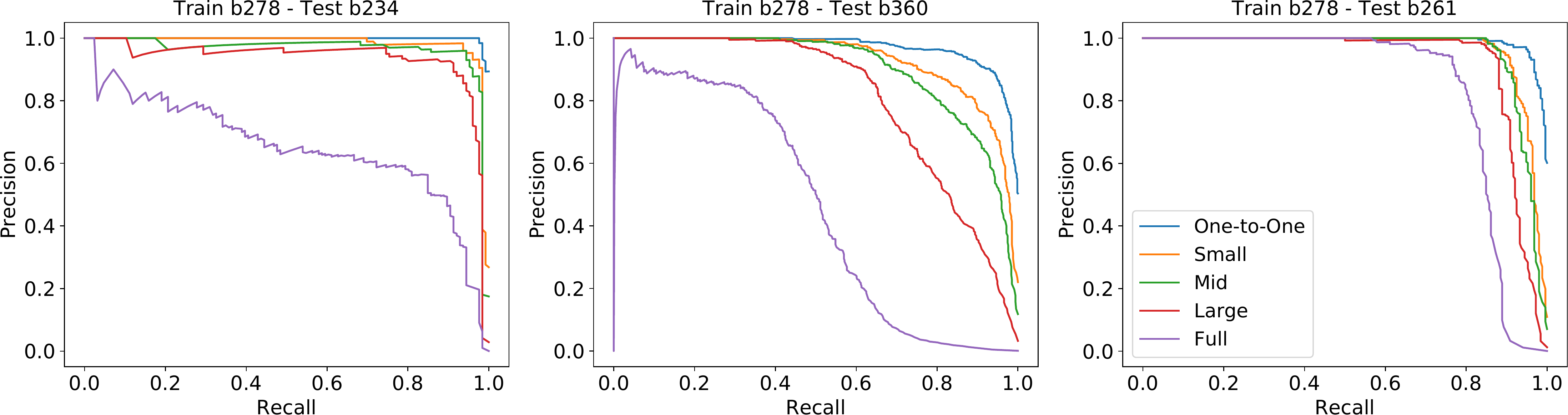}
    \caption{Recall vs. precision curves for different imbalance levels. All models were trained on tile $b278$. The left, center, and right panels are for the test tiles: $b234$, $b261,$ and $b360$, respectively. Test and corresponding training sets have the same imbalance level.
    \label{fig:s_unbalance:same_size_curve}}
\end{figure*}

The treatment of highly imbalanced datasets posses several problems to learning, as is discussed, for example, in a recent work by \citet{hosenie2020imbalance}. In particular, there is a problem that is usually not considered by practitioners which considers whether learning be improved by changing the balance between classes. Or whether, by changing the imbalance, we can validate a real performance boost.

Among the various strategies in the literature, \citet{japkowicz2002class} provides a scheme that is appropriate to our problem, achieved by subsampling on the negative class, as we already implemented in the first two experiments. Aside from a windfall in gained sensitivity towards the objective class, reducing the sample size is desirable as it always leads to a considerable decrease in the computational burden for any ML solution. This scheme is implicitly used in most previous works; only a small sample of the negative class is considered \citep{pashchenko2017MNRAS, Mackenzie2016ApJ}.

In our third experiment, we evaluate the relationship between the size of the negative class and the performance of the classifiers. We use RF as classifier, with the same setup as the previous section. We also use the same four tiles and evaluation method (we train RF on one tile and test it on the other three, using precision--recall curves).

We considered five different sizes for the negative class. First, we used all sources available ("full sample"). Then we considered three fixed size samples: 20000 sources ("large sample"), 5000 sources ("mid sample"), and 2500 sources ("small sample"). Finally, we considered a completely balanced sample, equal to the number of RRLs in the tile ("one--to--one sample"). Each sample was taken from the precedent one so for example, the small sample for tile $b278$ was taken from the mid sample of the same tile. In all cases, we used all the RRLs in the corresponding tile.

Figure~\ref{fig:s_unbalance:same_size_curve} shows the results of the experiment for tile $b278$ (results for all datasets in all the experiments in this section can be found in Appendix~\ref{appendix:unbalance}). The figure shows that the one-to-one sample is clearly superior in performance than are the imbalanced samples: there is a clear advantage in using balanced datasets over highly imbalanced ones and not only in performance, but also in running time.

There is a hidden flaw in our last results. We compared curves taken from datasets with different imbalances. If we apply a one--to--one model to predict a new tile, it will have to classify all the unknown sources, and not only a sample taken from them. In that situation, the number of FP will certainly increase from what the model estimated using a reduced sample. A fairer comparison involves the use of a complete tile to estimate our performance metrics, or at least the use of a corrected estimate of the performance that takes into account the proportions in the evaluated sample and on the full sample.

If we assume that the training sample is a fair sample (even though the RRL number densities may vary from tile to tile) within a specific tile, say b278, we could argue that local densities will be conserved and, as a consequence, for each FP in the sample, there should be a higher number of incorrect sources in the full dataset, with a proportion inverse to the sampling proportion. This reasoning leads to the equation:

\begin{equation*}
    FP^* = FP \times \frac{S_F}{S_R} ,
\end{equation*}

\noindent where FP\textsuperscript{*} is the estimate of the real number of FP, FP is the number measured on the reduced sample, \textit{S\textsubscript{F}} is the size of the Full sample, and \textit{S\textsubscript{R}} is the size of the reduced sample. Using this value, we can produce a corrected estimate of the Precision on the sample:

\begin{equation}
    P^* = \frac{TP}{TP + FP^*} .
    \label{eq:P*}
\end{equation}

\noindent where P\textsuperscript{*} is the estimation of the corrected precision and \textit{TP} is the value estimated in the reduced sample. On the other side, recall values only use the positive class for its estimation and do not need a sample size correction.

\begin{figure*}
\centering
    \includegraphics[width=.99\textwidth]{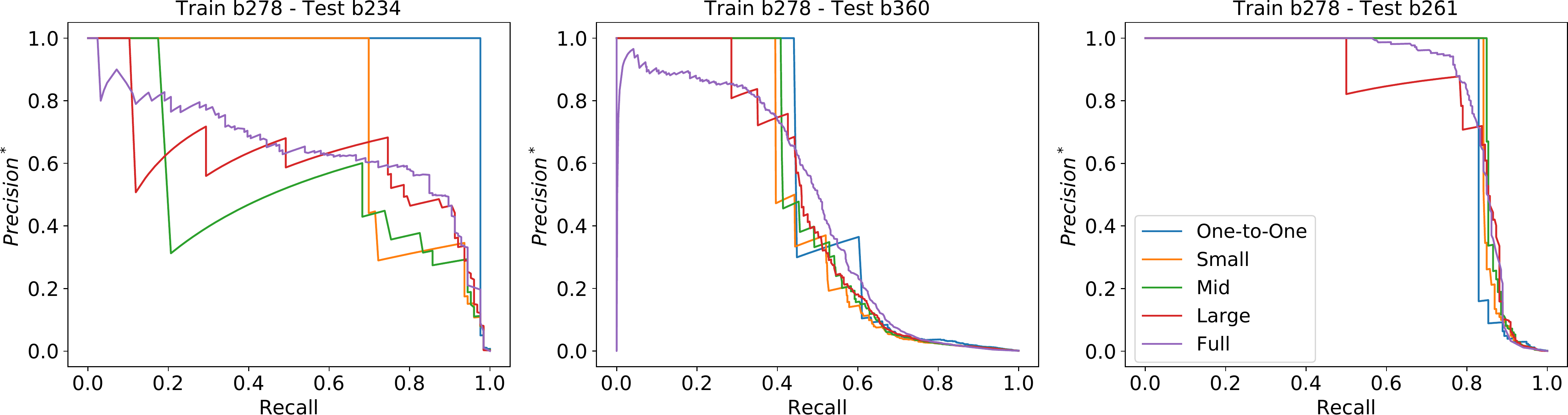}
    \caption{Recall vs. corrected precision (P*) curves for different imbalance levels. In the left, center, and right panels, the test tiles are $b234$, $b261,$ and $b360$, respectively. All models were trained on tile $b278$. Test sets have the same imbalance level than the corresponding training set.
    \label{fig:s_unbalance:same_size__pstar_curve}}
\end{figure*}

In Fig.~\ref{fig:s_unbalance:same_size__pstar_curve}, we show "corrected precision" recall curves, using Equation \ref{eq:P*}, for the same classifiers as in the previous figure. The correction seems to be a good approximation only in the low Recall regions, where the number of FP and TP is always bigger. For higher thresholds, when TP and FP become smaller, there is a clear "discrete" effect in the correction and it becomes useless.

\begin{figure*}
\centering
    \includegraphics[width=.99\textwidth]{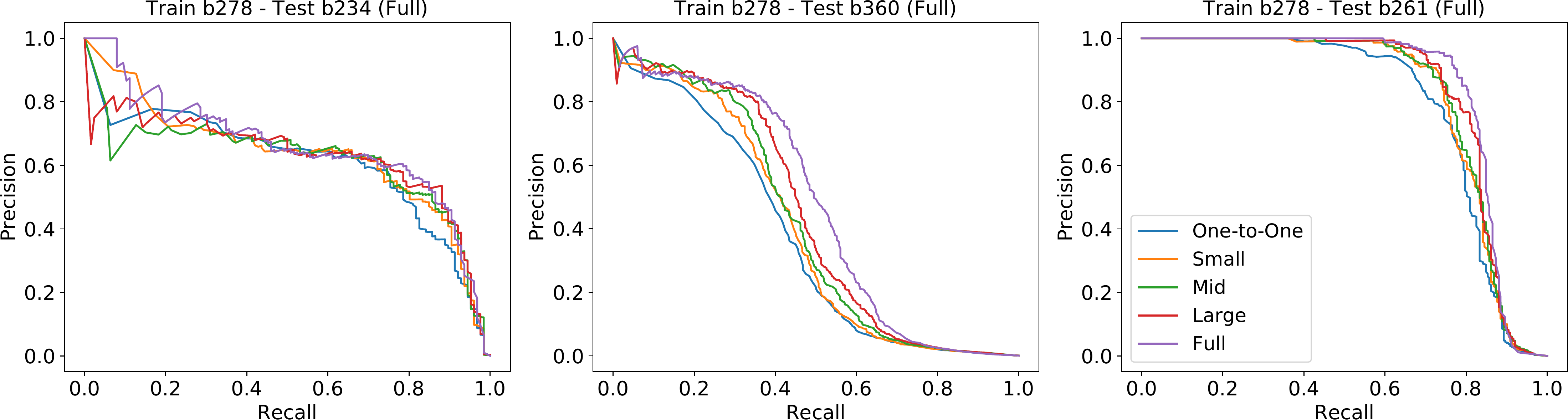}
    \caption{Recall vs Precision curves for different imbalance levels, using always test sets of full size. In the left, center and right panel the test tiles are $b234$, $b261$ and $b360$. All models were trained on tile $b278$.
    \label{fig:s_unbalance:diff_size_curve}}
\end{figure*}

Therefore, the only valid method to compare the different samplings is to estimate performances using the complete tiles, with the associated computational burden. Figure~\ref{fig:s_unbalance:diff_size_curve} shows the corresponding results. The performances of the diverse sampling strategies are mostly similar, as the only full sample results that are consistently better than the rest. Our results show that when there are enough computational resources, the best modeling strategy is to use all data available.

\section{Setting the final classifier}
\label{section:final}

\begin{figure*}
\centering
    \includegraphics[width=.95\textwidth]{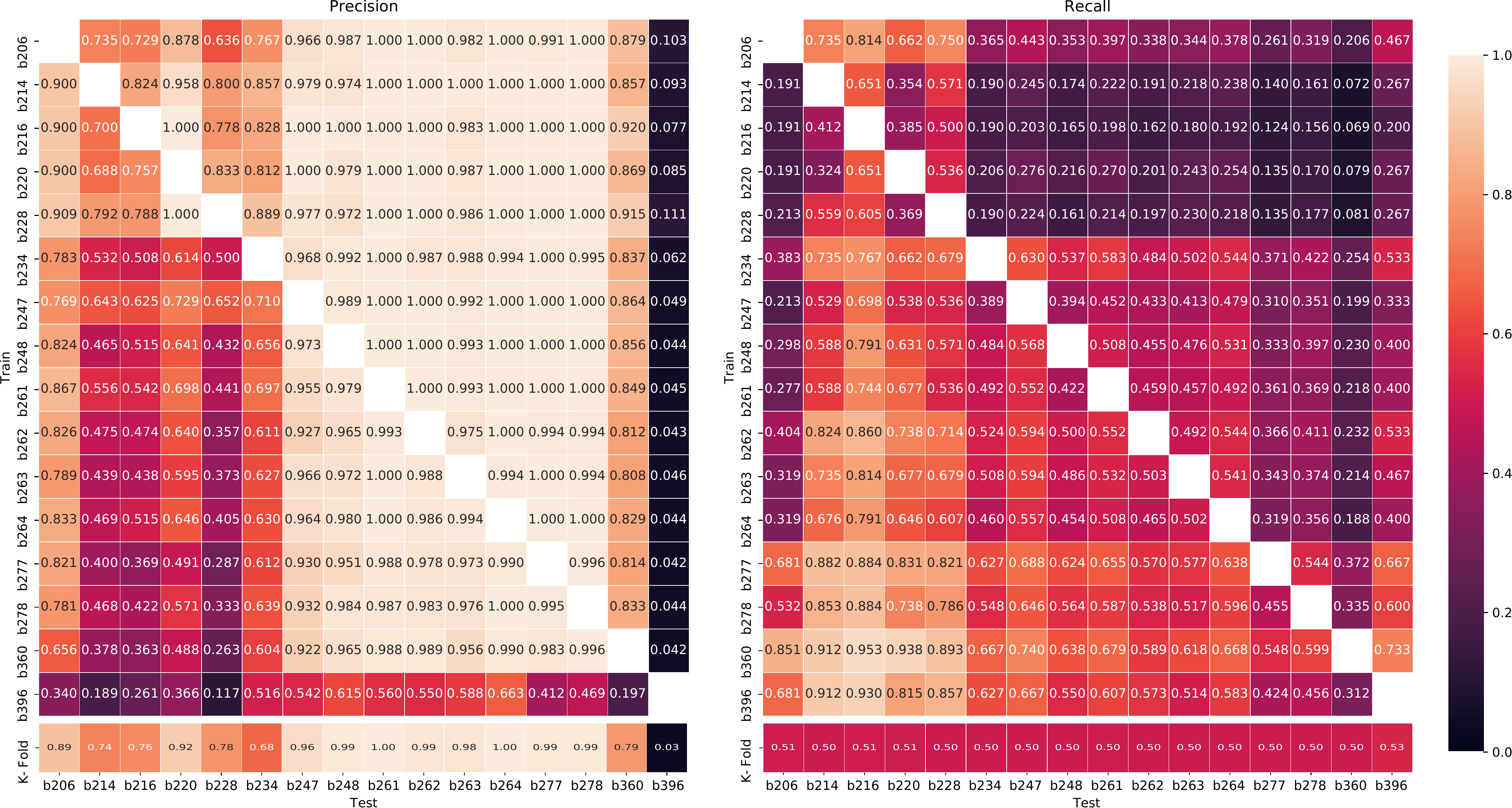}
    \caption{Tables of precision and recall, training and testing with complete data
of the 16 tiles. The last (separated) row shows the estimated precision (left)
achieved by selecting a threshold value that sets the recall to $\sim0.5$ (right), training and testing on each
tile with a 10 K-Folds procedure. The upper rows shows the recall and precision values observed on the test set (columns) when training on diverse tiles (rows), using the thresholds selected by   10 K-Folds (used on the last row).}
    \label{fig:s_kfold:kfolds_heatmap}
\end{figure*}

Throughout the previous sections, we show that the best modeling strategy is to use a RF classifier with all the features available and training and testing on complete tiles. We applied this recipe to all the 16 tiles included in our study. Fig.~\ref{fig:s_kfold:kfolds_heatmap} shows the corresponding results. There are very diverse behaviours, depending on the training and test tiles. Overall, it seems that we can select a recall value of 0.5 and find precision values over 0.5 in almost all cases. It means that we can expect, in a completely automatic way, to recover at least half of the RRLs in a tile paying the price of manually checking a maximum of a false alarm for each correct new finding.

\begin{figure*}
\centering
    \includegraphics[width=.95\textwidth]{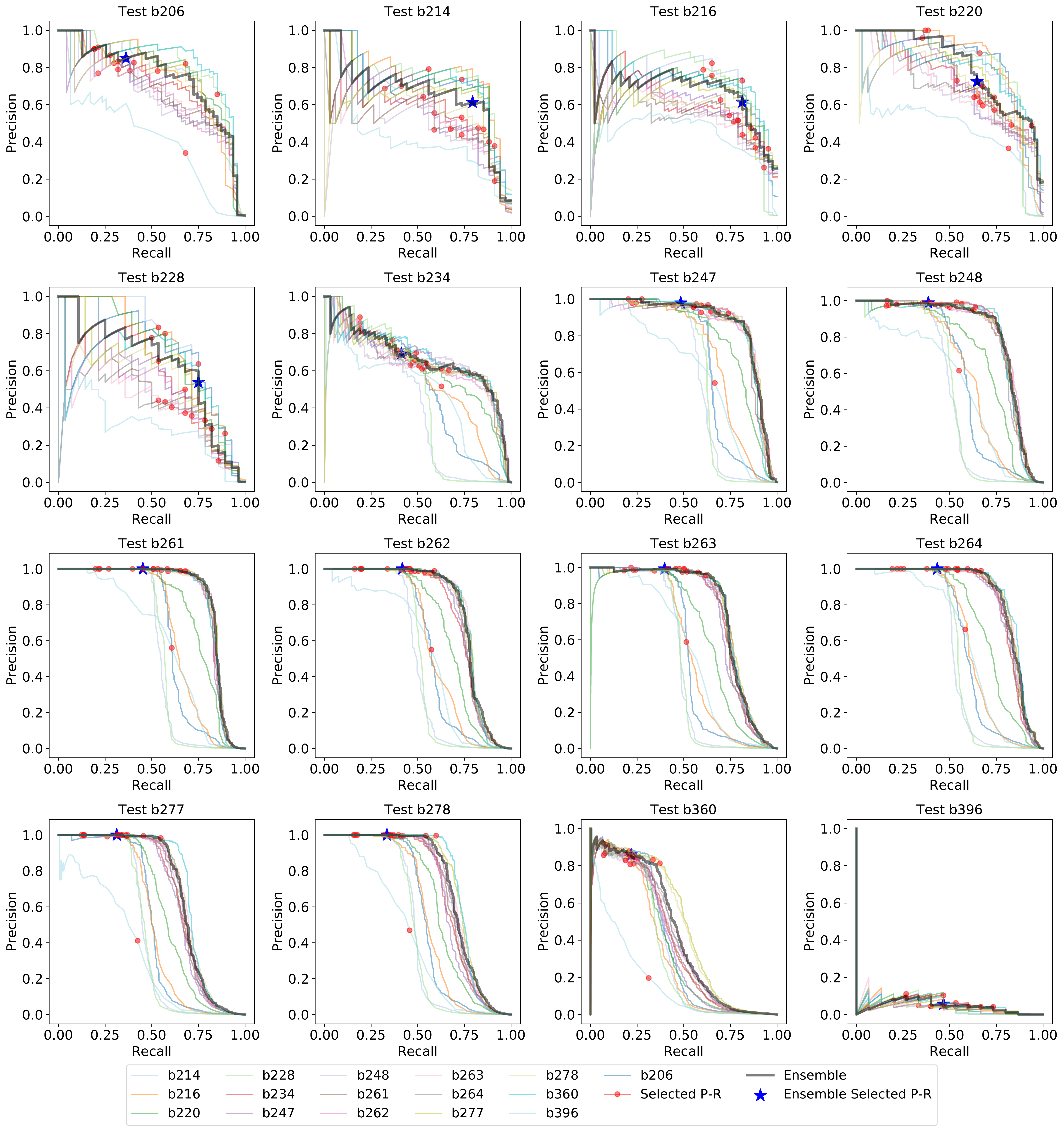}
    \caption{Precision vs. recall curves for the final classifiers on each test tile. In each panel, the colored curves represent a classifier trained with a tile, while the black curve is the ensemble classifier. Red dots correspond to a precision and recall fixed at a threshold equivalent to the recall $\sim0.5$ when training and testing on the same tile with 10 k-folds. The blue stars represent the working point selected by averaging the thresholds corresponding to the red points.}
    \label{fig:s_kfold:ensemble_curves}
\end{figure*}

There is a last problem to be solved. In a real situation, we do not have the correct labels and we cannot fix  the threshold directly to secure the desired recall level. We need to set the expected recall value using the training data, and assume it will be similar to the recall that is actually observed in the test data. We used a 10-fold CV procedure to estimate the performance of our classifiers on training data only and selected the corresponding thresholds that leads to a recall of $\sim0.5$. In Fig.~\ref{fig:s_kfold:ensemble_curves}, we show the full results of our final individual classifiers. Each panel shows the prediction of a different tile using all the classifiers trained on the other tiles. We added a red point to each curve showing the observed recall value on each test set for the thresholds set by 10 folds CV. We find that in most situations, our procedure leads to real recall values in the 0.3 to 0.7 zone.

On the other hand, another clear result from Figs.~\ref{fig:s_kfold:kfolds_heatmap} and \ref{fig:s_kfold:ensemble_curves} is that classifiers trained over different tiles produce very diverse results on a given tile and, of course, we cannot know in advance which classifier will work better for each tile. A practical solution to this problem is the use of ensemble classifiers \citep{rokach2010ensemble, granitto2005neural}. An ensemble classifier is a new type of classifier formed by the composition of a set of (already trained) individual classifiers. In order to predict the class of a new source, the ensemble combines the prediction of all the individual classifiers that form it.
The ensemble is in fact using information from all tiles at the same time with no additional computational cost. This additional information can produce improved performances in some cases and, more importantly, it usually produces more uniform results over diverse situations (diverse tiles).
In our case, we combined all classifiers trained on the diverse tiles in order to make predictions of a new tile. In this evaluation phase, we exclude the classifier trained on the tile that is being tested. Our combination method is to average over all classifiers (with equal weight) the estimated probabilities of being an RRL. The threshold for this ensemble classifiers is taken as the average of individual thresholds.

\begin{figure*}
\centering
    \includegraphics[width=.95\textwidth]{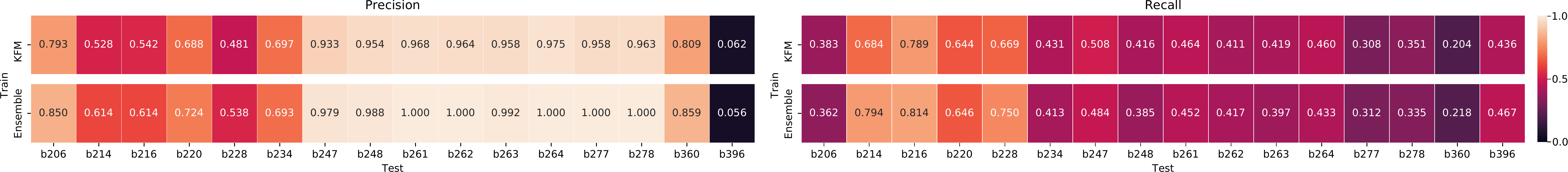}
    \caption{Precision (left) and recall (right) values of the ensemble classifier (see text). For comparison, we include in the first row the average of the values or each individual classifier in the ensemble, as shown in Fig.~\ref{fig:s_kfold:kfolds_heatmap}.}
    \label{fig:s_kfold:ensemble_heatmap}
\end{figure*}

Figure~\ref{fig:s_kfold:ensemble_curves} also shows the results of the ensemble for all tiles in the paper as a bold black line. It is clear from the figure that the ensemble is equal or better, on average, than any of the individual classifiers. In Fig.~\ref{fig:s_kfold:ensemble_heatmap}, we show the specific recall and accuracy values for each tile using the ensemble classifier with the previous setup. We also include the average of each column of Fig.~\ref{fig:s_kfold:kfolds_heatmap} as a comparison (the average of all classifiers). It is clear from the figure that our automatic procedure produce a classifier that works well in all tiles and produce the expected result of recall and precision (all but one recall values are bigger than 0.3 and all but one precision values are bigger than 0.5).

\section{Analysis of misidentifications}
\label{section:fpfn}

\begin{figure*}
\centering
    \includegraphics[width=.99\textwidth]{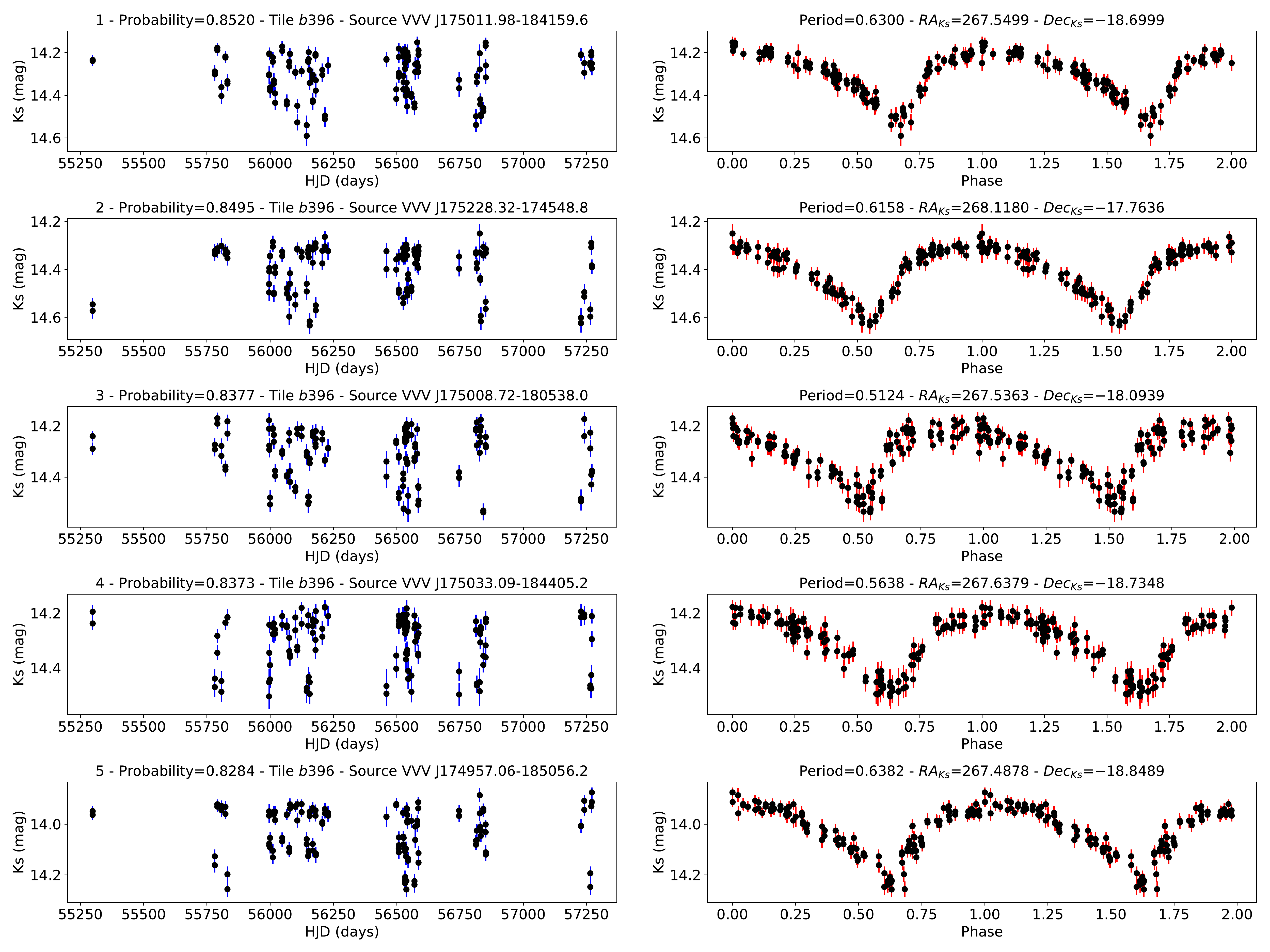}
    \caption{Light curves in time (blue on the left) and in two phases (red on the right) for the five FPs with the highest probability of being an RRL. The titles of each left panel shows probabilities, tile, and identifier of the source. The titles on the right panel shows the estimated period, right ascension ($RA_{Ks}$), and declination ($Dec_{Ks}$).}
    \label{fig:s_lcs:fp}
\end{figure*}

\begin{figure*}
\centering
    \includegraphics[width=.99\textwidth]{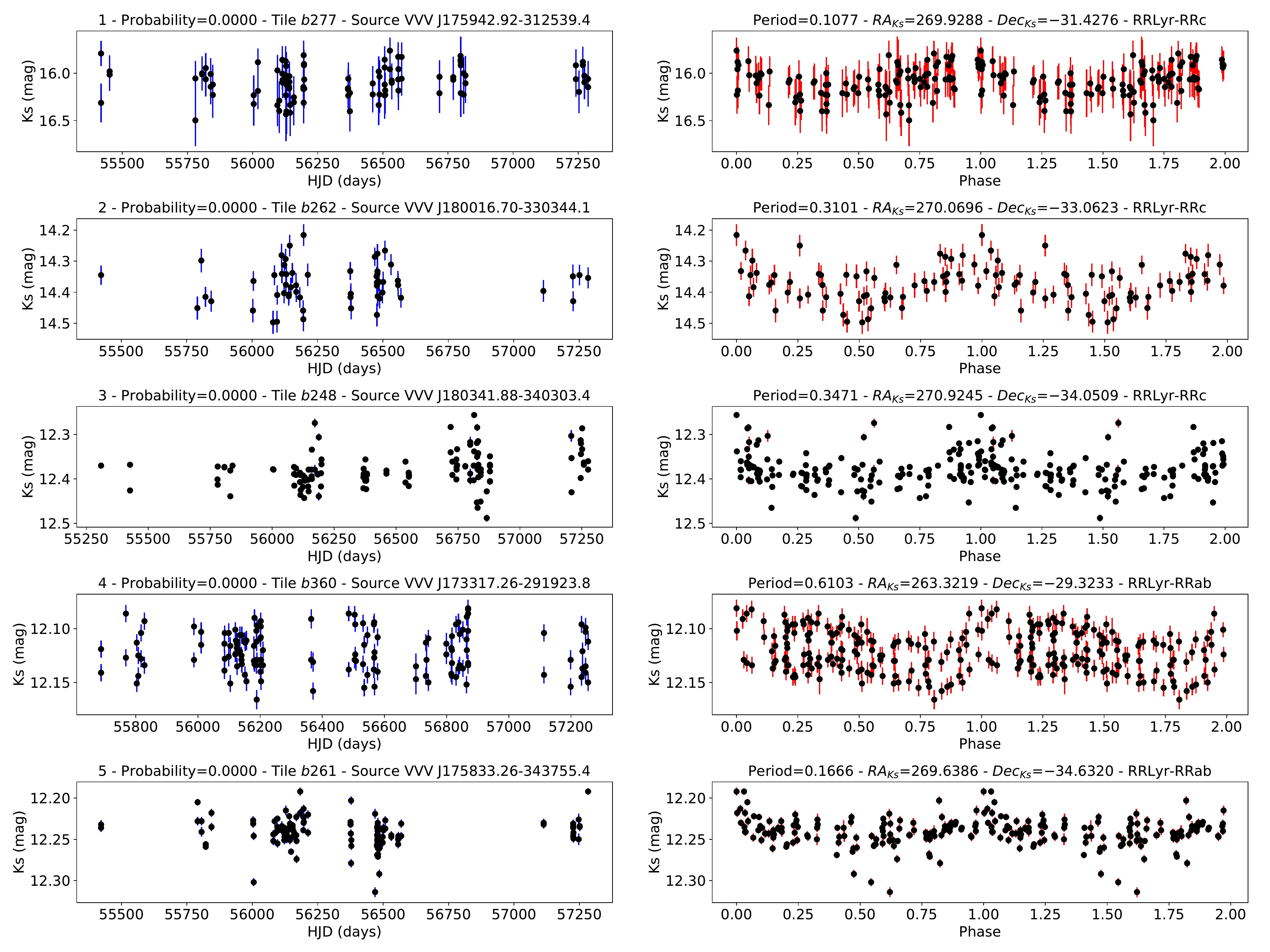}
    \caption{
    Same as Fig.~\ref{fig:s_lcs:fp} for the five FNs with the lowest probability of being an RRL}
        \label{fig:s_lcs:fn}
\end{figure*}

There are two classes of possible errors (misidentifications) for our ensemble classifier. First, there are FPs. These are sources that are identified as RRLs by our method but are not registered as so in the catalogs (variable stars discarded in another survey or present in new catalogs that were not taken into account in our work). In Fig.~\ref{fig:s_lcs:fp}, we show the light curves corresponding to the five FPs identified with the highest probabilities. It is clear that all five are potential RRLs, with clear periodicity. Most FPs are similar to these and based on this evidence, we consider them as RRL candidates more than errors.

Second, there are FNs. These are already known RRLs that our pipeline assigned to the unknown-star class. In Fig.~\ref{fig:s_lcs:fn} we show the light curves corresponding to the five FNs identified with the lowest probabilities.
These low-probability FNs are stars for which available observations on VVV were not adequate as to find the periodicity in the light curves for a correct features extraction. Typically, these errors are more related to the limitations on the feature extraction process than to limitations of the classifier itself.

We also found another distinct group of FNs with the help of an innovative use of ML methods. We devised a new experiment looking for explanations about why some RRLs are misidentified. Our hypothesis is that if there are some patterns that clearly distinguishes FNs and TPs (RRLs that are correctly and incorrectly identified), a good classifier will be able to learn the patterns.

We started by creating a dataset to learn. To achieve this we trained a RF with the complete data of all available tiles. RF has a simple and unbiased way to estimate the classification probabilities of training samples that is equivalent to a K-fold procedure, called Out-of-Bag (OOB) estimation \citep{breiman2001random}. We set the threshold of our RF to obtain an OOB Recall $\sim 0.9$. Next, we used this RF to OOB-classify all RRLs in our dataset and split them in two classes: TPs or FNs. At this point, we create a reduced dataset of only RRLs, with a new positive class of those correctly identified in the last step, and a new negative class with the rest of the RRLs. Finally, we train a new RF on this reduced dataset, trying to differentiate between these two kind of RRLs. We find that this RF has an OOB error level well below random guessing, showing that there are patterns that discriminate these two classes.

\begin{figure}
\centering
    \includegraphics[width=.99\columnwidth]{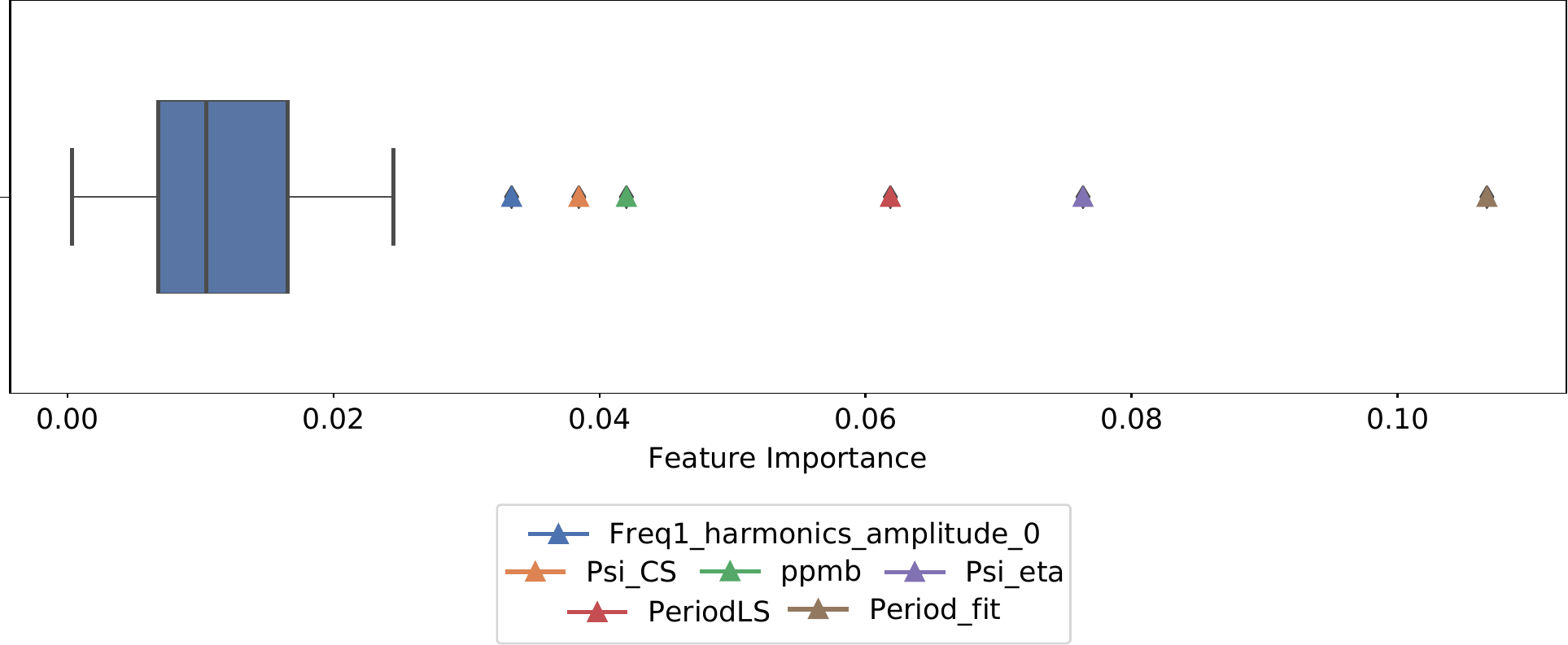}
    \caption{
    Distribution of the relative feature importance of a RF classifier trained with all the RRLs and the objective class of TP and FN; see text for more.
    \label{fig:s_fn:box}}
\end{figure}

The RF method is also capable of estimating the relative importance of each feature in the dataset for the classification task (again, see \citealt{breiman2001random} for a detailed explanation of the procedure). Figure~\ref{fig:s_fn:box} presents in a boxplot the relative importance of all our features for the problem of splitting between the FN and TP RRLs. We can see that there are six features that are outliers from the entire set, which means that these six features are highly related to the learned patterns. The six features are \texttt{freq1\_harmonic\_amplitude\_0} (the first amplitude of the Fourier component), \texttt{Psi\_CS} (folded range cumulative sum), \texttt{ppmb} (pseudo-Phase Multi-Band), \texttt{PeriodLS} (the period extracted with the Lomb-Scargle method), \texttt{Psi\_eta} (phase folded dependency of the observations respect or the previous one), and \texttt{Period\_fit} (the period false alarm probability). It is interesting to note that all are period-based.

\begin{figure}
\centering
    \includegraphics[width=.99\columnwidth]{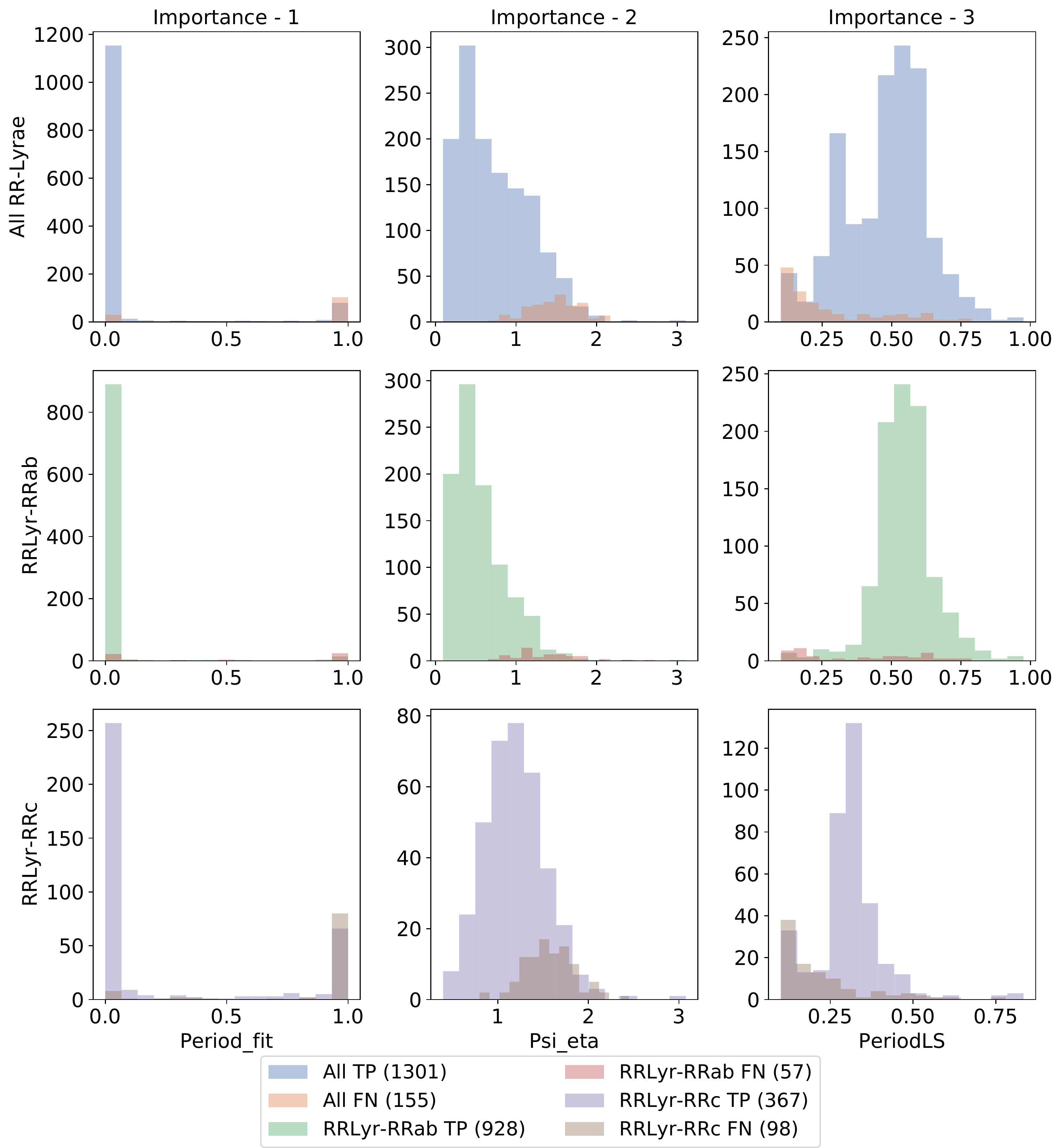}
    \caption{
    Distributions of the TP and FN classes for the RRL stars in all tiles according to the three most important features to differentiate them. In the top row all the stars are compared, in the mid row only the RRL-AB type, and in the low row the RRL-C type. Left column corresponds to \texttt{period\_fit}, mid column to \texttt{Psi\_eta} and right column to \texttt{PeriodLS}. All features were converted to z-scores and are dimensionless.
    \label{fig:s_fn:body_hist}
    }
\end{figure}

For a more detailed analysis, we compared how the FN and TP classes are distributed in our sample of RRLs according to these six features. The complete results can be found in Appendix~\ref{appendix:fn}, but we show in Fig.~\ref{fig:s_fn:body_hist} the results for the three more relevant features. In the figure, we also divided the sample into RRLs-AB and RRLs-C types because we found with these plots that the shoulders in the distribution are related to RRL subtypes. In the lower panel of the figure, there is an evident relationship between subtypes and FNs. In fact, we find that from the 155 FN in our dataset, 98 are RRL-C.
The complete analysis shows that we have two main sources of FN; first some sources are poorly observed in VVV, leading to poor estimates of periods and all derived features; second, RRL-C stars are much more difficult to identify than AB type RRL.

\section{Data release}
\label{section:carpyncho}

All the data produced in this work, light curves, features for each source and our catalog of RRL candidates are available to the community via the Carpyncho tool-set facility \footnote{\url{https://carpyncho.github.io/}}  \citep{2020ascl.soft05007C}. At this time, we share $\sim14.3$ million $K_s$ band light curves and $\sim11.2$ million features, all for curves that have at least 30 epochs.


We also publish a list of $242$ RRL candidates distributed in 11 tiles as shown in Table ~\ref{tab:cat_candidates} (The full list of candidates can be found in Appendix~\ref{appendix:fn}). We find using our final classifier a large sample of 117 candidate RRL in tile b396 and it is our estimate that more than half of these candidates are bona fide RRL stars. This number is between one and two orders of magnitude higher than the RRL that were previously known, in part because this region lies mostly outside of the OGLE-IV sensitivity footprint for RRL and our study was concluded prior to the Gaia data Release 2 and RRL catalogs of \citet{clem_2019}. Nevertheless a comparison of our RRL candidates and those of \citet{clem_2019} can be used to refine our method (as we hope to do after more data is released onto the Carpyncho facility). Given the importance of RRL as dynamical test particles for the formation and evolution modeling of the bulge in the Galaxy, we promptly present our first data release of RRL, as described above,  to the community.

\begin{table}[tb]
\caption[]{Number of RRL candidates per tile.}
\centering
\begin{tabular}{cc}
\toprule
\textbf{Tile} &   \textbf{Candidates} \\
\midrule
b206 &    3 \\
b214 &   17 \\
b216 &   21 \\
b220 &   15 \\
b228 &   18 \\
b234 &   23 \\
b247 &    2 \\
b248 &    1 \\
b263 &    1 \\
b360 &   24 \\
b396 &  117 \\
\bottomrule
\end{tabular}
\label{tab:cat_candidates}
\end{table}

\section{Summary, conclusions, and future work}
\label{section:conclusions}

In this work, we derive a method for the automatic classification of RRL stars. We begin by discussing the context of RRL as keystones for stellar evolution and pulsation astrophysics and  their importance as rungs on the intra- and extragalactic distance scale ladder, as well as for galaxy formation models. We  base our models on RRL that have previously been classified in the literature prior to Gaia DR2. We match VVV data to those stars, and extract features using the feets package affiliated to astropy, presented in \citet{cabral_fats_2018}. We explore the difficulty inherent in existing semi-automatic methods as found in the literature and set out to test some of these pitfalls to learn from them to build a more robust classifier of RRL for the VVV survey based on a newly crafted ML tool.

We start by considering a set of traditional period- and intensity-based features, plus a subset of new pseudo-color and pseudo-magnitudes-based features. We show that period-based features are the most relevant for RRLs identification and that the incorporation of color features also add information that improves performance. We next discuss the choice of classifier and find that as in previous studies, classifiers based on multiple classification trees are the best for this particular task. We used RF in this work because of its reliability and internal OOB estimations, but boosting is a valid option, as shown by \citet{elorrieta2016machine}.

We discuss extensively the need for appropriate quality metrics for these highly imbalanced problems and the issues related to the use of reduced samples. We show that precision-recall tables and curves are well suited to these problems and that the best strategy for obtaining good estimations of performance is to work on the complete, highly imbalanced dataset.

The last step of the pipeline is the model selection process. We show that the threshold of our classifiers can be correctly selected using the internal K-fold method and that the use of an ensemble classifier can overcome the problem of selecting of the appropriate tile for modeling.

In the last sections of this work, we discuss how this analysis led us to a reliable and completely automatic selection of RRL candidates, with a precision$\geq0.6$ and a recall$\geq0.3 $ in almost all tiles. Averaging over all tiles except b396 we have a recall of 0.48 and a precision of 0.86, or what is the same, we expect to recover the $48\% $ of RRLs candidate stars in any new tile, and more than 8 out of 10 of them will be confirmed as RRLs. We left tile b396 aside because there were previously no classified RRL in this region.

We also discuss the most common errors of the automatic procedure, related to the low quality of some observations and the more difficult identification of the RRL-C type. All data produced in this work, including light curves, features, and the catalog of candidates has been released and made public trough the Carpyncho tool.

In terms of future work, we propose to extend the data collection contained in Carpyncho to more tiles in the VVV, along with the complete publication of the internal database (Tile-Catalogs and Pawprint-Stacks). We wish to cross-check our candidate RRL with those of Gaia-2 and determine metallicities for our RRL based on our derived light curves to make spatial metallicity maps for our RRL and independent dust extinction maps. We hope to tune our classifier for other variables. In particular, to delta-Scuti type variables that have historically been misclassified as RRL, and to determine confusion thresholds for our model RRL candidates. Although the color features used for this work are based on the first epoch VVV data we hope to be able to incorporate more color epochs from a subsample of tiles for which that information exists in later epochs. This will allow for the introduction of color features that are not fixed in time. We also wish to tune our classifier to other populations of RRL, for example, the disk population and the Sag dSph population of RRL that both fall within the foot-print of the VVV and VVV(X) data. We could hope to tune our classifier for halo RRL, given that some halo RRL are also within the VVV and VVV(X) data. We also hope to explore the change of the photometric base of our VVV RRL data, from one that is based on aperture magnitudes to one that is based on PSF magnitudes, as derived from the work of \cite{alonso18}. This will permit a higher level of precision in feature observables for future studies of RRL.


\begin{acknowledgements}
The authors would like to thank to their families and friends, and also IATE
astronomers for useful comments and suggestions.

This work was partially supported by the Consejo Nacional
de Investigaciones Cient\'ificas y T\'ecnicas (CONICET, Argentina)
and the Secretar\'ia de Ciencia y Tecnolog\'ia de la Universidad
Nacional de C\'ordoba (SeCyT-UNC, Argentina).
F.R and J.B.C are supported by fellowships from CONICET. P.M.G. acknowledge partial support from PIP0066-CONICET.
Some processing was achieved with Argentine VO (NOVA) infrastructure,
for which the authors express their gratitude. SG would like to thank Ken Freeman for infecting him
with the RRL Bug and Marcio Catelan, Dante Minniti, Roberto Saito,
Javier Alonso and other members of the VVV scientific team for their genuine
intellectual generosity. We gratefully acknowledge data from the ESO Public Survey program ID 179.B-2002 taken with the VISTA telescope and products from the Cambridge Astronomical Survey Unit (CASU).
J.B.C. thanks to Jim Emerson for extend explanations about the VISTA instruments, and finally Bruno Sánchez, Laura Baravalle and Martín Beroiz for the continuous support and friendship.

This research has made use of the
\url{http://adsabs.harvard.edu/}, Cornell University xxx.arxiv.org repository, adstex (\url{https://github.com/yymao/adstex}), astropy and
the Python programming language.
\end{acknowledgements}

%
%

\label{biblio}
\bibliographystyle{aa}
\bibliography{main.bib}

\begin{thebibliography}{92}
\expandafter\ifx\csname natexlab\endcsname\relax\def\natexlab#1{#1}\fi

\bibitem[{{Alonso-Garc{\'\i}a} {et~al.}(2018){Alonso-Garc{\'\i}a}, {Saito},
  {Hempel}, {Minniti}, {Pullen}, {Catelan}, {Ramos}, {Cross}, {Gonzalez},
  {Lucas}, {Palma}, {Valenti}, \& {Zoccali}}]{alonso18}
{Alonso-Garc{\'\i}a}, J., {Saito}, R.~K., {Hempel}, M., {et~al.} 2018, \aap,
  619, A4

\bibitem[{{Armstrong} {et~al.}(2015){Armstrong}, {Kirk}, {Lam}, {McCormac},
  {Walker}, {Brown}, {Osborn}, {Pollacco}, \& {Spake}}]{armstrong2015k2}
{Armstrong}, D.~J., {Kirk}, J., {Lam}, K.~W.~F., {et~al.} 2015, \aap, 579, A19

\bibitem[{{Astropy Collaboration} {et~al.}(2013){Astropy Collaboration},
  {Robitaille}, {Tollerud}, {Greenfield}, {Droettboom}, {Bray}, {Aldcroft},
  {Davis}, {Ginsburg}, {Price-Whelan}, {Kerzendorf}, {Conley}, {Crighton},
  {Barbary}, {Muna}, {Ferguson}, {Grollier}, {Parikh}, {Nair}, {Unther},
  {Deil}, {Woillez}, {Conseil}, {Kramer}, {Turner}, {Singer}, {Fox}, {Weaver},
  {Zabalza}, {Edwards}, {Azalee Bostroem}, {Burke}, {Casey}, {Crawford},
  {Dencheva}, {Ely}, {Jenness}, {Labrie}, {Lim}, {Pierfederici}, {Pontzen},
  {Ptak}, {Refsdal}, {Servillat}, \& {Streicher}}]{robitaille2013astropy}
{Astropy Collaboration}, {Robitaille}, T.~P., {Tollerud}, E.~J., {et~al.} 2013,
  \aap, 558, A33

\bibitem[{{Baade}(1946)}]{baade1946search}
{Baade}, W. 1946, \pasp, 58, 249

\bibitem[{{Bailey}(1902)}]{bailey1902discussion}
{Bailey}, S.~I. 1902, Annals of Harvard College Observatory, 38, 1

\bibitem[{{Bell} \& {de Jong}(2001)}]{2001ApJ...550..212B}
{Bell}, E.~F. \& {de Jong}, R.~S. 2001, \apj, 550, 212

\bibitem[{Bowley(1901)}]{bowley1901elements}
Bowley, A. 1901

\bibitem[{Breiman(2001)}]{breiman2001random}
Breiman, L. 2001, Machine learning, 45, 5

\bibitem[{{Brough} {et~al.}(2020){Brough}, {Collins}, {Demarco}, {Ferguson},
  {Galaz}, {Holwerda}, {Martinez-Lombilla}, {Mihos}, \&
  {Montes}}]{brough2020vera}
{Brough}, S., {Collins}, C., {Demarco}, R., {et~al.} 2020, arXiv e-prints,
  arXiv:2001.11067

\bibitem[{{Burnham}(1978)}]{burnham1978burnham}
{Burnham}, Robert, J. 1978, {Burnham's Celestial Handbook: An Observer's Guide
  to the Universe Beyond the Solar System, in three volumes.}

\bibitem[{Cabral {et~al.}(2016)Cabral, Gurovich, Gran, \&
  Minnitti}]{cabral_carpyncho_2016}
Cabral, J., Gurovich, S., Gran, F., \& Minnitti, D. 2016, in 7th {VVV}
  {Science} {Workshop}

\bibitem[{{Cabral} {et~al.}(2020){Cabral}, {Ramos}, {Gurovich}, \&
  {Granitto}}]{2020ascl.soft05007C}
{Cabral}, J.~B., {Ramos}, F., {Gurovich}, S., \& {Granitto}, P. 2020,
  {Carpyncho: VVV Catalog browser toolkit}

\bibitem[{{Cabral} {et~al.}(2017){Cabral}, {S{\'a}nchez}, {Beroiz},
  {Dom{\'\i}nguez}, {Lares}, {Gurovich}, \& {Granitto}}]{cabral2017corral}
{Cabral}, J.~B., {S{\'a}nchez}, B., {Beroiz}, M., {et~al.} 2017, Astronomy and
  Computing, 20, 140

\bibitem[{{Cabral} {et~al.}(2018){Cabral}, {S{\'a}nchez}, {Ramos}, {Gurovich},
  {Granitto}, \& {Vanderplas}}]{cabral_fats_2018}
{Cabral}, J.~B., {S{\'a}nchez}, B., {Ramos}, F., {et~al.} 2018, Astronomy and
  Computing, 25, 213

\bibitem[{{Cardelli} {et~al.}(1989){Cardelli}, {Clayton}, \&
  {Mathis}}]{cardelli1989relationship}
{Cardelli}, J.~A., {Clayton}, G.~C., \& {Mathis}, J.~S. 1989, \apj, 345, 245

\bibitem[{{Catelan} {et~al.}(2011){Catelan}, {Minniti}, {Lucas},
  {Alonso-Garc{\'\i}a}, {Angeloni}, {Beam{\'\i}n}, {Bonatto}, {Borissova},
  {Contreras}, {Cross}, {D{\'e}k{\'a}any}, {Emerson}, {Eyheramendy}, {Geisler},
  {Gonz{\'a}lez-Solares}, {Helminiak}, {Hempel}, {Irwin}, {Ivanov},
  {Jord{\'a}n}, {Kerins}, {Kurtev}, {Mauro}, {Moni Bidin}, {Navarrete},
  {P{\'e}rez}, {Pichara}, {Read}, {Rejkuba}, {Saito}, {Sale}, \&
  {Toledo}}]{catelan2011vista}
{Catelan}, M., {Minniti}, D., {Lucas}, P.~W., {et~al.} 2011, in RR Lyrae Stars,
  Metal-Poor Stars, and the Galaxy, ed. A.~{McWilliam}, Vol.~5, 145

\bibitem[{{Clement}(2017)}]{2017yCat.5150....0C}
{Clement}, C.~M. 2017, VizieR Online Data Catalog, V/150

\bibitem[{{Clementini} {et~al.}(2001){Clementini}, {Federici}, {Corsi},
  {Cacciari}, {Bellazzini}, \& {Smith}}]{clementini2001rr}
{Clementini}, G., {Federici}, L., {Corsi}, C., {et~al.} 2001, \apjl, 559, L109

\bibitem[{{Clementini} {et~al.}(2019){Clementini}, {Ripepi}, {Molinaro},
  {Garofalo}, {Muraveva}, {Rimoldini}, {Guy}, {Jevardat de Fombelle},
  {Nienartowicz}, {Marchal}, {Audard}, {Holl}, {Leccia}, {Marconi}, {Musella},
  {Mowlavi}, {Lecoeur-Taibi}, {Eyer}, {De Ridder}, {Regibo}, {Sarro},
  {Szabados}, {Evans}, \& {Riello}}]{clem_2019}
{Clementini}, G., {Ripepi}, V., {Molinaro}, R., {et~al.} 2019, \aap, 622, A60

\bibitem[{{Collinge} {et~al.}(2006){Collinge}, {Sumi}, \&
  {Fabrycky}}]{2006ApJ...651..197C}
{Collinge}, M.~J., {Sumi}, T., \& {Fabrycky}, D. 2006, \apj, 651, 197

\bibitem[{{Czesla} {et~al.}(2019){Czesla}, {Schr{\"o}ter}, {Schneider},
  {Huber}, {Pfeifer}, {Andreasen}, \& {Zechmeister}}]{czesla2019pya}
{Czesla}, S., {Schr{\"o}ter}, S., {Schneider}, C.~P., {et~al.} 2019, {PyA:
  Python astronomy-related packages}

\bibitem[{{de Grijs} {et~al.}(2017){de Grijs}, {Courbin},
  {Mart{\'\i}nez-V{\'a}zquez}, {Monelli}, {Oguri}, \& {Suyu}}]{de2017toward}
{de Grijs}, R., {Courbin}, F., {Mart{\'\i}nez-V{\'a}zquez}, C.~E., {et~al.}
  2017, \ssr, 212, 1743

\bibitem[{{Dwek} {et~al.}(1995){Dwek}, {Arendt}, {Hauser}, {Kelsall}, {Lisse},
  {Moseley}, {Silverberg}, {Sodroski}, \& {Weiland}}]{dwek1995morphology}
{Dwek}, E., {Arendt}, R.~G., {Hauser}, M.~G., {et~al.} 1995, \apj, 445, 716

\bibitem[{{Eastman} {et~al.}(2010){Eastman}, {Siverd}, \&
  {Gaudi}}]{eastman_achieving_2010}
{Eastman}, J., {Siverd}, R., \& {Gaudi}, B.~S. 2010, \pasp, 122, 935

\bibitem[{{Elorrieta} {et~al.}(2016){Elorrieta}, {Eyheramendy}, {Jord{\'a}n},
  {D{\'e}k{\'a}ny}, {Catelan}, {Angeloni}, {Alonso-Garc{\'\i}a},
  {Contreras-Ramos}, {Gran}, {Hajdu}, {Espinoza}, {Saito}, \&
  {Minniti}}]{elorrieta2016machine}
{Elorrieta}, F., {Eyheramendy}, S., {Jord{\'a}n}, A., {et~al.} 2016, \aap, 595,
  A82

\bibitem[{{Emerson} {et~al.}(2004){Emerson}, {Irwin}, {Lewis}, {Hodgkin},
  {Evans}, {Bunclark}, {McMahon}, {Hambly}, {Mann}, {Bond}, {Sutorius}, {Read},
  {Williams}, {Lawrence}, \& {Stewart}}]{emerson_vista_2004}
{Emerson}, J.~P., {Irwin}, M.~J., {Lewis}, J., {et~al.} 2004, in Society of
  Photo-Optical Instrumentation Engineers (SPIE) Conference Series, Vol. 5493,
  \procspie, ed. P.~J. {Quinn} \& A.~{Bridger}, 401--410

\bibitem[{{Gaia Collaboration} {et~al.}(2018){Gaia Collaboration}, {Brown},
  {Vallenari}, {Prusti}, {de Bruijne}, {Babusiaux}, {Bailer-Jones}, {Biermann},
  {Evans}, {Eyer}, \& et~al.}]{2018A&A...616A...1G}
{Gaia Collaboration}, {Brown}, A.~G.~A., {Vallenari}, A., {et~al.} 2018, \aap,
  616, A1

\bibitem[{{Gaia Collaboration} {et~al.}(2016){Gaia Collaboration}, {Prusti},
  {de Bruijne}, {Brown}, {Vallenari}, {Babusiaux}, {Bailer-Jones}, {Bastian},
  {Biermann}, {Evans}, \& et~al.}]{2016A&A...595A...1G}
{Gaia Collaboration}, {Prusti}, T., {de Bruijne}, J.~H.~J., {et~al.} 2016,
  \aap, 595, A1

\bibitem[{{Gavrilchenko} {et~al.}(2014){Gavrilchenko}, {Klein}, {Bloom}, \&
  {Richards}}]{2014MNRAS.441..715G}
{Gavrilchenko}, T., {Klein}, C.~R., {Bloom}, J.~S., \& {Richards}, J.~W. 2014,
  \mnras, 441, 715

\bibitem[{{Gonzalez} {et~al.}(2011){Gonzalez}, {Rejkuba}, {Zoccali}, {Valenti},
  \& {Minniti}}]{gonzalez2011reddening}
{Gonzalez}, O.~A., {Rejkuba}, M., {Zoccali}, M., {Valenti}, E., \& {Minniti},
  D. 2011, \aap, 534, A3

\bibitem[{{Gonzalez} {et~al.}(2012){Gonzalez}, {Rejkuba}, {Zoccali}, {Valenti},
  {Minniti}, {Schultheis}, {Tobar}, \& {Chen}}]{gonzalez2012reddening}
{Gonzalez}, O.~A., {Rejkuba}, M., {Zoccali}, M., {et~al.} 2012, \aap, 543, A13

\bibitem[{{Gonz{\'a}lez-Fern{\'a}ndez}
  {et~al.}(2018){Gonz{\'a}lez-Fern{\'a}ndez}, {Hodgkin}, {Irwin},
  {Gonz{\'a}lez-Solares}, {Koposov}, {Lewis}, {Emerson}, {Hewett},
  {Yolda{\textcommabelow s}}, \& {Riello}}]{gonzalez2017vista}
{Gonz{\'a}lez-Fern{\'a}ndez}, C., {Hodgkin}, S.~T., {Irwin}, M.~J., {et~al.}
  2018, \mnras, 474, 5459

\bibitem[{{Gran} {et~al.}(2015){Gran}, {Minniti}, {Saito}, {Navarrete},
  {D{\'e}k{\'a}ny}, {McDonald}, {Contreras Ramos}, \&
  {Catelan}}]{gran_bulge_2015}
{Gran}, F., {Minniti}, D., {Saito}, R.~K., {et~al.} 2015, \aap, 575, A114

\bibitem[{{Gran} {et~al.}(2016){Gran}, {Minniti}, {Saito}, {Zoccali},
  {Gonzalez}, {Navarrete}, {Catelan}, {Contreras Ramos}, {Elorrieta},
  {Eyheramendy}, \& {Jord{\'a}n}}]{2016A&A...591A.145G}
{Gran}, F., {Minniti}, D., {Saito}, R.~K., {et~al.} 2016, \aap, 591, A145

\bibitem[{Granitto {et~al.}(2005)Granitto, Verdes, \&
  Ceccatto}]{granitto2005neural}
Granitto, P.~M., Verdes, P.~F., \& Ceccatto, H.~A. 2005, Artificial
  Intelligence, 163, 139

\bibitem[{{Gurovich} {et~al.}(2010){Gurovich}, {Freeman}, {Jerjen},
  {Staveley-Smith}, \& {Puerari}}]{2010AJ....140..663G}
{Gurovich}, S., {Freeman}, K., {Jerjen}, H., {Staveley-Smith}, L., \&
  {Puerari}, I. 2010, \aj, 140, 663

\bibitem[{{Hanisch} {et~al.}(2001){Hanisch}, {Farris}, {Greisen}, {Pence},
  {Schlesinger}, {Teuben}, {Thompson}, \& {Warnock}}]{hanisch2001definition}
{Hanisch}, R.~J., {Farris}, A., {Greisen}, E.~W., {et~al.} 2001, \aap, 376, 359

\bibitem[{{Hosenie} {et~al.}(2020){Hosenie}, {Lyon}, {Stappers}, {Mootoovaloo},
  \& {McBride}}]{hosenie2020imbalance}
{Hosenie}, Z., {Lyon}, R., {Stappers}, B., {Mootoovaloo}, A., \& {McBride}, V.
  2020, \mnras, 493, 6050

\bibitem[{Hunter(2007)}]{hunter2007matplotlib}
Hunter, J.~D. 2007, Computing in science \& engineering, 9, 90

\bibitem[{{Ivezi{\'c}} {et~al.}(2019){Ivezi{\'c}}, {Kahn}, {Tyson}, {Abel},
  {Acosta}, {Allsman}, {Alonso}, {AlSayyad}, {Anderson}, {Andrew}, \&
  et~al.}]{ivezic2019lsst}
{Ivezi{\'c}}, {\v{Z}}., {Kahn}, S.~M., {Tyson}, J.~A., {et~al.} 2019, \apj,
  873, 111

\bibitem[{Japkowicz \& Stephen(2002)}]{japkowicz2002class}
Japkowicz, N. \& Stephen, S. 2002, Intelligent data analysis, 6, 429

\bibitem[{{Kim} {et~al.}(2014){Kim}, {Protopapas}, {Bailer-Jones}, {Byun},
  {Chang}, {Marquette}, \& {Shin}}]{kim2014epoch}
{Kim}, D.-W., {Protopapas}, P., {Bailer-Jones}, C. A.~L., {et~al.} 2014, \aap,
  566, A43

\bibitem[{{Kim} {et~al.}(2011){Kim}, {Protopapas}, {Byun}, {Alcock}, {Khardon},
  \& {Trichas}}]{kim2011quasi}
{Kim}, D.-W., {Protopapas}, P., {Byun}, Y.-I., {et~al.} 2011, \apj, 735, 68

\bibitem[{{Kov{\'a}cs} \& {Walker}(2001)}]{kovacs_walk_01}
{Kov{\'a}cs}, G. \& {Walker}, A.~R. 2001, \aap, 371, 579

\bibitem[{{Kunder} \& {Chaboyer}(2008)}]{2008AJ....136.2441K}
{Kunder}, A. \& {Chaboyer}, B. 2008, \aj, 136, 2441

\bibitem[{{Kunder} {et~al.}(2008){Kunder}, {Popowski}, {Cook}, \&
  {Chaboyer}}]{2008AJ....135..631K}
{Kunder}, A., {Popowski}, P., {Cook}, K.~H., \& {Chaboyer}, B. 2008, \aj, 135,
  631

\bibitem[{{Lee}(1992)}]{lee92_aj}
{Lee}, Y.-W. 1992, \aj, 104, 1780

\bibitem[{{Li{\v{s}}ka} {et~al.}(2016){Li{\v{s}}ka}, {Skarka}, {Zejda},
  {Mikul{\'a}{\v{s}}ek}, \& {de Villiers}}]{2016MNRAS.459.4360L}
{Li{\v{s}}ka}, J., {Skarka}, M., {Zejda}, M., {Mikul{\'a}{\v{s}}ek}, Z., \& {de
  Villiers}, S.~N. 2016, \mnras, 459, 4360

\bibitem[{{Lomb}(1976)}]{lomb_least-squares_1976}
{Lomb}, N.~R. 1976, \apss, 39, 447

\bibitem[{{Mackenzie} {et~al.}(2016){Mackenzie}, {Pichara}, \&
  {Protopapas}}]{Mackenzie2016ApJ}
{Mackenzie}, C., {Pichara}, K., \& {Protopapas}, P. 2016, \apj, 820, 138

\bibitem[{{Majaess} {et~al.}(2018){Majaess}, {D{\'e}k{\'a}ny}, {Hajdu},
  {Minniti}, {Turner}, \& {Gieren}}]{2018Ap&SS.363..127M}
{Majaess}, D., {D{\'e}k{\'a}ny}, I., {Hajdu}, G., {et~al.} 2018, \apss, 363,
  127

\bibitem[{{Minniti} {et~al.}(2010){Minniti}, {Lucas}, {Emerson}, {Saito},
  {Hempel}, {Pietrukowicz}, {Ahumada}, {Alonso}, {Alonso-Garcia}, {Arias},
  {Bandyopadhyay}, {Barb{\'a}}, {Barbuy}, {Bedin}, {Bica}, {Borissova},
  {Bronfman}, {Carraro}, {Catelan}, {Clari{\'a}}, {Cross}, {de Grijs},
  {D{\'e}k{\'a}ny}, {Drew}, {Fari{\~n}a}, {Feinstein}, {Fern{\'a}ndez
  Laj{\'u}s}, {Gamen}, {Geisler}, {Gieren}, {Goldman}, {Gonzalez}, {Gunthardt},
  {Gurovich}, {Hambly}, {Irwin}, {Ivanov}, {Jord{\'a}n}, {Kerins}, {Kinemuchi},
  {Kurtev}, {L{\'o}pez-Corredoira}, {Maccarone}, {Masetti}, {Merlo},
  {Messineo}, {Mirabel}, {Monaco}, {Morelli}, {Padilla}, {Palma}, {Parisi},
  {Pignata}, {Rejkuba}, {Roman-Lopes}, {Sale}, {Schreiber}, {Schr{\"o}der},
  {Smith}, {}, {Soto}, {Tamura}, {Tappert}, {Thompson}, {Toledo}, {Zoccali}, \&
  {Pietrzynski}}]{minniti2010vista}
{Minniti}, D., {Lucas}, P.~W., {Emerson}, J.~P., {et~al.} 2010, \na, 15, 433

\bibitem[{Mitchell(1997)}]{mitchell1997machine}
Mitchell, T.~M. 1997, Machine learning (McGraw-hill New York)

\bibitem[{{Moretti} {et~al.}(2018){Moretti}, {Hatzidimitriou}, {Karampelas},
  {Sokolovsky}, {Bonanos}, {Gavras}, \& {Yang}}]{Moretti2018MNRAS}
{Moretti}, M.~I., {Hatzidimitriou}, D., {Karampelas}, A., {et~al.} 2018,
  \mnras, 477, 2664

\bibitem[{{Nishiyama} {et~al.}(2009){Nishiyama}, {Tamura}, {Hatano}, {Kato},
  {Tanab{\'e}}, {Sugitani}, \& {Nagata}}]{nishiyama2009interstellar}
{Nishiyama}, S., {Tamura}, M., {Hatano}, H., {et~al.} 2009, \apj, 696, 1407

\bibitem[{{Nun} {et~al.}(2015){Nun}, {Protopapas}, {Sim}, {Zhu}, {Dave},
  {Castro}, \& {Pichara}}]{nun2015fats}
{Nun}, I., {Protopapas}, P., {Sim}, B., {et~al.} 2015, arXiv e-prints,
  arXiv:1506.00010

\bibitem[{{Ochsenbein} {et~al.}(2000){Ochsenbein}, {Bauer}, \&
  {Marcout}}]{ochsenbein2000vizier}
{Ochsenbein}, F., {Bauer}, P., \& {Marcout}, J. 2000, \aaps, 143, 23

\bibitem[{Oliphant(2007)}]{jones_scipy:_2014}
Oliphant, T.~E. 2007, Computing in Science and Engineering, 9, 10

\bibitem[{{Pashchenko} {et~al.}(2018){Pashchenko}, {Sokolovsky}, \&
  {Gavras}}]{pashchenko2017MNRAS}
{Pashchenko}, I.~N., {Sokolovsky}, K.~V., \& {Gavras}, P. 2018, \mnras, 475,
  2326

\bibitem[{Pedregosa {et~al.}(2011)Pedregosa, Varoquaux, Gramfort, Michel,
  Thirion, Grisel, Blondel, Prettenhofer, Weiss, Dubourg,
  {et~al.}}]{pedregosa2011scikit}
Pedregosa, F., Varoquaux, G., Gramfort, A., {et~al.} 2011, Journal of Machine
  Learning Research, 12, 2825

\bibitem[{{Prudil} \& {Skarka}(2017)}]{2017MNRAS.466.2602P}
{Prudil}, Z. \& {Skarka}, M. 2017, \mnras, 466, 2602

\bibitem[{Ragan-Kelley {et~al.}(2014)Ragan-Kelley, Perez, Granger, Kluyver,
  Ivanov, Frederic, \& Bussonier}]{ipython_2014}
Ragan-Kelley, M., Perez, F., Granger, B., {et~al.} 2014, in {AGU} {Fall}
  {Meeting} {Abstracts}, Vol.~1, 07

\bibitem[{{Richards} {et~al.}(2011){Richards}, {Starr}, {Butler}, {Bloom},
  {Brewer}, {Crellin-Quick}, {Higgins}, {Kennedy}, \&
  {Rischard}}]{richards2011machine}
{Richards}, J.~W., {Starr}, D.~L., {Butler}, N.~R., {et~al.} 2011, \apj, 733,
  10

\bibitem[{{Richards} {et~al.}(2012){Richards}, {Starr}, {Miller}, {Bloom},
  {Butler}, {Brink}, \& {Crellin-Quick}}]{richards2012construction}
{Richards}, J.~W., {Starr}, D.~L., {Miller}, A.~A., {et~al.} 2012, \apjs, 203,
  32

\bibitem[{Rokach(2010)}]{rokach2010ensemble}
Rokach, L. 2010, Artificial Intelligence Review, 33, 1

\bibitem[{{Sakai} \& {Madore}(2001)}]{sakai2001observation}
{Sakai}, S. \& {Madore}, B.~F. 2001, \apj, 555, 280

\bibitem[{{Samus'} {et~al.}(2017){Samus'}, {Kazarovets}, {Durlevich},
  {Kireeva}, \& {Pastukhova}}]{2017ARep...61...80S}
{Samus'}, N.~N., {Kazarovets}, E.~V., {Durlevich}, O.~V., {Kireeva}, N.~N., \&
  {Pastukhova}, E.~N. 2017, Astronomy Reports, 61, 80

\bibitem[{{Scargle}(1982)}]{scargle_studies_1982}
{Scargle}, J.~D. 1982, \apj, 263, 835

\bibitem[{{Schultheis} {et~al.}(2014){Schultheis}, {Chen}, {Jiang}, {Gonzalez},
  {Enokiya}, {Fukui}, {Torii}, {Rejkuba}, \& {Minniti}}]{2014A&A...566A.120S}
{Schultheis}, M., {Chen}, B.~Q., {Jiang}, B.~W., {et~al.} 2014, \aap, 566, A120

\bibitem[{{Seares} \& {Shapley}(1914)}]{seares1914color}
{Seares}, F.~H. \& {Shapley}, H. 1914, \pasp, 26, 202

\bibitem[{{Sesar} {et~al.}(2017){Sesar}, {Hernitschek}, {Mitrovi{\'c}},
  {Ivezi{\'c}}, {Rix}, {Cohen}, {Bernard}, {Grebel}, {Martin}, {Schlafly},
  {Burgett}, {Draper}, {Flewelling}, {Kaiser}, {Kudritzki}, {Magnier},
  {Metcalfe}, {Tonry}, \& {Waters}}]{2017AJ....153..204S}
{Sesar}, B., {Hernitschek}, N., {Mitrovi{\'c}}, S., {et~al.} 2017, \aj, 153,
  204

\bibitem[{{Shapley}(1918)}]{shapley1918studies}
{Shapley}, H. 1918, \apj, 48, 154

\bibitem[{{Shin} {et~al.}(2009){Shin}, {Sekora}, \& {Byun}}]{Shin2009MNRAS}
{Shin}, M.-S., {Sekora}, M., \& {Byun}, Y.-I. 2009, \mnras, 400, 1897

\bibitem[{{Shin} {et~al.}(2012){Shin}, {Yi}, {Kim}, {Chang}, \&
  {Byun}}]{Shin2012AJ}
{Shin}, M.-S., {Yi}, H., {Kim}, D.-W., {Chang}, S.-W., \& {Byun}, Y.-I. 2012,
  \aj, 143, 65

\bibitem[{{Silbermann} \& {Smith}(1995)}]{silbermann1995rr}
{Silbermann}, N.~A. \& {Smith}, H.~A. 1995, \aj, 110, 704

\bibitem[{{Skrutskie} {et~al.}(2006){Skrutskie}, {Cutri}, {Stiening},
  {Weinberg}, {Schneider}, {Carpenter}, {Beichman}, {Capps}, {Chester},
  {Elias}, {Huchra}, {Liebert}, {Lonsdale}, {Monet}, {Price}, {Seitzer},
  {Jarrett}, {Kirkpatrick}, {Gizis}, {Howard}, {Evans}, {Fowler}, {Fullmer},
  {Hurt}, {Light}, {Kopan}, {Marsh}, {McCallon}, {Tam}, {Van Dyk}, \&
  {Wheelock}}]{skrutskie2006two}
{Skrutskie}, M.~F., {Cutri}, R.~M., {Stiening}, R., {et~al.} 2006, \aj, 131,
  1163

\bibitem[{{Smith}(2004)}]{smith2004rr}
{Smith}, H.~A. 2004, {RR Lyrae Stars}

\bibitem[{{Soszy{\'n}ski} {et~al.}(2011){Soszy{\'n}ski}, {Dziembowski},
  {Udalski}, {Poleski}, {Szyma{\'n}ski}, {Kubiak}, {Pietrzy{\'n}ski},
  {Wyrzykowski}, {Ulaczyk}, {Koz{\l}owski}, \&
  {Pietrukowicz}}]{2011AcA....61....1S}
{Soszy{\'n}ski}, I., {Dziembowski}, W.~A., {Udalski}, A., {et~al.} 2011,
  \actaa, 61, 1

\bibitem[{{Soszy{\'n}ski} {et~al.}(2014){Soszy{\'n}ski}, {Udalski},
  {Szyma{\'n}ski}, {Pietrukowicz}, {Mr{\'o}z}, {Skowron}, {Koz{\l}owski},
  {Poleski}, {Skowron}, {Pietrzy{\'n}ski}, {Wyrzykowski}, {Ulaczyk}, \&
  {Kubiak}}]{2014AcA....64..177S}
{Soszy{\'n}ski}, I., {Udalski}, A., {Szyma{\'n}ski}, M.~K., {et~al.} 2014,
  \actaa, 64, 177

\bibitem[{{Soszy{\'n}ski} {et~al.}(2017){Soszy{\'n}ski}, {Udalski},
  {Szyma{\'n}ski}, {Wyrzykowski}, {Ulaczyk}, {Poleski}, {Pietrukowicz},
  {Koz{\l}owski}, {Skowron}, {Skowron}, {Mr{\'o}z}, {Pawlak}, {Rybicki}, \&
  {Jacyszyn-Dobrzeniecka}}]{2017AcA....67..297S}
{Soszy{\'n}ski}, I., {Udalski}, A., {Szyma{\'n}ski}, M.~K., {et~al.} 2017,
  \actaa, 67, 297

\bibitem[{{Soszy{\'n}ski} {et~al.}(2019){Soszy{\'n}ski}, {Udalski}, {Wrona},
  {Szyma{\'n}ski}, {Pietrukowicz}, {Skowron}, {Skowron}, {Poleski},
  {Koz{\l}owski}, {Mr{\'o}z}, {Ulaczyk}, {Rybicki}, {Iwanek}, \&
  {Gromadzki}}]{2019AcA....69..321S}
{Soszy{\'n}ski}, I., {Udalski}, A., {Wrona}, M., {et~al.} 2019, \actaa, 69, 321

\bibitem[{Sutherland {et~al.}(2015)Sutherland, Emerson, Dalton, Atad-Ettedgui,
  Beard, Bennett, Bezawada, Born, Caldwell, Clark,
  {et~al.}}]{sutherland2015visible}
Sutherland, W., Emerson, J., Dalton, G., {et~al.} 2015, Astronomy \&
  Astrophysics, 575, A25

\bibitem[{{Tollerud}(2012)}]{tollerud2012astropysics}
{Tollerud}, E. 2012, {Astropysics: Astrophysics utilities for python}

\bibitem[{{Udalski}(2003)}]{2003AcA....53..291U}
{Udalski}, A. 2003, \actaa, 53, 291

\bibitem[{{Udalski} {et~al.}(1994){Udalski}, {Kubiak}, {Szymanski}, {Kaluzny},
  {Mateo}, \& {Krzeminski}}]{1994AcA....44..317U}
{Udalski}, A., {Kubiak}, M., {Szymanski}, M., {et~al.} 1994, \actaa, 44, 317

\bibitem[{{Udalski} {et~al.}(2015){Udalski}, {Szyma{\'n}ski}, \&
  {Szyma{\'n}ski}}]{2015AcA....65....1U}
{Udalski}, A., {Szyma{\'n}ski}, M.~K., \& {Szyma{\'n}ski}, G. 2015, \actaa, 65,
  1

\bibitem[{Van Der~Walt {et~al.}(2011)Van Der~Walt, Colbert, \&
  Varoquaux}]{van2011numpy}
Van Der~Walt, S., Colbert, S.~C., \& Varoquaux, G. 2011, Computing in Science
  \& Engineering, 13, 22

\bibitem[{{VanderPlas}(2018)}]{vanderplas2018understanding}
{VanderPlas}, J.~T. 2018, \apjs, 236, 16

\bibitem[{Vapnik(2013)}]{vapnik2013nature}
Vapnik, V. 2013, The nature of statistical learning theory (Springer science \&
  business media)

\bibitem[{{Watson} {et~al.}(2006){Watson}, {Henden}, \&
  {Price}}]{2006SASS...25...47W}
{Watson}, C.~L., {Henden}, A.~A., \& {Price}, A. 2006, Society for Astronomical
  Sciences Annual Symposium, 25, 47

\bibitem[{{Wozniak}(2000)}]{2000AcA....50..421W}
{Wozniak}, P.~R. 2000, \actaa, 50, 421

\bibitem[{{Zejda} {et~al.}(2012){Zejda}, {Paunzen}, {Baumann},
  {Mikul{\'a}{\v{s}}ek}, \& {Li{\v{s}}ka}}]{2012A&A...548A..97Z}
{Zejda}, M., {Paunzen}, E., {Baumann}, B., {Mikul{\'a}{\v{s}}ek}, Z., \&
  {Li{\v{s}}ka}, J. 2012, \aap, 548, A97

\end{thebibliography}


\appendix

\section{List of catalogs accessed in this work}
\label{appendix:vizier}

\begin{enumerate}
    \item The Optical Gravitational Lensing Experiment. The OGLE-III Catalog of Variable Stars. XI. RR Lyrae Stars in the Galactic Bulge \citep{2011AcA....61....1S}.
    \item Gaia Data Release 2. Summary of the contents and survey properties \citep{2018A&A...616A...1G}.
    \item Metallicity Analysis of MACHO Galactic Bulge RR0 Lyrae Stars from their Light Curves \citep{2008AJ....136.2441K}.
    \item VizieR Online Data Catalog: Updated catalog of variable stars in globular clusters (Clement+ 2017) \citep{2017yCat.5150....0C}.
    \item The International Variable Star Index (VSX) \citep{2006SASS...25...47W}.
    \item The Extinction Toward the Galactic Bulge from RR Lyrae Stars \citep{2008AJ....135..631K}.
    \item A mid-infrared study of RR Lyrae stars with the Wide-field Infrared Survey Explorer all-sky data release \citep{2014MNRAS.441..715G}.
    \item Blazhko effect in the Galactic bulge fundamental mode RR Lyrae stars - I. Incidence rate and differences between modulated and non-modulated stars \citep{2017MNRAS.466.2602P}.
    \item The Gaia mission \citep{2016A&A...595A...1G}.
    \item Mapping the outer bulge with RRab stars from the VVV Survey \citep{2016A&A...591A.145G}.
    \item Establishing the Galactic Centre distance using VVV Bulge RR Lyrae variables \citep{2018Ap&SS.363..127M}.
    \item Over 38000 RR Lyrae Stars in the OGLE Galactic Bulge Fields \citep{2014AcA....64..177S}.
    \item General catalogue of variable stars: Version GCVS 5.1 \citep{2017ARep...61...80S}.
    \item Machine-learned Identification of RR Lyrae Stars from Sparse, Multi-band Data: The PS1 Sample \citep{2017AJ....153..204S}.
    \item Catalogue of variable stars in open cluster fields \citep{2012A&A...548A..97Z}.
    \item Cyclic variations in O-C diagrams of field RR Lyrae stars as a result of LiTE \citep{2016MNRAS.459.4360L}.
    \item Catalog of Fundamental-Mode RR Lyrae Stars in the Galactic Bulge from the Optical Gravitational Lensing Experiment \citep{2006ApJ...651..197C}.
    \item The Optical Gravitational Lensing Experiment. The Catalog of Periodic Variable Stars in the Galactic Bulge. I. Periodic Variables in the Center of the Baade's Window \citep{1994AcA....44..317U}.
    \item The OGLE Collection of Variable Stars. Classical, Type II, and Anomalous Cepheids toward the Galactic Center \citep{2017AcA....67..297S}.
\end{enumerate}

\section{Features related to \textit{feets}  }
\label{appendixa}

Most of these descriptions are adapted from the feets documentation (\url{https://feets.readthedocs.io}).

\begin{description}

    \item[\texttt{Amplitude} ($A_{K_s}$)] the half of difference between the median of the $5\%$ upper values and
    the median of the $5\%$ of the lower values of the $K_s$ band \citep{richards2011machine}.

    \item[\texttt{Autocor\_length}] Auto-Correlation Length
    is the cross correlation of the signal with itself. Is useful to find patterns like noise hidden periodic signal \citep{kim2011quasi}.

    \item[\texttt{Beyond1Std}] Percentage of points beyond one standard deviation from the weighted mean. For a normal distribution, it should take a value close to $0.32$ \citep{richards2011machine}.

    \item[\texttt{Con}] ``\textit{Consecutive points}'' . number of three consecutive points greater or lesser than $2\sigma$ (normalized by $N-200$). This index was created by the OGLE survey to find variable stars \citep{2000AcA....50..421W, kim2011quasi}.

    \item[\texttt{Eta\_e} ($\eta^e$)] Verify the dependency of the observations respect or the previous ones \citep{kim2014epoch}.

    \item[\texttt{FluxPercentileRatioMid}]
    The ``\textit{Middle flux percentiles Ratio}'' characterizes the distributions of sorted magnitudes based on their  fluxes percentiles\footnote{Flux or luminosity is the amount of energy given by some star in one unit of time}. If
    $F_{i,j}$ is the difference between the flux percentil $j$ and the flux percentile $i$ we can calculate:
    \begin{itemize}
        \item $FluxPercentileRatioMid20 = F_{40,60}/F_{5,95}$
        \item $FluxPercentileRatioMid35 = F_{32.5,67.5}/F_{5,95}$
        \item $FluxPercentileRatioMid50 = F_{25,75}/F_{5,95}$
        \item $FluxPercentileRatioMid65 = F_{17.5,82.5}/F_{5,95}$
        \item $FluxPercentileRatioMid80 = F_{10,90}/F_{5,95}$
    \end{itemize}

    \item[\texttt{Fourier components}] The first three Fourier components for the first three  period candidates.
    \textit{Freq{k}\_harmonics\_amplitude\_i} y \textit{Freq{k}\_harmonics\_rel\_phase\_j} represents the \textit{i-nth} amplitude ant the \textit{j-nth} phase of the signal for the \textit{k-nth} period. \citep{richards2011machine}.

    \item[\texttt{Gskew}] Median-of-magnitudes based measure of the skew

    $$Gskew = m_{q3} + m_{q97} - 2m$$

    Where: $m_{q3}$ is the median of magnitudes lesser or equal than the quantile 3, $m_{q97}$ is the median of magnitudes greater or equal than the quantile 97; and $m$ is the median of magnitudes \citep{bowley1901elements}.

    \item[\texttt{LinearTrend}] Linear tendency. the slope of the linear regression of the data \citep{richards2011machine}.

    \item[\texttt{MaxSlope}] Maximum absolute slope between two consecutive observations\citep{richards2011machine}.

    \item[\texttt{Mean}] Mean of magnitudes \citep{kim2014epoch}.

    \item[\texttt{MedianAbsDev}] The Median of the absolute deviations, is defined as the median of the difference of every observed magnitude with the median of the entire time-serie. In simbols $Median Absolute Deviation = median(|mag - median(mag)|)$ \citep{richards2011machine}.

    \item[\texttt{MedianBRP}] - The ``\textit{Median Buffer Range Percentage}'' can be interpreted as the percentage closest to the median. Proportion of the magnitudes between the 10th part of the total range around the median \citep{richards2011machine}.

    \item[\texttt{PairSlopeTrend}] Considering the last 30 (time-sorted) measurements of source magnitude, the fraction of increasing first differences minus the fraction of decreasing first differences \citep{richards2011machine}.

    \item[\texttt{PercentAmplitude}] Largest percentage difference between either the max or min magnitude and the median. \citep{richards2011machine}.

    \item[\texttt{PercentDifferenceFluxPercentile}] Ratio of $F_{5,95}$ (converted to magnitude) over the median magnitude \citep{richards2011machine}.

    \item[\texttt{PeriodLS}] \textit{Lomb-Scargle Period}. Is the first period extracted with the \textit{Lomb-Scargle} method \citep{kim2011quasi, kim2014epoch, vanderplas2018understanding}.

    \item[\texttt{Period\_fit}] ``\textit{Period Fitness}''. Measure of the reliability of the \texttt{PeriodLS} value. The reliability falls as it approaches the value of $1$ \citep{kim2011quasi, kim2014epoch, vanderplas2018understanding}.

    \item[\texttt{Psi\_CS ($\Psi_{CS}$)}] $R_CS$ applied to the phase-folded light curve (generated using the period as estimated from the Lomb-Scargle method) \citep{kim2011quasi, kim2014epoch}.

    \item[\texttt{Psi\_eta ($\Psi_{\eta})$}] $\eta^e$ index calculated from the folded light curve  (generated using the period estimate from the Lomb-Scargle method) \citep{kim2011quasi, kim2014epoch}.

    \item[\texttt{Q31 ($Q_{3-1}$)}] Is the difference between the third quartile, $Q_3$ , and the first quartile,  $Q_1$ of magnitudes \citep{kim2014epoch}.

    \item[\texttt{Rcs ($R_{CS}$})] ``\textit{Range Cumulative Sum}'' is literally the range of the cumulative sum of magnitudes normalized by $1/{N\sigma}$, where $N$ is the number of observations and $\sigma$ is the standard deviation of magnitudes.\citep{kim2011quasi}.

    \item[\texttt{Skew}] The skewness of magnitudes \citep{kim2011quasi}.

    \item[\texttt{SmallKurtosis}] Small sample kurtosis of the magnitudes. For a normal distribution this values is $\sim 0$ \citep{richards2011machine}.

    \item[\texttt{Std}] The standard deviation of magnitudes \citep{richards2011machine}.

\end{description}


\onecolumn
\section{Feature analysis}
\label{appendix:feat_analysis}

\begin{figure*}[h!]
\centering
    \includegraphics[width=.99\textwidth]{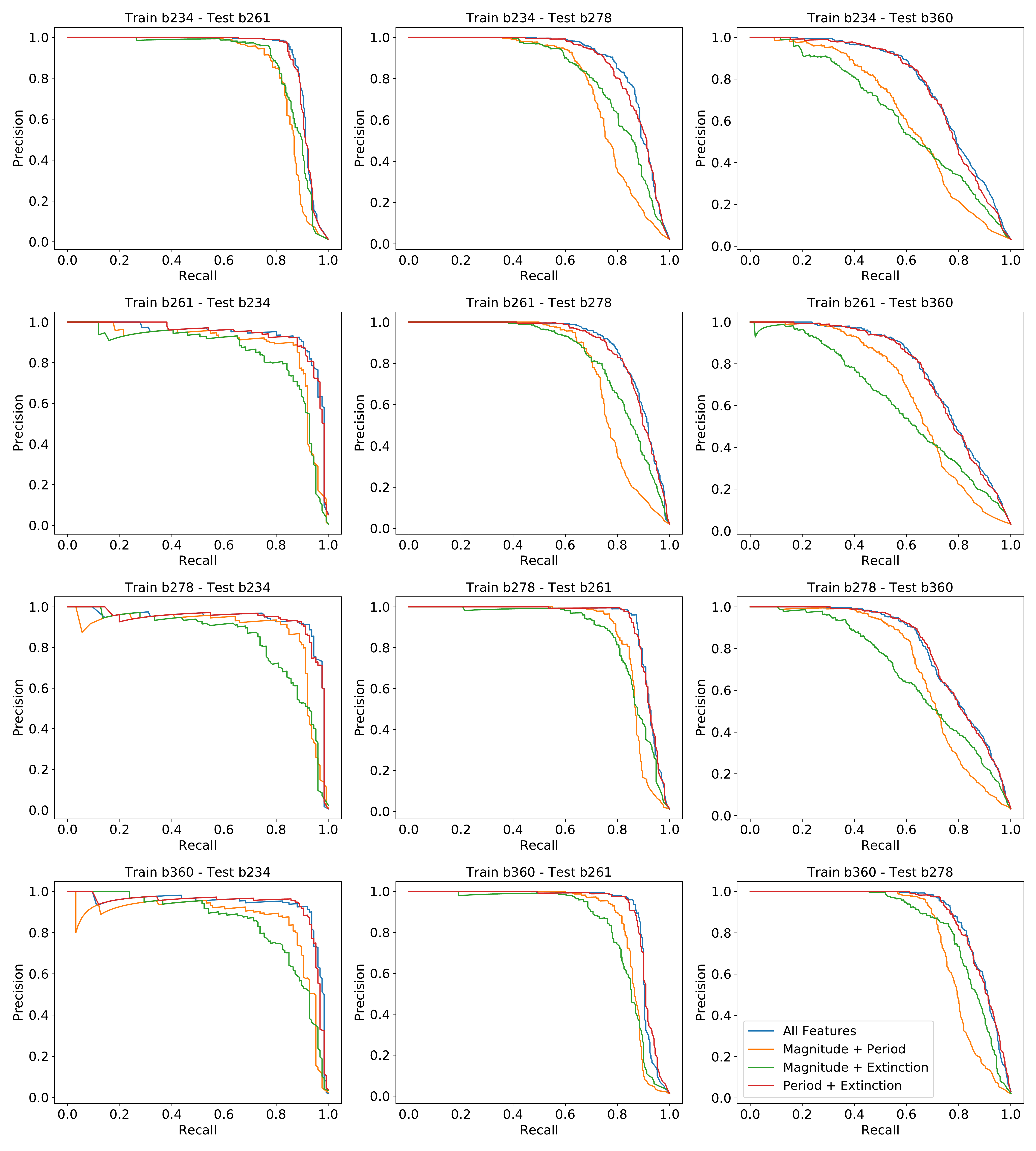}
    \caption{Precision vs. recall curves for different features sub-sets for different tile combinations in training and testing. Every plot is one combination of train and test and every curve is one feature subset.}
\end{figure*}


\onecolumn
\section{Model selection}
\label{appendix:model_selection}

\begin{figure*}[h!]
\centering
    \includegraphics[width=.99\textwidth]{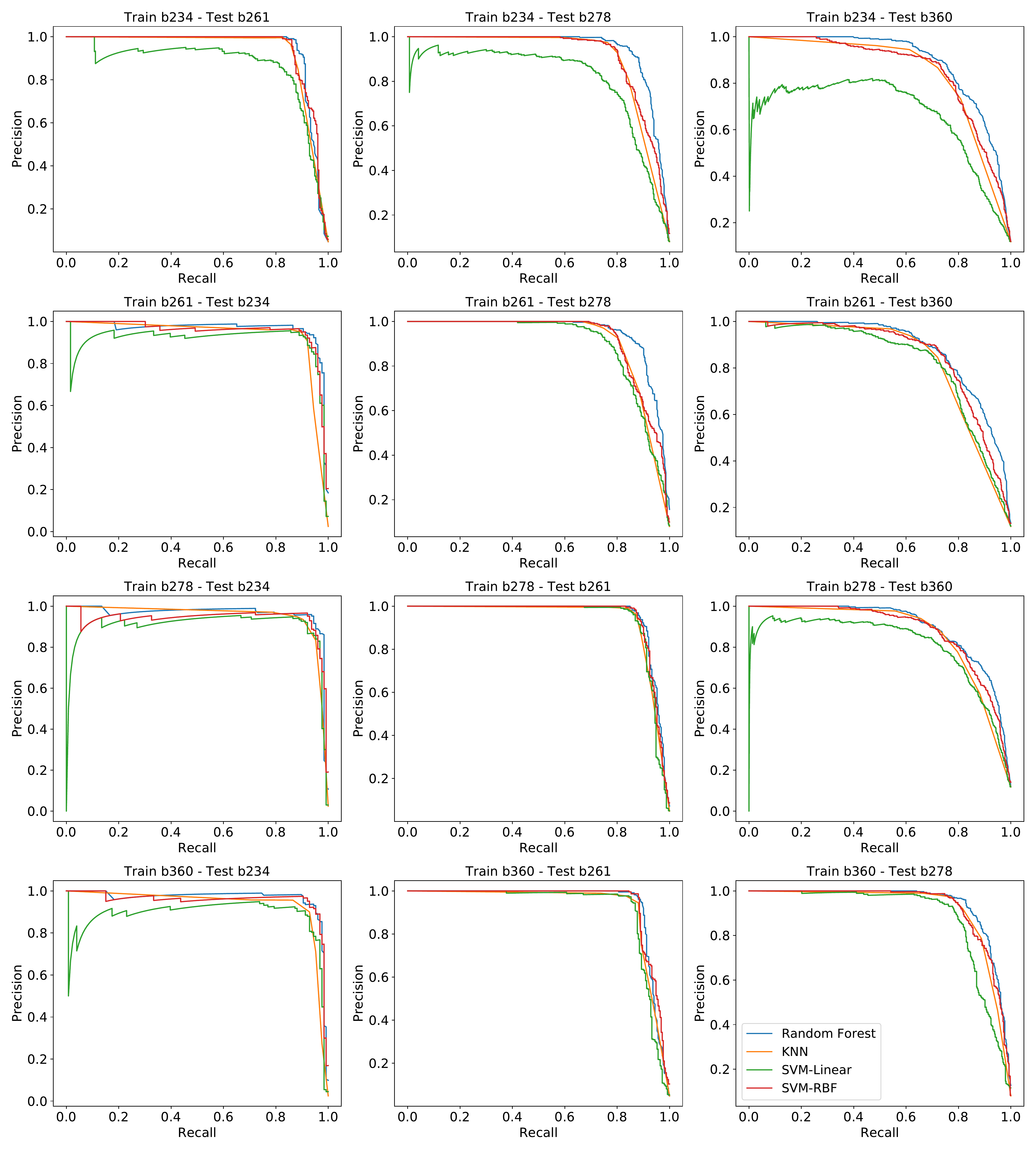}
    \caption{Precision vs. recall curves for different models for different tile combinations in training and testing. Every plot is one combination of training and testing and every curve is one model.}
\end{figure*}


\onecolumn
\section{Unbalance analysis}
\label{appendix:unbalance}

\begin{figure*}[h!]
\centering
    \includegraphics[width=.99\textwidth]{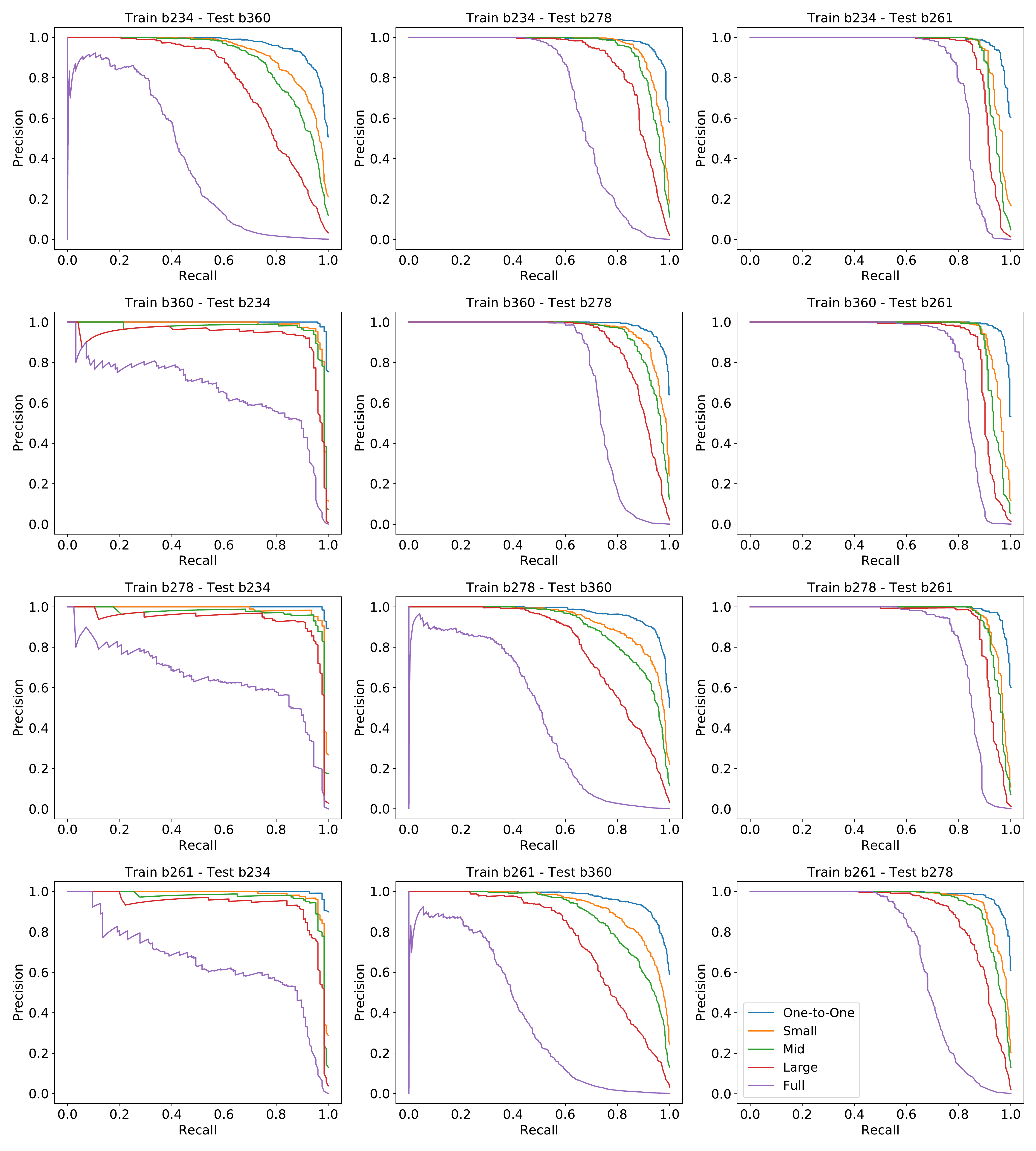}
    \caption{
        Precision vs. recall curves training and testing on the same sample sizes (one-to-one, small, mid, large, and full) for different tile combinations in training and testing. Every plot is one combination of training and testing and every curve is one sample size.
    }
\end{figure*}


\begin{figure*}[h!]
\centering
    \includegraphics[width=.99\textwidth]{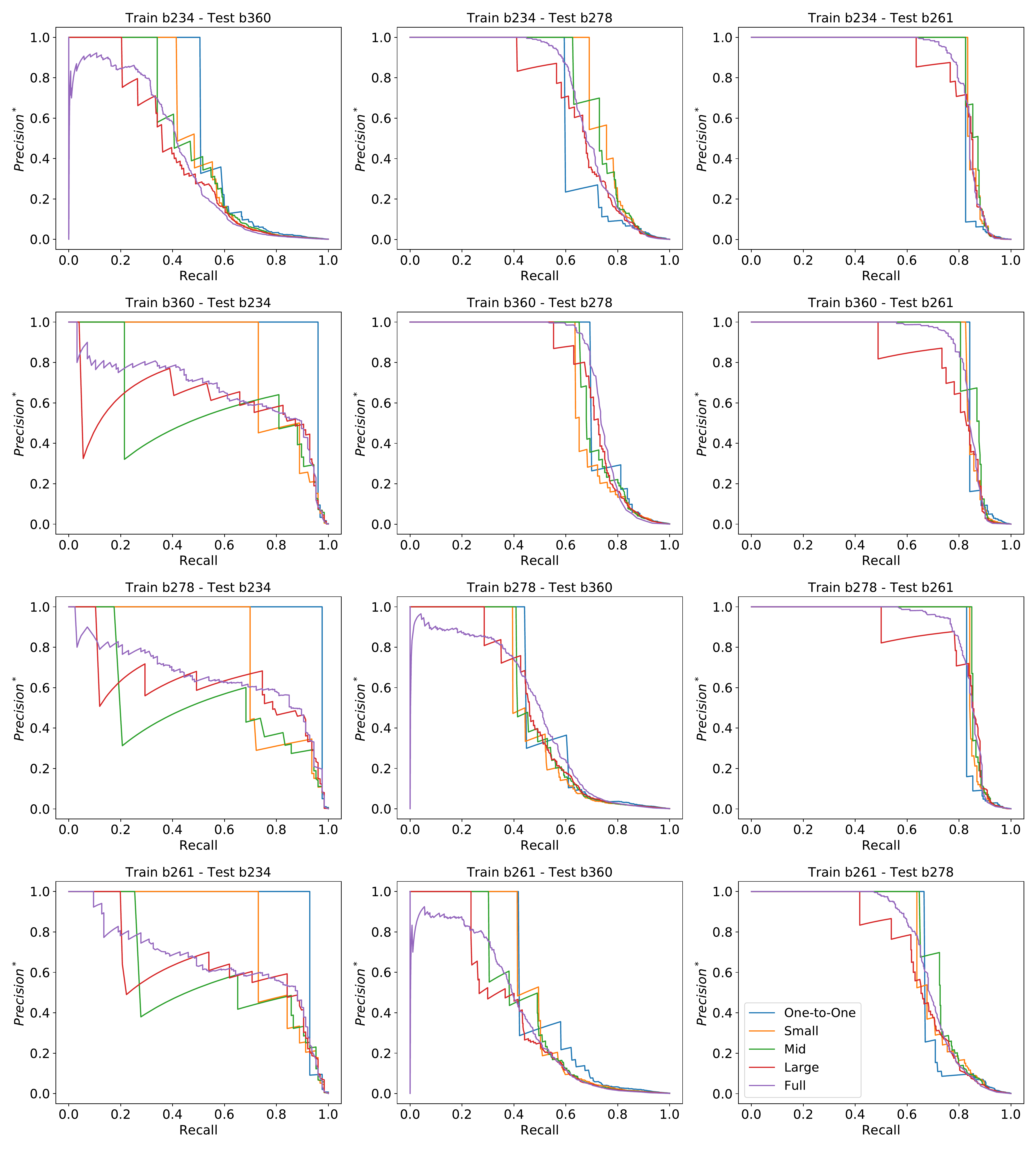}
    \caption{
        Precision$^*$ vs. recall curves training on different sample sizes (one-to-one, small, mid, large, and full) and testing on the full sample for different tile combinations in training and testing. Every plot is one combination of train and test and every curve is one sample size
    }
\end{figure*}


\begin{figure*}[h!]
\centering
    \includegraphics[width=.99\textwidth]{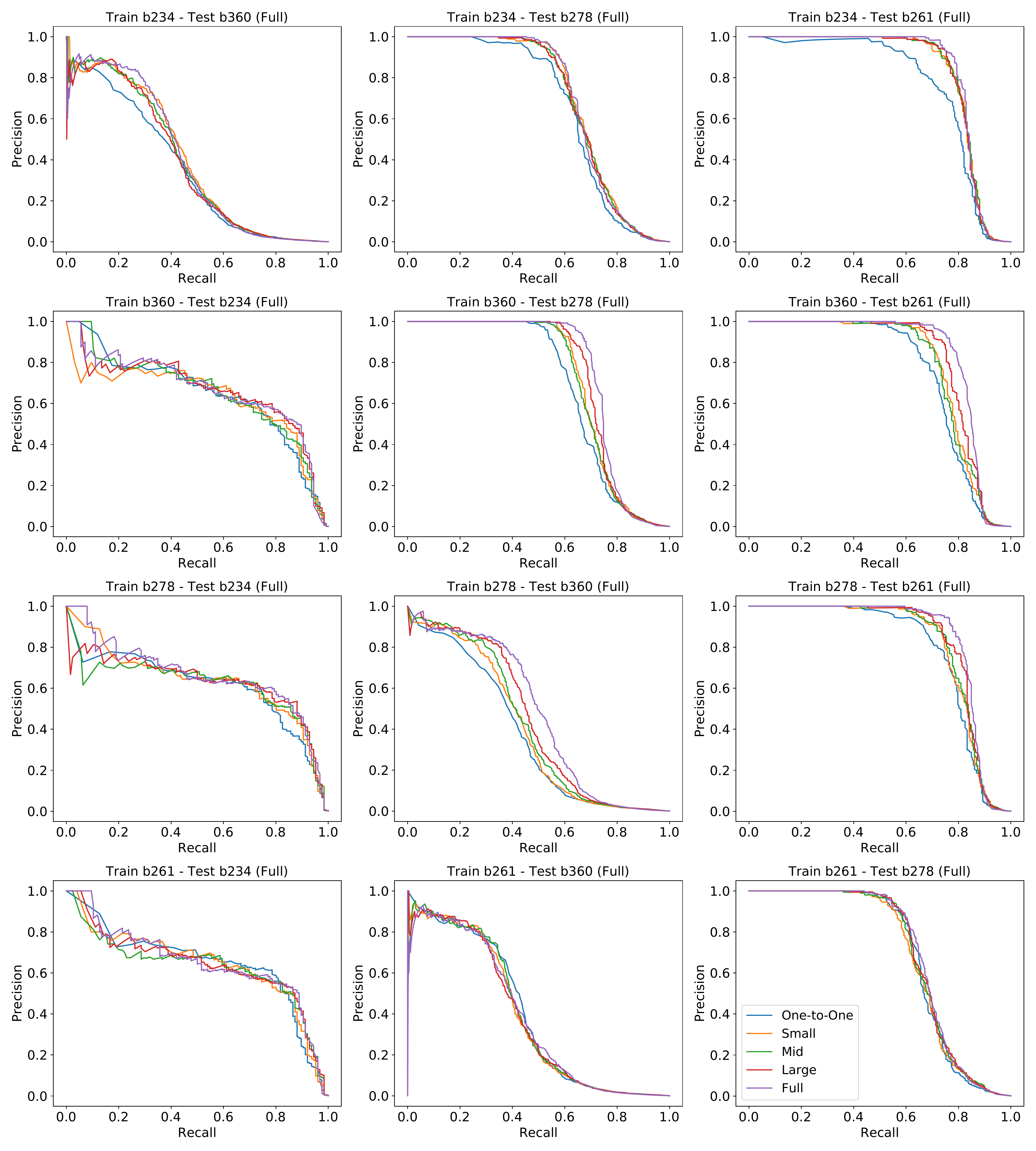}
    \caption{
   Precision vs. recall curves training on different sample sizes (one-to-one, small, mid, large, and full) and testing on the full sample for different tile combinations in training and testing. Every plot is one combination of train and test and every curve is one sample size
    }
\end{figure*}


\onecolumn
\section{Analysis of misidentifications}
\label{appendix:fn}

\begin{figure*}[h!]
\centering
    \includegraphics[width=.99\textwidth]{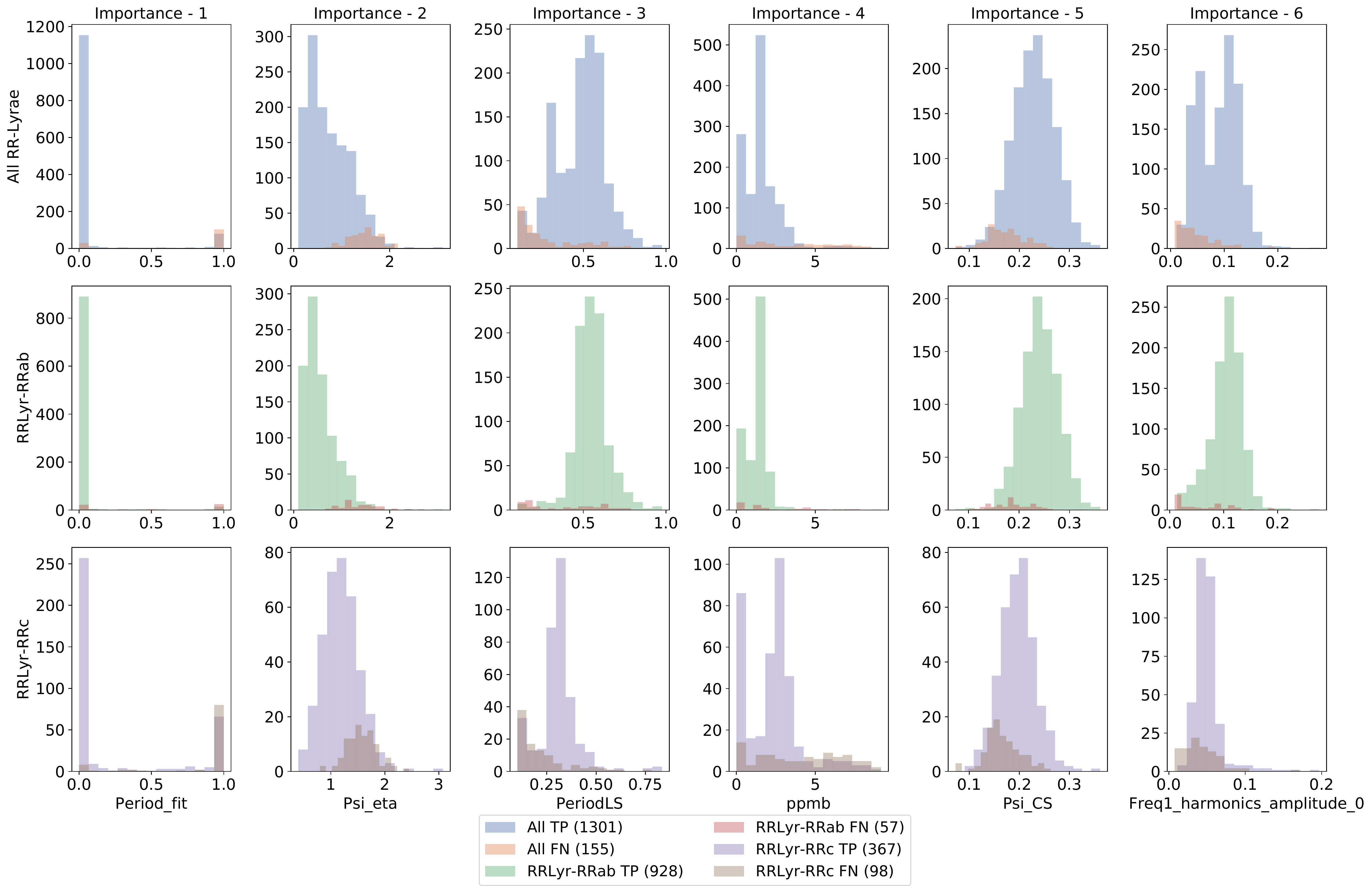}
    \caption{
            Distributions of the TP and FN classes for the RRL stars in all tiles according to the six most important features to differentiate them. In the top row, all the stars are compared, in the mid row, only the RRL-AB type, and in the low row, the RRL-C type. From right to left, the columns corresponds to: \texttt{period\_fit}, \texttt{Psi\_eta}, \texttt{PeriodLS}, \texttt{ppmb}, \texttt{Psi\_CS,} and \texttt{Freq1\_harmonics\_amplitude\_0}. All features were converted to z-scores and are dimensionless.
    }
\end{figure*}


\onecolumn
\section{Catalog}
\label{appendix:catalog}

\begin{table}[tbh!]
\caption{Candidates 1--61 candidates to RRL sorted by probability of being  an RRL.}
\begin{center}
\resizebox{.81\textwidth}{!} {
  \begin{tabular}{l|llrrrrrr}
\toprule
{} &                        ID &  Tile &               $RA_{K_s}$ &           $Dec_{K_s}$  &           Period &  Mean Mag. & Amplitude & Prob. \\
{} &                        {} &    {} &    \small{(J2000, Deg.)} &  \small{(J2000, Deg.)} &   \small{(Days)} &         {} &        {} &    {} \\
\toprule
1  &  VVV J175011.98-184159.6 &  b396 & 267.54992 & -18.69989 & 0.62997 &  14.306558 & 0.17800 & 0.852 \\
2  &  VVV J175228.32-174548.8 &  b396 & 268.11802 & -17.76356 & 0.61577 &  14.396361 & 0.16400 & 0.849 \\
3  &  VVV J175008.72-180538.0 &  b396 & 267.53635 & -18.09389 & 0.51236 &  14.313710 & 0.17350 & 0.838 \\
4  &  VVV J175033.09-184405.2 &  b396 & 267.63789 & -18.73477 & 0.56379 &  14.307435 & 0.15550 & 0.837 \\
5  &  VVV J174957.06-185056.2 &  b396 & 267.48776 & -18.84894 & 0.63820 &  14.029275 & 0.16350 & 0.828 \\
6  &  VVV J175125.96-181823.4 &  b396 & 267.85817 & -18.30650 & 0.56330 &  13.940565 & 0.15800 & 0.827 \\
7  &  VVV J175110.05-181645.5 &  b396 & 267.79188 & -18.27930 & 0.55656 &  14.007191 & 0.14900 & 0.825 \\
8  &  VVV J175137.35-182130.6 &  b396 & 267.90562 & -18.35851 & 0.55214 &  14.010037 & 0.17300 & 0.825 \\
9  &  VVV J175252.40-183041.0 &  b396 & 268.21834 & -18.51140 & 0.55839 &  14.610723 & 0.15575 & 0.825 \\
10 &  VVV J180447.32-401217.9 &  b216 & 271.19715 & -40.20497 & 0.60829 &  14.435616 & 0.16500 & 0.812 \\
11 &  VVV J173339.15-285501.0 &  b360 & 263.41313 & -28.91696 & 0.55995 &  14.503172 & 0.16450 & 0.808 \\
12 &  VVV J174806.73-183056.2 &  b396 & 267.02806 & -18.51561 & 0.49765 &  13.864885 & 0.17500 & 0.804 \\
13 &  VVV J175153.98-175623.5 &  b396 & 267.97493 & -17.93985 & 0.64240 &  13.658129 & 0.14800 & 0.804 \\
14 &  VVV J175244.49-181004.1 &  b396 & 268.18538 & -18.16781 & 0.64917 &  13.571275 & 0.16850 & 0.801 \\
15 &  VVV J174910.42-182413.2 &  b396 & 267.29342 & -18.40367 & 0.79211 &  13.726371 & 0.16125 & 0.796 \\
16 &  VVV J175222.29-181747.2 &  b396 & 268.09285 & -18.29646 & 0.56563 &  13.610202 & 0.15900 & 0.791 \\
17 &  VVV J174739.05-182535.8 &  b396 & 266.91271 & -18.42661 & 0.57735 &  14.112968 & 0.16725 & 0.791 \\
18 &  VVV J181233.16-345705.0 &  b234 & 273.13815 & -34.95137 & 0.56021 &  14.337519 & 0.16450 & 0.787 \\
19 &  VVV J175017.49-180054.8 &  b396 & 267.57288 & -18.01521 & 0.46292 &  14.436433 & 0.16850 & 0.786 \\
20 &  VVV J174901.70-180909.8 &  b396 & 267.25710 & -18.15271 & 0.54296 &  13.406229 & 0.16800 & 0.785 \\
21 &  VVV J175055.31-183812.4 &  b396 & 267.73045 & -18.63678 & 0.48131 &  14.234667 & 0.16350 & 0.785 \\
22 &  VVV J175221.02-174952.7 &  b396 & 268.08760 & -17.83132 & 0.69004 &  14.610855 & 0.15650 & 0.776 \\
23 &  VVV J175115.83-174827.0 &  b396 & 267.81597 & -17.80751 & 0.70771 &  13.750252 & 0.13600 & 0.775 \\
24 &  VVV J180201.49-395209.4 &  b216 & 270.50620 & -39.86927 & 0.73553 &  14.096023 & 0.15350 & 0.766 \\
25 &  VVV J175001.34-181438.8 &  b396 & 267.50558 & -18.24411 & 0.62780 &  14.558194 & 0.17975 & 0.764 \\
26 &  VVV J175010.14-183116.9 &  b396 & 267.54227 & -18.52135 & 0.47182 &  14.711811 & 0.17300 & 0.762 \\
27 &  VVV J174924.53-180746.9 &  b396 & 267.35220 & -18.12969 & 0.43561 &  14.542198 & 0.16450 & 0.761 \\
28 &  VVV J175123.43-185840.2 &  b396 & 267.84763 & -18.97782 & 0.46822 &  13.648694 & 0.17150 & 0.759 \\
29 &  VVV J175305.84-175213.9 &  b396 & 268.27435 & -17.87053 & 0.70312 &  14.216122 & 0.12600 & 0.757 \\
30 &  VVV J184352.31-245246.4 &  b214 & 280.96797 & -24.87954 & 0.74086 &  14.267529 & 0.15825 & 0.755 \\
31 &  VVV J175038.85-174859.9 &  b396 & 267.66187 & -17.81664 & 0.48545 &  14.521467 & 0.19125 & 0.755 \\
32 &  VVV J184022.72-235648.3 &  b228 & 280.09466 & -23.94674 & 0.46551 &  14.685971 & 0.16550 & 0.752 \\
33 &  VVV J175339.18-175938.2 &  b396 & 268.41323 & -17.99396 & 0.47261 &  14.356423 & 0.19000 & 0.751 \\
34 &  VVV J175129.85-175305.1 &  b396 & 267.87437 & -17.88475 & 0.55569 &  14.347469 & 0.16850 & 0.746 \\
35 &  VVV J175039.31-183449.6 &  b396 & 267.66379 & -18.58044 & 0.53109 &  14.455916 & 0.14850 & 0.744 \\
36 &  VVV J180212.50-394734.6 &  b216 & 270.55208 & -39.79296 & 0.48912 &  13.581348 & 0.12550 & 0.740 \\
37 &  VVV J175134.20-181323.8 &  b396 & 267.89250 & -18.22328 & 0.45851 &  14.495703 & 0.16800 & 0.733 \\
38 &  VVV J180301.33-402052.2 &  b216 & 270.75555 & -40.34783 & 0.59245 &  14.108386 & 0.12500 & 0.732 \\
39 &  VVV J184645.80-243125.1 &  b214 & 281.69085 & -24.52364 & 0.45093 &  14.274632 & 0.16700 & 0.731 \\
40 &  VVV J174942.51-173958.2 &  b396 & 267.42712 & -17.66616 & 0.57721 &  14.137341 & 0.16500 & 0.727 \\
41 &  VVV J180136.63-395020.7 &  b216 & 270.40262 & -39.83907 & 0.65193 &  14.284261 & 0.13950 & 0.727 \\
42 &  VVV J174941.08-183449.9 &  b396 & 267.42116 & -18.58052 & 0.50584 &  14.818811 & 0.18050 & 0.723 \\
43 &  VVV J181058.84-335034.6 &  b234 & 272.74518 & -33.84294 & 0.52978 &  14.666947 & 0.17350 & 0.721 \\
44 &  VVV J184441.70-241120.2 &  b214 & 281.17373 & -24.18896 & 0.50501 &  14.437030 & 0.16300 & 0.721 \\
45 &  VVV J175205.44-183754.7 &  b396 & 268.02266 & -18.63185 & 0.46989 &  14.627549 & 0.19850 & 0.719 \\
46 &  VVV J184400.27-252157.8 &  b214 & 281.00111 & -25.36606 & 0.60092 &  14.390530 & 0.14450 & 0.717 \\
47 &  VVV J180649.57-315928.9 &  b263 & 271.70656 & -31.99136 & 0.52976 &  14.088983 & 0.16700 & 0.717 \\
48 &  VVV J184056.31-244819.8 &  b228 & 280.23464 & -24.80550 & 0.51275 &  14.201942 & 0.14650 & 0.716 \\
49 &  VVV J181338.52-341911.1 &  b234 & 273.41051 & -34.31974 & 0.58462 &  14.159077 & 0.12650 & 0.714 \\
50 &  VVV J174756.74-182743.2 &  b396 & 266.98641 & -18.46201 & 0.41117 &  14.609176 & 0.17600 & 0.709 \\
51 &  VVV J175042.93-181324.5 &  b396 & 267.67887 & -18.22347 & 0.48632 &  14.535879 & 0.20100 & 0.708 \\
52 &  VVV J174923.39-180950.5 &  b396 & 267.34745 & -18.16404 & 0.42305 &  14.362825 & 0.16550 & 0.705 \\
53 &  VVV J181059.73-345617.5 &  b234 & 272.74887 & -34.93819 & 0.42151 &  14.415356 & 0.14900 & 0.704 \\
54 &  VVV J172934.42-293248.5 &  b360 & 262.39341 & -29.54682 & 0.60060 &  14.289349 & 0.16000 & 0.700 \\
55 &  VVV J173346.21-290936.8 &  b360 & 263.44255 & -29.16023 & 0.74245 &  14.218582 & 0.15050 & 0.698 \\
56 &  VVV J174803.95-181602.0 &  b396 & 267.01645 & -18.26724 & 0.58704 &  14.803940 & 0.15400 & 0.692 \\
57 &  VVV J181301.46-342837.9 &  b234 & 273.25609 & -34.47721 & 0.58798 &  14.180629 & 0.14250 & 0.683 \\
58 &  VVV J181748.88-345945.4 &  b220 & 274.45365 & -34.99594 & 0.43784 &  14.434854 & 0.16200 & 0.680 \\
59 &  VVV J183750.44-241844.8 &  b228 & 279.46016 & -24.31244 & 0.45987 &  14.459764 & 0.17775 & 0.679 \\
60 &  VVV J184520.08-243150.0 &  b214 & 281.33366 & -24.53054 & 0.48765 &  13.647779 & 0.12800 & 0.679 \\
61 &  VVV J175144.83-185532.0 &  b396 & 267.93678 & -18.92556 & 0.67278 &  14.511625 & 0.14775 & 0.679 \\
\bottomrule
\end{tabular}
}\end{center}
{\textbf{Note:}
Column (1) is the order; column (2) the tile where the candidate is located; columns (3) and (4), RA and Dec in J2000 coordinates; column (5) is the calculated Period; column (6) the mean of magnitudes, column (7) is the amplitude; and finally column (5) is the probability to be a RRL. Th Appendix~\ref{appendixa}
describes in detail how Amplitude ($A_{KS}$) and Period ($PeriodLS$) are calculated.}
\label{tab:catalog1}
\end{table}

\begin{table}[tbh!]
\caption{Candidates 62--122 to RRL sorted by probability to be an RRL.
}
\begin{center}
\resizebox{.81\textwidth}{!} {
  \begin{tabular}{l|llrrrrrr}
\toprule
{} &                        ID &  Tile &               $RA_{K_s}$ &           $Dec_{K_s}$  &           Period &  Mean Mag. & Amplitude & Prob. \\
{} &                        {} &    {} &    \small{(J2000, Deg.)} &  \small{(J2000, Deg.)} &   \small{(Days)} &         {} &        {} &    {} \\
\toprule
62  &  VVV J175147.56-181433.8 &  b396 & 267.94816 & -18.24273 & 0.52625 &  14.224069 & 0.15100 & 0.679 \\
63  &  VVV J180146.84-401017.7 &  b216 & 270.44515 & -40.17158 & 0.52174 &  14.466260 & 0.15850 & 0.677 \\
64  &  VVV J175227.50-183530.5 &  b396 & 268.11458 & -18.59182 & 0.54910 &  14.802849 & 0.17700 & 0.676 \\
65  &  VVV J180443.52-400343.2 &  b216 & 271.18132 & -40.06201 & 0.72308 &  13.555667 & 0.14750 & 0.676 \\
66  &  VVV J175055.99-175233.4 &  b396 & 267.73330 & -17.87594 & 0.46532 &  14.872555 & 0.17500 & 0.674 \\
67  &  VVV J184422.22-241105.0 &  b214 & 281.09258 & -24.18472 & 0.60786 &  14.318809 & 0.11650 & 0.673 \\
68  &  VVV J175113.40-180449.9 &  b396 & 267.80582 & -18.08053 & 0.60792 &  14.606886 & 0.12600 & 0.673 \\
69  &  VVV J180418.77-394938.4 &  b216 & 271.07822 & -39.82732 & 0.47581 &  14.746172 & 0.17100 & 0.666 \\
70  &  VVV J175014.44-181511.2 &  b396 & 267.56017 & -18.25312 & 0.61917 &  14.003946 & 0.11600 & 0.665 \\
71  &  VVV J174948.48-181052.8 &  b396 & 267.45200 & -18.18134 & 0.60773 &  13.766404 & 0.13100 & 0.665 \\
72  &  VVV J175140.73-175300.8 &  b396 & 267.91971 & -17.88357 & 0.50270 &  14.285242 & 0.16900 & 0.663 \\
73  &  VVV J184136.04-235324.2 &  b228 & 280.40016 & -23.89006 & 0.41126 &  14.430500 & 0.15275 & 0.661 \\
74  &  VVV J175251.58-174457.5 &  b396 & 268.21493 & -17.74931 & 0.53431 &  14.240718 & 0.14200 & 0.661 \\
75  &  VVV J181227.59-344031.0 &  b234 & 273.11497 & -34.67529 & 0.68363 &  14.420188 & 0.12550 & 0.660 \\
76  &  VVV J184714.69-245037.1 &  b214 & 281.81120 & -24.84363 & 0.48505 &  14.464294 & 0.13850 & 0.657 \\
77  &  VVV J175339.54-180642.2 &  b396 & 268.41476 & -18.11172 & 0.55641 &  14.699515 & 0.13850 & 0.651 \\
78  &  VVV J173152.91-300840.5 &  b360 & 262.97044 & -30.14458 & 0.61330 &  14.432890 & 0.18150 & 0.649 \\
79  &  VVV J174933.60-182711.6 &  b396 & 267.38999 & -18.45321 & 0.44022 &  14.449957 & 0.17450 & 0.648 \\
80  &  VVV J180935.02-335151.8 &  b248 & 272.39593 & -33.86439 & 0.52426 &  14.511369 & 0.16475 & 0.647 \\
81  &  VVV J173331.56-293241.3 &  b360 & 263.38151 & -29.54482 & 0.63007 &  14.545403 & 0.20550 & 0.642 \\
82  &  VVV J175232.88-174312.6 &  b396 & 268.13700 & -17.72017 & 0.39186 &  14.471931 & 0.13750 & 0.641 \\
83  &  VVV J180614.11-351239.5 &  b247 & 271.55880 & -35.21098 & 0.54152 &  14.269311 & 0.15500 & 0.640 \\
84  &  VVV J175106.12-182433.3 &  b396 & 267.77550 & -18.40924 & 0.46868 &  14.784901 & 0.15750 & 0.634 \\
85  &  VVV J173141.82-293139.8 &  b360 & 262.92425 & -29.52772 & 0.44879 &  14.908584 & 0.18700 & 0.634 \\
86  &  VVV J181129.86-345913.9 &  b234 & 272.87441 & -34.98718 & 0.54909 &  14.575766 & 0.14350 & 0.632 \\
87  &  VVV J174928.62-183945.2 &  b396 & 267.36927 & -18.66256 & 0.54239 &  13.419081 & 0.17200 & 0.631 \\
88  &  VVV J175152.88-183749.0 &  b396 & 267.97035 & -18.63028 & 0.49317 &  14.854412 & 0.18050 & 0.631 \\
89  &  VVV J175052.39-175426.6 &  b396 & 267.71830 & -17.90740 & 0.77854 &  14.433668 & 0.12800 & 0.630 \\
90  &  VVV J180034.35-402401.8 &  b216 & 270.14313 & -40.40049 & 0.44718 &  14.312008 & 0.20100 & 0.629 \\
91  &  VVV J175031.50-182011.6 &  b396 & 267.63123 & -18.33656 & 0.61785 &  14.412606 & 0.11700 & 0.629 \\
92  &  VVV J174838.44-181758.4 &  b396 & 267.16017 & -18.29957 & 0.57399 &  14.631496 & 0.12800 & 0.627 \\
93  &  VVV J180445.81-401231.1 &  b216 & 271.19087 & -40.20865 & 0.55307 &  14.232366 & 0.11900 & 0.626 \\
94  &  VVV J175058.63-183908.0 &  b396 & 267.74430 & -18.65222 & 0.61298 &  14.228781 & 0.15500 & 0.625 \\
95  &  VVV J175225.27-180705.0 &  b396 & 268.10530 & -18.11804 & 0.75172 &  13.901447 & 0.15350 & 0.622 \\
96  &  VVV J181436.10-343718.5 &  b220 & 273.65043 & -34.62180 & 0.45274 &  14.497057 & 0.12100 & 0.622 \\
97  &  VVV J181417.48-342508.0 &  b234 & 273.57282 & -34.41888 & 0.46647 &  14.367053 & 0.18750 & 0.621 \\
98  &  VVV J175153.50-175251.9 &  b396 & 267.97290 & -17.88109 & 0.32565 &  14.150061 & 0.14500 & 0.621 \\
99  &  VVV J182201.32-351345.9 &  b206 & 275.50550 & -35.22942 & 0.50528 &  13.985881 & 0.18275 & 0.615 \\
100 &  VVV J173148.72-290052.4 &  b360 & 262.95300 & -29.01456 & 1.19559 &  14.766105 & 0.19850 & 0.611 \\
101 &  VVV J173109.12-285519.9 &  b360 & 262.78799 & -28.92219 & 0.39245 &  13.379934 & 0.15350 & 0.610 \\
102 &  VVV J173121.82-300816.1 &  b360 & 262.84090 & -30.13782 & 0.54021 &  14.708926 & 0.21350 & 0.610 \\
103 &  VVV J174943.68-183005.5 &  b396 & 267.43200 & -18.50152 & 0.33120 &  13.979621 & 0.14050 & 0.608 \\
104 &  VVV J175216.33-184401.9 &  b396 & 268.06803 & -18.73387 & 0.35511 &  14.137985 & 0.13850 & 0.608 \\
105 &  VVV J180123.10-351918.6 &  b247 & 270.34624 & -35.32184 & 0.49902 &  14.580500 & 0.12550 & 0.607 \\
106 &  VVV J180606.03-393827.8 &  b216 & 271.52513 & -39.64106 & 0.70484 &  14.824082 & 0.19150 & 0.607 \\
107 &  VVV J181825.02-341206.7 &  b220 & 274.60423 & -34.20187 & 0.50807 &  14.838024 & 0.18050 & 0.605 \\
108 &  VVV J175232.89-181925.8 &  b396 & 268.13705 & -18.32383 & 0.45452 &  14.444980 & 0.18550 & 0.605 \\
109 &  VVV J175051.64-190024.2 &  b396 & 267.71517 & -19.00673 & 0.70033 &  14.043564 & 0.13525 & 0.596 \\
110 &  VVV J173132.57-293253.4 &  b360 & 262.88571 & -29.54816 & 0.65042 &  14.655571 & 0.18800 & 0.593 \\
111 &  VVV J175033.71-182745.1 &  b396 & 267.64046 & -18.46252 & 0.56814 &  13.757696 & 0.15200 & 0.591 \\
112 &  VVV J184302.38-242841.9 &  b214 & 280.75990 & -24.47829 & 0.50483 &  14.915676 & 0.17425 & 0.591 \\
113 &  VVV J174837.69-182243.3 &  b396 & 267.15703 & -18.37871 & 0.33390 &  14.533334 & 0.14800 & 0.591 \\
114 &  VVV J175109.70-185006.0 &  b396 & 267.79043 & -18.83501 & 0.61546 &  14.330465 & 0.11000 & 0.589 \\
115 &  VVV J175005.18-180035.5 &  b396 & 267.52157 & -18.00987 & 0.36925 &  14.595371 & 0.17600 & 0.585 \\
116 &  VVV J180305.11-402803.8 &  b216 & 270.77130 & -40.46773 & 0.63171 &  14.711038 & 0.16000 & 0.584 \\
117 &  VVV J184422.12-243004.0 &  b214 & 281.09219 & -24.50112 & 0.64824 &  14.122191 & 0.09525 & 0.581 \\
118 &  VVV J175135.60-185156.8 &  b396 & 267.89834 & -18.86577 & 0.44414 &  13.379947 & 0.19450 & 0.580 \\
119 &  VVV J173138.32-293219.7 &  b360 & 262.90968 & -29.53882 & 0.53319 &  14.716767 & 0.20625 & 0.577 \\
120 &  VVV J175256.82-174505.0 &  b396 & 268.23677 & -17.75140 & 0.39167 &  14.240672 & 0.10350 & 0.577 \\
121 &  VVV J181017.83-335048.2 &  b234 & 272.57428 & -33.84673 & 0.49756 &  14.588391 & 0.13550 & 0.577 \\
122 &  VVV J184236.63-241216.9 &  b228 & 280.65264 & -24.20471 & 0.35425 &  14.707500 & 0.15325 & 0.576 \\
\bottomrule
\end{tabular}
}
\end{center}
{\textbf{Note:} All details are the same as in Table~\ref{tab:catalog1}.}
\label{tab:catalog2}
\end{table}

\begin{table}[tbh!]
\caption{Candidates 123--182 to RRL sorted by probability to be an RRL.
}
\begin{center}
\resizebox{.81\textwidth}{!} {
  \begin{tabular}{l|llrrrrrr}
\toprule
{} &                        ID &  Tile &               $RA_{K_s}$ &           $Dec_{K_s}$  &           Period &  Mean Mag. & Amplitude & Prob. \\
{} &                        {} &    {} &    \small{(J2000, Deg.)} &  \small{(J2000, Deg.)} &   \small{(Days)} &         {} &        {} &    {} \\
\toprule
123 &  VVV J184402.60-243818.9 &  b214 & 281.01084 & -24.63858 & 0.42983 &  14.758015 & 0.12200 & 0.574 \\
124 &  VVV J175029.05-182011.5 &  b396 & 267.62105 & -18.33653 & 0.37973 &  14.448916 & 0.12900 & 0.573 \\
125 &  VVV J175223.79-175532.0 &  b396 & 268.09910 & -17.92557 & 0.52771 &  14.435710 & 0.11800 & 0.572 \\
126 &  VVV J181254.17-342725.3 &  b234 & 273.22572 & -34.45702 & 0.44624 &  14.966774 & 0.19450 & 0.572 \\
127 &  VVV J175959.30-402059.7 &  b216 & 269.99709 & -40.34992 & 0.64158 &  14.301632 & 0.12250 & 0.570 \\
128 &  VVV J175151.29-185422.0 &  b396 & 267.96372 & -18.90612 & 0.39547 &  13.960515 & 0.09925 & 0.569 \\
129 &  VVV J181514.67-341758.0 &  b220 & 273.81113 & -34.29944 & 0.60604 &  14.253613 & 0.11000 & 0.567 \\
130 &  VVV J173104.93-293408.7 &  b360 & 262.77054 & -29.56908 & 0.45803 &  14.737644 & 0.20200 & 0.567 \\
131 &  VVV J184353.62-251759.5 &  b214 & 280.97344 & -25.29986 & 0.56023 &  14.644209 & 0.16525 & 0.567 \\
132 &  VVV J183959.38-244738.9 &  b228 & 279.99743 & -24.79414 & 0.53939 &  13.037118 & 0.10975 & 0.565 \\
133 &  VVV J173514.52-291508.9 &  b360 & 263.81048 & -29.25249 & 0.60347 &  14.150329 & 0.19300 & 0.565 \\
134 &  VVV J184134.85-250632.3 &  b228 & 280.39522 & -25.10898 & 0.59658 &  14.105015 & 0.12575 & 0.565 \\
135 &  VVV J180038.10-402217.3 &  b216 & 270.15877 & -40.37148 & 0.63669 &  14.317283 & 0.10650 & 0.564 \\
136 &  VVV J181612.09-351210.7 &  b220 & 274.05039 & -35.20296 & 0.63611 &  13.779581 & 0.11650 & 0.563 \\
137 &  VVV J182050.39-350506.0 &  b206 & 275.20996 & -35.08501 & 0.54185 &  14.600896 & 0.15525 & 0.562 \\
138 &  VVV J173121.98-291332.8 &  b360 & 262.84160 & -29.22578 & 0.34461 &  14.369184 & 0.17950 & 0.554 \\
139 &  VVV J181022.76-342819.9 &  b234 & 272.59482 & -34.47220 & 0.44854 &  14.647595 & 0.15500 & 0.554 \\
140 &  VVV J184149.94-235645.8 &  b228 & 280.45808 & -23.94604 & 0.62714 &  14.046736 & 0.08225 & 0.553 \\
141 &  VVV J175116.73-181532.6 &  b396 & 267.81971 & -18.25906 & 1.20681 &  14.490817 & 0.18550 & 0.551 \\
142 &  VVV J175147.25-183024.2 &  b396 & 267.94687 & -18.50673 & 0.56745 &  14.233916 & 0.12000 & 0.551 \\
143 &  VVV J175234.11-183947.4 &  b396 & 268.14212 & -18.66317 & 0.98240 &  14.363797 & 0.19450 & 0.549 \\
144 &  VVV J180330.66-404605.5 &  b216 & 270.87773 & -40.76819 & 0.70014 &  13.499436 & 0.09650 & 0.549 \\
145 &  VVV J174822.20-182630.9 &  b396 & 267.09251 & -18.44192 & 0.72387 &  14.495015 & 0.11400 & 0.548 \\
146 &  VVV J173334.49-293202.7 &  b360 & 263.39370 & -29.53409 & 0.45166 &  14.863846 & 0.20450 & 0.547 \\
147 &  VVV J180122.05-394751.2 &  b216 & 270.34186 & -39.79756 & 0.49847 &  14.106507 & 0.15600 & 0.545 \\
148 &  VVV J175010.71-185558.2 &  b396 & 267.54461 & -18.93284 & 0.49126 &  14.976538 & 0.19000 & 0.542 \\
149 &  VVV J183827.60-243516.4 &  b228 & 279.61501 & -24.58788 & 0.35679 &  14.186794 & 0.13650 & 0.541 \\
150 &  VVV J174851.35-183557.1 &  b396 & 267.21395 & -18.59920 & 0.47823 &  14.828702 & 0.22250 & 0.540 \\
151 &  VVV J181357.00-343414.0 &  b220 & 273.48750 & -34.57056 & 0.55161 &  14.387049 & 0.10675 & 0.540 \\
152 &  VVV J174818.47-184240.3 &  b396 & 267.07695 & -18.71118 & 0.65717 &  14.957061 & 0.20900 & 0.538 \\
153 &  VVV J175055.40-174744.1 &  b396 & 267.73082 & -17.79559 & 0.47883 &  14.617106 & 0.21250 & 0.537 \\
154 &  VVV J174729.91-183118.1 &  b396 & 266.87463 & -18.52169 & 0.54615 &  14.372712 & 0.10175 & 0.536 \\
155 &  VVV J184302.38-242841.9 &  b228 & 280.75993 & -24.47831 & 0.50485 &  14.931221 & 0.18625 & 0.534 \\
156 &  VVV J175335.32-181114.5 &  b396 & 268.39718 & -18.18737 & 0.25673 &  12.874030 & 0.16425 & 0.534 \\
157 &  VVV J174858.77-182134.1 &  b396 & 267.24489 & -18.35946 & 0.63239 &  14.226136 & 0.21800 & 0.533 \\
158 &  VVV J175311.35-175231.5 &  b396 & 268.29730 & -17.87541 & 0.60246 &  14.833838 & 0.14625 & 0.532 \\
159 &  VVV J183725.74-243751.1 &  b228 & 279.35726 & -24.63087 & 0.31666 &  13.241353 & 0.12375 & 0.532 \\
160 &  VVV J181254.45-341459.0 &  b234 & 273.22689 & -34.24972 & 0.48941 &  14.722361 & 0.16850 & 0.532 \\
161 &  VVV J181159.60-343500.1 &  b234 & 272.99835 & -34.58335 & 0.63124 &  13.644068 & 0.10050 & 0.532 \\
162 &  VVV J175007.04-184545.3 &  b396 & 267.52931 & -18.76258 & 0.33112 &  14.772765 & 0.15050 & 0.530 \\
163 &  VVV J175044.68-183310.8 &  b396 & 267.68615 & -18.55301 & 0.38966 &  14.333492 & 0.09600 & 0.527 \\
164 &  VVV J181522.12-350716.6 &  b220 & 273.84219 & -35.12128 & 0.28828 &  13.992228 & 0.12800 & 0.526 \\
165 &  VVV J175012.91-184052.0 &  b396 & 267.55378 & -18.68112 & 0.29193 &  14.171443 & 0.19950 & 0.524 \\
166 &  VVV J181759.92-345805.5 &  b220 & 274.49967 & -34.96821 & 0.65434 &  14.148250 & 0.14200 & 0.523 \\
167 &  VVV J175208.24-175314.8 &  b396 & 268.03433 & -17.88744 & 0.42989 &  15.173374 & 0.19400 & 0.522 \\
168 &  VVV J181246.66-343549.4 &  b234 & 273.19443 & -34.59704 & 0.54813 &  14.803782 & 0.12500 & 0.522 \\
169 &  VVV J174936.98-180337.6 &  b396 & 267.40408 & -18.06046 & 0.41483 &  14.933797 & 0.16425 & 0.521 \\
170 &  VVV J175106.82-181234.8 &  b396 & 267.77842 & -18.20966 & 0.43832 &  14.792894 & 0.16100 & 0.521 \\
171 &  VVV J180506.41-401414.1 &  b216 & 271.27669 & -40.23724 & 0.64023 &  13.677761 & 0.11450 & 0.518 \\
172 &  VVV J181646.84-352508.5 &  b220 & 274.19518 & -35.41902 & 0.50942 &  14.424503 & 0.19825 & 0.518 \\
173 &  VVV J174914.37-183922.0 &  b396 & 267.30988 & -18.65612 & 0.41291 &  15.073359 & 0.17400 & 0.516 \\
174 &  VVV J173212.93-285453.9 &  b360 & 263.05386 & -28.91498 & 0.53633 &  14.733316 & 0.20700 & 0.514 \\
175 &  VVV J181320.86-342628.8 &  b234 & 273.33691 & -34.44133 & 0.45602 &  14.793616 & 0.19550 & 0.514 \\
176 &  VVV J175251.71-174540.4 &  b396 & 268.21545 & -17.76123 & 0.47730 &  13.572054 & 0.07950 & 0.513 \\
177 &  VVV J181826.64-344401.2 &  b220 & 274.61100 & -34.73367 & 0.52195 &  14.418227 & 0.11250 & 0.513 \\
178 &  VVV J173230.95-285727.1 &  b360 & 263.12895 & -28.95752 & 0.68201 &  14.479752 & 0.18100 & 0.511 \\
179 &  VVV J181935.38-344801.2 &  b220 & 274.89743 & -34.80032 & 0.32298 &  14.337214 & 0.10075 & 0.507 \\
180 &  VVV J174816.23-182243.9 &  b396 & 267.06761 & -18.37887 & 0.48778 &  14.747733 & 0.12450 & 0.506 \\
181 &  VVV J184124.57-242949.1 &  b228 & 280.35238 & -24.49697 & 0.49971 &  14.895250 & 0.18825 & 0.506 \\
182 &  VVV J180126.41-401920.6 &  b216 & 270.36006 & -40.32240 & 0.64301 &  13.891366 & 0.10400 & 0.506 \\
\bottomrule
\end{tabular}
}
\end{center}
{\textbf{Note:} All details are the same as in Table~\ref{tab:catalog1}.}
\label{tab:catalog3}
\end{table}

\begin{table}[tbh!]
\caption{Candidates 183--242 to RRL sorted by probability to be an RRL.
}
\begin{center}
\resizebox{.81\textwidth}{!} {
  \begin{tabular}{l|llrrrrrr}
\toprule
{} &                        ID &  Tile &               $RA_{K_s}$ &           $Dec_{K_s}$  &           Period &  Mean Mag. & Amplitude & Prob. \\
{} &                        {} &    {} &    \small{(J2000, Deg.)} &  \small{(J2000, Deg.)} &   \small{(Days)} &         {} &        {} &    {} \\
\toprule
183 &  VVV J175013.79-183031.1 &  b396 & 267.55744 & -18.50862 & 0.34924 &  14.675773 & 0.11650 & 0.505 \\
184 &  VVV J175059.00-183434.7 &  b396 & 267.74583 & -18.57631 & 0.44657 &  12.109485 & 0.14300 & 0.505 \\
185 &  VVV J174748.65-182609.1 &  b396 & 266.95272 & -18.43586 & 0.59155 &  13.216329 & 0.08250 & 0.504 \\
186 &  VVV J180333.03-394105.2 &  b216 & 270.88765 & -39.68478 & 0.37988 &  14.838353 & 0.15650 & 0.504 \\
187 &  VVV J180951.71-342819.1 &  b234 & 272.46545 & -34.47197 & 0.49563 &  14.171205 & 0.20600 & 0.503 \\
188 &  VVV J182113.23-344729.2 &  b206 & 275.30511 & -34.79144 & 0.42821 &  14.373597 & 0.16075 & 0.501 \\
189 &  VVV J175115.23-183311.3 &  b396 & 267.81348 & -18.55315 & 0.37318 &  15.065302 & 0.18700 & 0.500 \\
190 &  VVV J175113.82-183107.5 &  b396 & 267.80758 & -18.51876 & 0.44981 &  14.559584 & 0.11400 & 0.500 \\
191 &  VVV J181342.66-342803.1 &  b234 & 273.42773 & -34.46752 & 0.46203 &  14.800813 & 0.15300 & 0.499 \\
192 &  VVV J173244.85-285440.1 &  b360 & 263.18688 & -28.91113 & 0.59835 &  14.608091 & 0.17000 & 0.499 \\
193 &  VVV J175213.79-180827.9 &  b396 & 268.05744 & -18.14107 & 0.65586 &  14.255504 & 0.14250 & 0.498 \\
194 &  VVV J175112.95-190345.5 &  b396 & 267.80394 & -19.06263 & 0.27137 &  13.806954 & 0.11950 & 0.497 \\
195 &  VVV J183751.54-242902.7 &  b228 & 279.46475 & -24.48409 & 0.64905 &  15.044247 & 0.17200 & 0.497 \\
196 &  VVV J174819.09-183741.8 &  b396 & 267.07953 & -18.62827 & 0.45167 &  14.658939 & 0.11150 & 0.496 \\
197 &  VVV J175404.34-175811.2 &  b396 & 268.51808 & -17.96977 & 0.35266 &  12.790197 & 0.20000 & 0.496 \\
198 &  VVV J175008.70-180601.2 &  b396 & 267.53625 & -18.10034 & 0.46554 &  13.881969 & 0.07250 & 0.493 \\
199 &  VVV J184000.98-244740.2 &  b228 & 280.00407 & -24.79450 & 0.38897 &  14.727768 & 0.14550 & 0.489 \\
200 &  VVV J173148.03-295554.0 &  b360 & 262.95015 & -29.93167 & 0.28216 &  14.507204 & 0.23100 & 0.489 \\
201 &  VVV J184124.06-245403.1 &  b228 & 280.35026 & -24.90086 & 0.31884 &  14.484309 & 0.15000 & 0.488 \\
202 &  VVV J175246.17-183139.2 &  b396 & 268.19237 & -18.52755 & 0.58697 &  14.545273 & 0.13425 & 0.487 \\
203 &  VVV J175027.88-180122.6 &  b396 & 267.61615 & -18.02295 & 0.54782 &  14.288711 & 0.18500 & 0.487 \\
204 &  VVV J173148.40-291123.4 &  b360 & 262.95166 & -29.18984 & 0.42390 &  13.387236 & 0.16650 & 0.487 \\
205 &  VVV J173316.61-285514.8 &  b360 & 263.31919 & -28.92079 & 0.34464 &  13.989234 & 0.21200 & 0.487 \\
206 &  VVV J181102.10-345941.2 &  b234 & 272.75876 & -34.99478 & 0.41287 &  13.929947 & 0.11550 & 0.486 \\
207 &  VVV J183940.52-242911.6 &  b228 & 279.91882 & -24.48655 & 0.48238 &  12.878464 & 0.12350 & 0.486 \\
208 &  VVV J175210.36-181822.1 &  b396 & 268.04315 & -18.30613 & 0.30762 &  13.991246 & 0.11900 & 0.486 \\
209 &  VVV J181347.03-350812.9 &  b220 & 273.44596 & -35.13691 & 0.26189 &  14.301592 & 0.16825 & 0.486 \\
210 &  VVV J175151.82-181311.0 &  b396 & 267.96592 & -18.21973 & 0.49593 &  13.722261 & 0.18550 & 0.485 \\
211 &  VVV J174858.70-184406.4 &  b396 & 267.24456 & -18.73512 & 0.17722 &  14.191008 & 0.19150 & 0.485 \\
212 &  VVV J184708.08-242541.1 &  b214 & 281.78366 & -24.42808 & 0.61992 &  13.572956 & 0.10350 & 0.483 \\
213 &  VVV J175234.27-174823.7 &  b396 & 268.14279 & -17.80658 & 0.28366 &  14.614862 & 0.08400 & 0.482 \\
214 &  VVV J181117.24-344747.3 &  b234 & 272.82184 & -34.79648 & 0.34690 &  12.191316 & 0.15100 & 0.481 \\
215 &  VVV J182028.27-342613.1 &  b220 & 275.11780 & -34.43698 & 0.51191 &  14.473590 & 0.17750 & 0.480 \\
216 &  VVV J181357.03-343414.2 &  b234 & 273.48763 & -34.57061 & 0.55163 &  14.394436 & 0.12000 & 0.480 \\
217 &  VVV J175107.70-185907.0 &  b396 & 267.78210 & -18.98529 & 0.63272 &  13.664555 & 0.12500 & 0.480 \\
218 &  VVV J180415.22-394522.6 &  b216 & 271.06343 & -39.75627 & 0.51927 &  14.897313 & 0.17375 & 0.480 \\
219 &  VVV J173022.12-291526.9 &  b360 & 262.59217 & -29.25747 & 0.15626 &  14.303645 & 0.15150 & 0.477 \\
220 &  VVV J175219.02-180707.9 &  b396 & 268.07925 & -18.11886 & 0.32261 &  14.891803 & 0.17300 & 0.477 \\
221 &  VVV J181323.20-343115.3 &  b234 & 273.34668 & -34.52092 & 0.62302 &  14.166402 & 0.09300 & 0.477 \\
222 &  VVV J174905.20-180158.6 &  b396 & 267.27168 & -18.03295 & 0.33799 &  14.726076 & 0.09450 & 0.476 \\
223 &  VVV J184536.67-241612.7 &  b214 & 281.40280 & -24.27018 & 0.84289 &  14.059838 & 0.08800 & 0.475 \\
224 &  VVV J175236.25-174554.4 &  b396 & 268.15106 & -17.76510 & 0.43096 &  12.355095 & 0.21250 & 0.475 \\
225 &  VVV J184712.36-244841.3 &  b214 & 281.80150 & -24.81147 & 0.47995 &  13.991441 & 0.13450 & 0.475 \\
226 &  VVV J184439.95-245216.2 &  b214 & 281.16644 & -24.87117 & 0.34216 &  13.976015 & 0.06650 & 0.474 \\
227 &  VVV J184630.34-251453.7 &  b214 & 281.62643 & -25.24826 & 0.42034 &  13.330500 & 0.10200 & 0.473 \\
228 &  VVV J181945.49-341939.4 &  b220 & 274.93955 & -34.32761 & 0.94775 &  13.411339 & 0.17300 & 0.472 \\
229 &  VVV J175151.57-180235.9 &  b396 & 267.96486 & -18.04329 & 0.33626 &  14.480251 & 0.08650 & 0.471 \\
230 &  VVV J174951.30-182038.2 &  b396 & 267.46375 & -18.34395 & 0.41009 &  14.992596 & 0.15350 & 0.469 \\
231 &  VVV J180531.65-400424.3 &  b216 & 271.38188 & -40.07343 & 0.92920 &  12.674581 & 0.15250 & 0.469 \\
232 &  VVV J181241.96-350513.8 &  b234 & 273.17484 & -35.08717 & 0.66180 &  14.033288 & 0.11700 & 0.468 \\
233 &  VVV J184238.41-244132.2 &  b214 & 280.66006 & -24.69229 & 0.34091 &  14.220500 & 0.09525 & 0.468 \\
234 &  VVV J181509.42-341210.4 &  b234 & 273.78926 & -34.20289 & 0.29957 &  13.641538 & 0.12900 & 0.468 \\
235 &  VVV J175120.85-173644.3 &  b396 & 267.83687 & -17.61230 & 1.27136 &  14.019130 & 0.09850 & 0.466 \\
236 &  VVV J184201.71-242900.8 &  b228 & 280.50713 & -24.48357 & 0.32113 &  13.079305 & 0.16625 & 0.466 \\
237 &  VVV J184100.22-245141.5 &  b228 & 280.25091 & -24.86152 & 0.78273 &  13.253191 & 0.08775 & 0.465 \\
238 &  VVV J173432.50-291813.8 &  b360 & 263.63542 & -29.30383 & 0.26678 &  13.976676 & 0.10700 & 0.463 \\
239 &  VVV J175006.12-181502.0 &  b396 & 267.52550 & -18.25055 & 0.39397 &  14.903939 & 0.11800 & 0.462 \\
240 &  VVV J174957.74-182458.3 &  b396 & 267.49059 & -18.41621 & 0.62848 &  13.894237 & 0.09000 & 0.461 \\
241 &  VVV J181854.12-341347.6 &  b220 & 274.72552 & -34.22988 & 0.35558 &  13.953846 & 0.07000 & 0.461 \\
242 &  VVV J173353.59-285439.9 &  b360 & 263.47329 & -28.91109 & 0.56931 &  14.740961 & 0.13800 & 0.461 \\
\bottomrule
\end{tabular}
}
\end{center}
{\textbf{Note:} All details are the same as in Table~\ref{tab:catalog1}.}
\label{tab:catalog4}
\end{table}

\begin{figure*}[h!]
\centering
    \includegraphics[width=.92\textwidth]{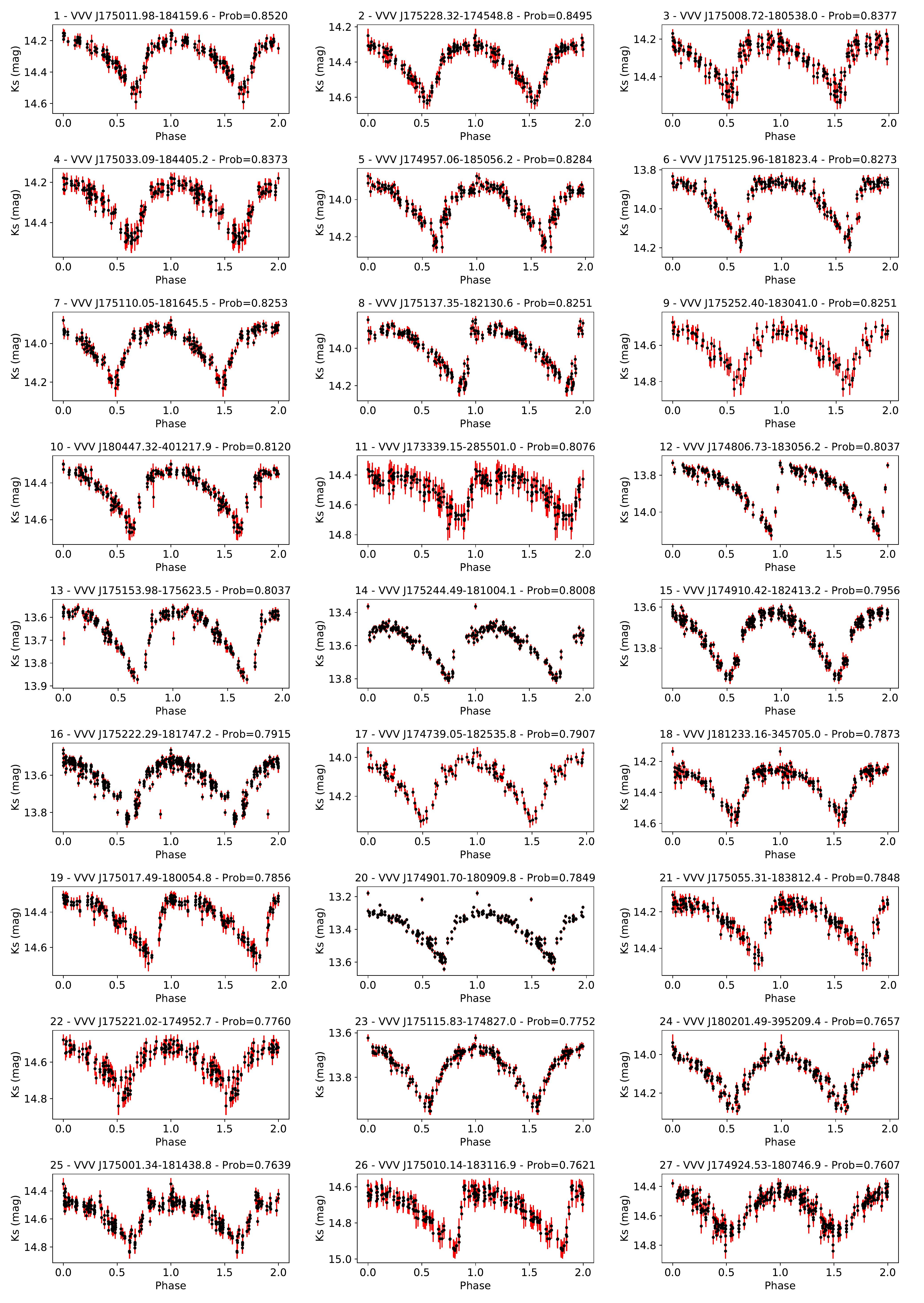}
    \caption{
        Folded light curve for the candidates 1 to 27 to RRL sorted by probability to be an RRL. In the title of every plot are displayed number of the source, the VVV identification and the probability to be a RRL. The horizontal axis is the phase, and the vertical shows the source magnitude. For visual ease, two periods are shown.
        \label{fig:cat_lc1}
        }
\end{figure*}

\begin{figure*}[h!]
\centering
    \includegraphics[width=.92\textwidth]{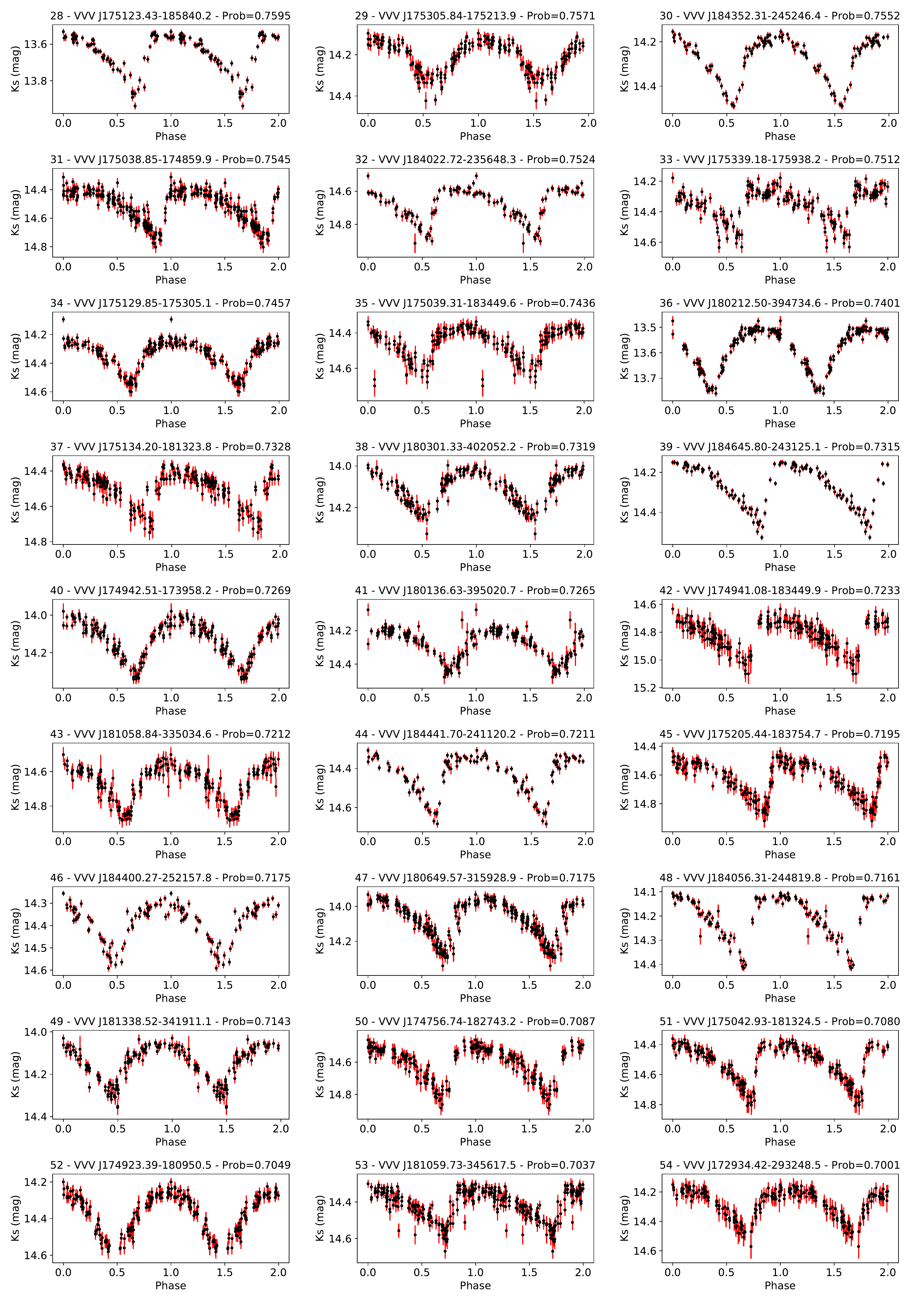}
    \caption{
        Folded light curve for the candidates 28 to 54 to RRL sorted by probability of being an RRL. All details are the same as in Fig.~\ref{fig:cat_lc1}.}
\end{figure*}

\begin{figure*}[h!]
\centering
    \includegraphics[width=.92\textwidth]{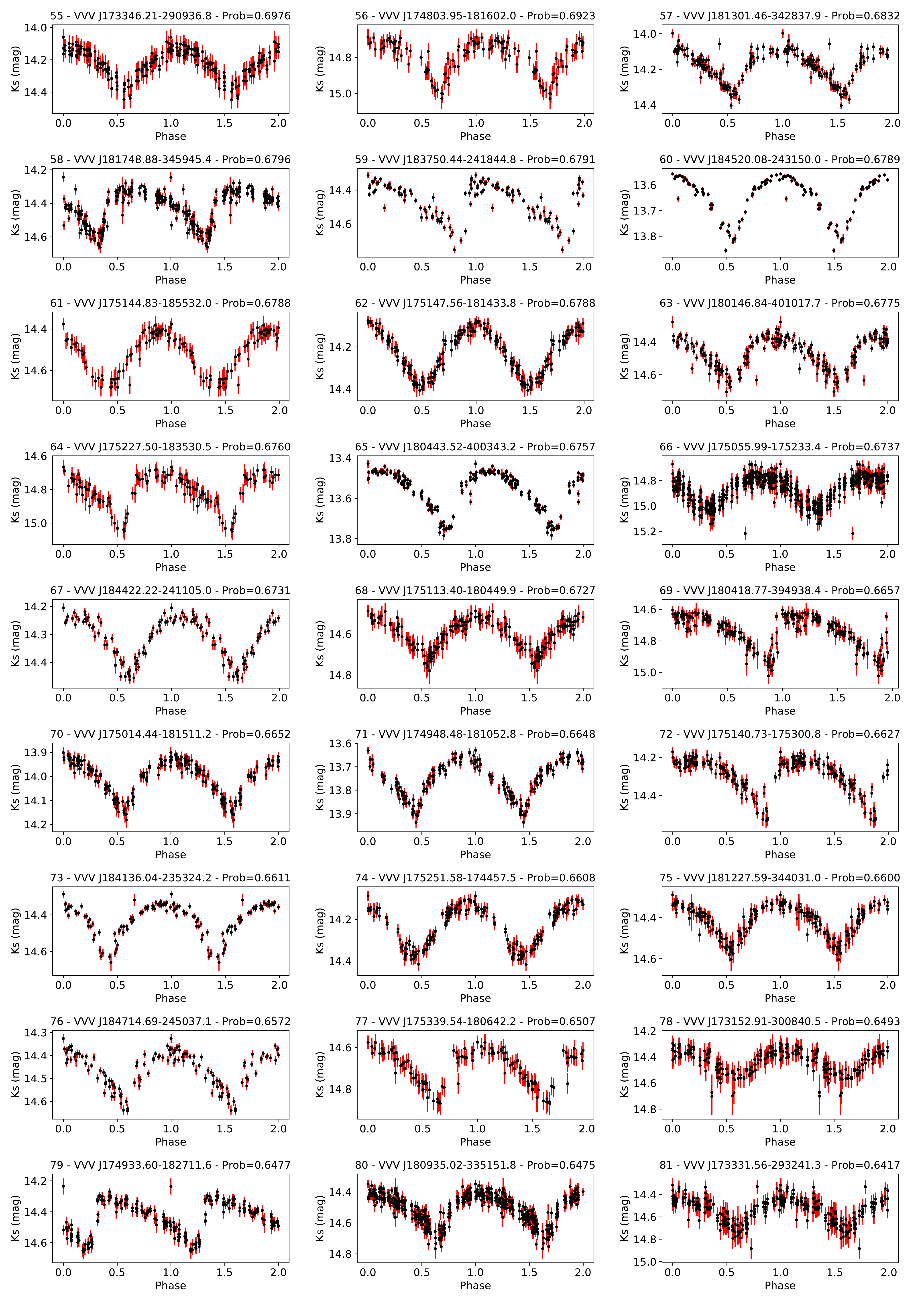}
    \caption{
        Folded light curve for the candidates 55 to 81 to RRL sorted by probability of being  an RRL. All details are the same as in Fig.~\ref{fig:cat_lc1}.}
\end{figure*}

\begin{figure*}[h!]
\centering
    \includegraphics[width=.92\textwidth]{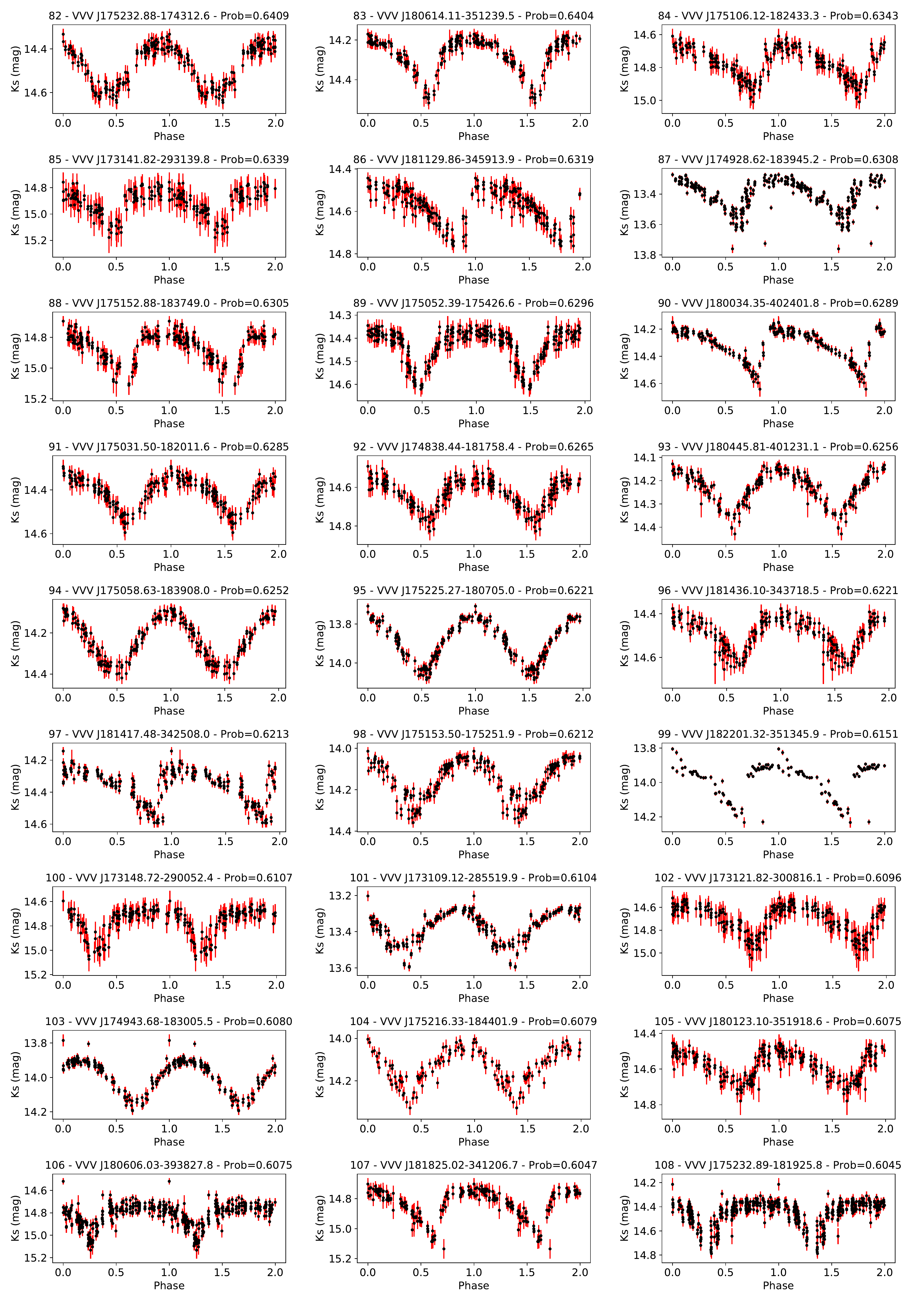}
    \caption{
        Folded light curve for the candidates 82 to 108 to RRL sorted by probability of being  an RRL. All details are the same as in Fig.~\ref{fig:cat_lc1}.}
\end{figure*}

\begin{figure*}[h!]
\centering
    \includegraphics[width=.92\textwidth]{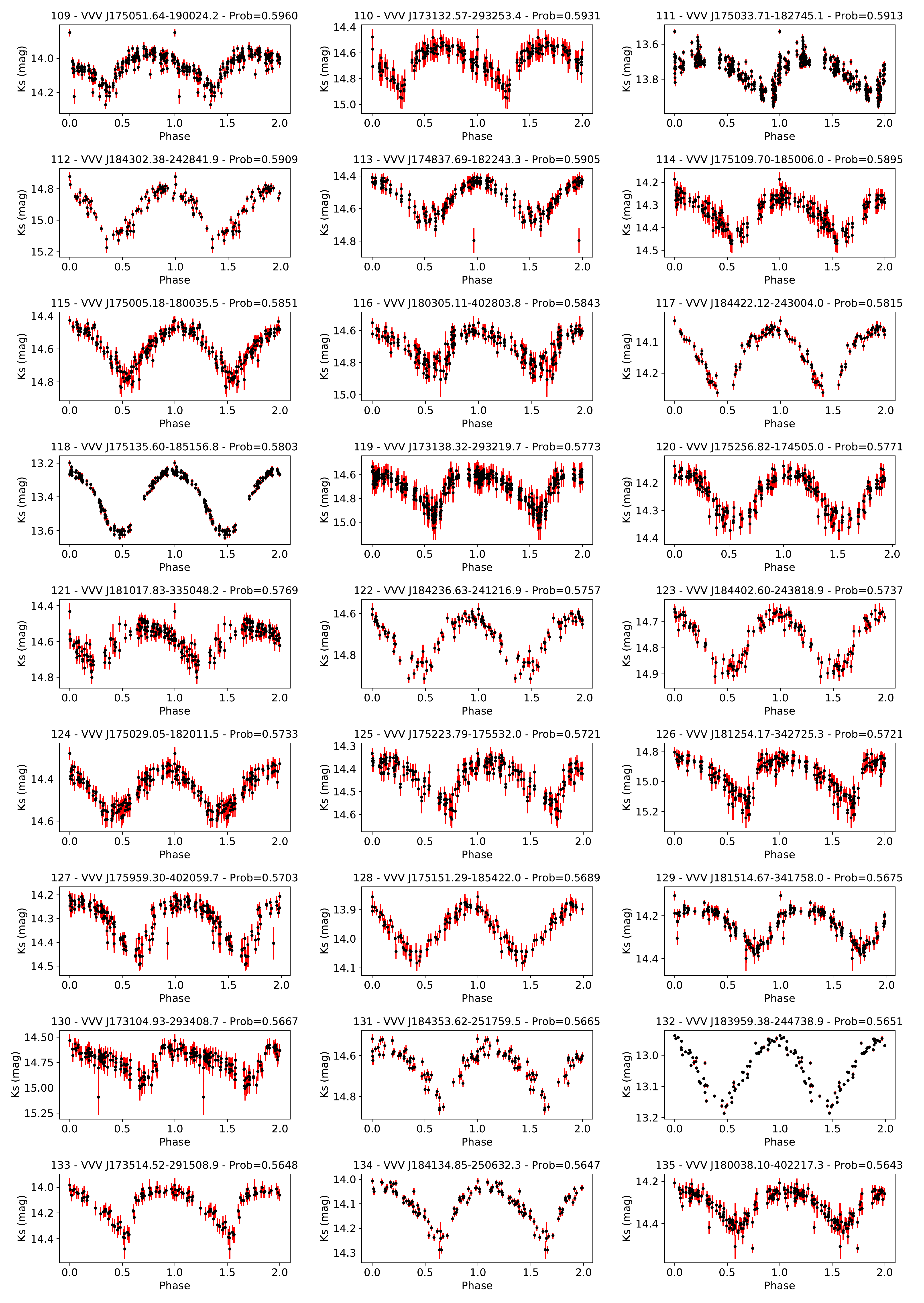}
    \caption{
        Folded lightcurve for the candidates 109 to 135 to RRL sorted by probability of being  an RRL. All details are the same as in Fig.~\ref{fig:cat_lc1}.}
\end{figure*}

\begin{figure*}[h!]
\centering
    \includegraphics[width=.92\textwidth]{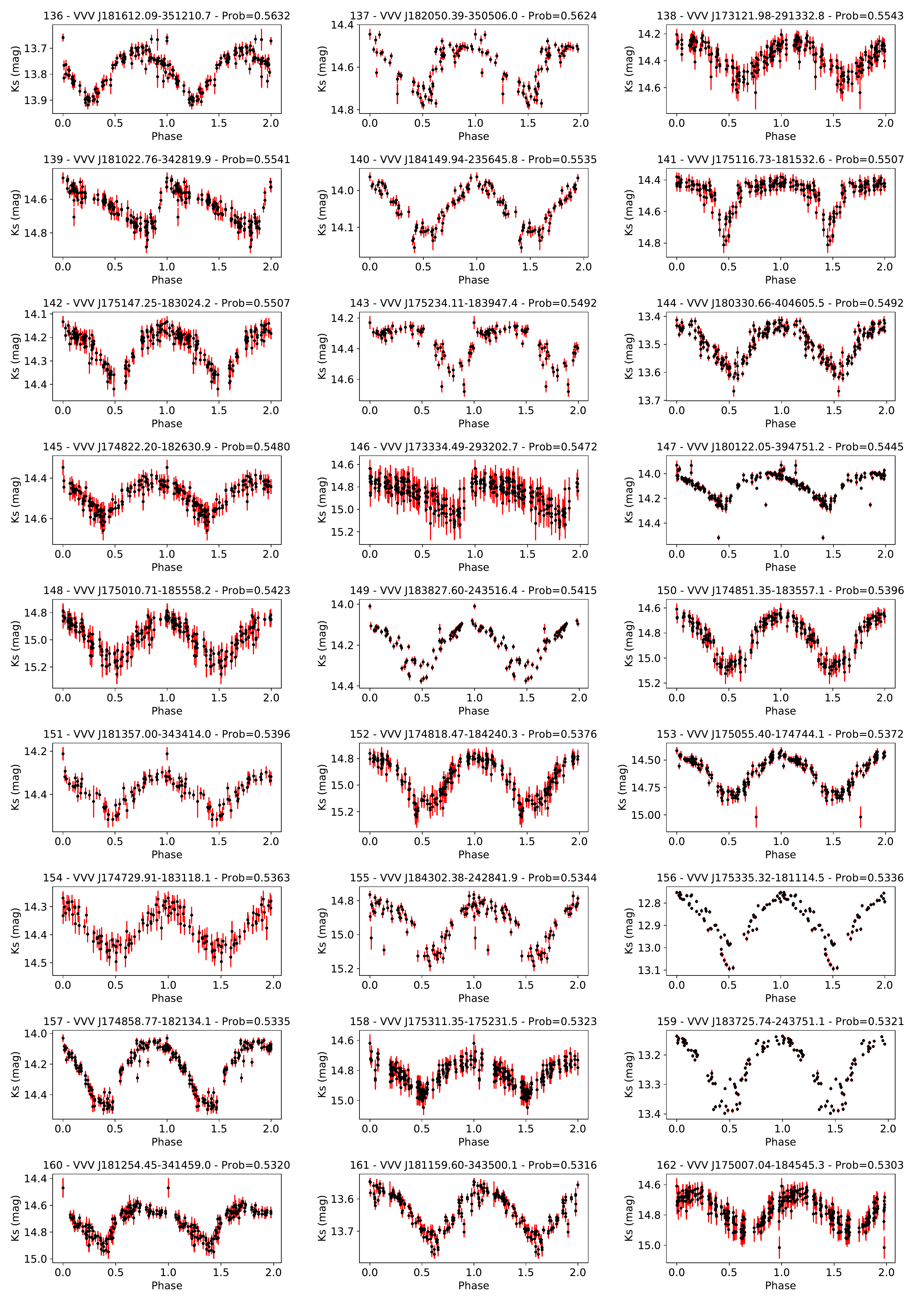}
    \caption{
        Folded light curve for the candidates 136 to 162 to RRL sorted by probability of being  an RRL. All details are the same as in Fig.~\ref{fig:cat_lc1}.}
\end{figure*}

\begin{figure*}[h!]
\centering
    \includegraphics[width=.92\textwidth]{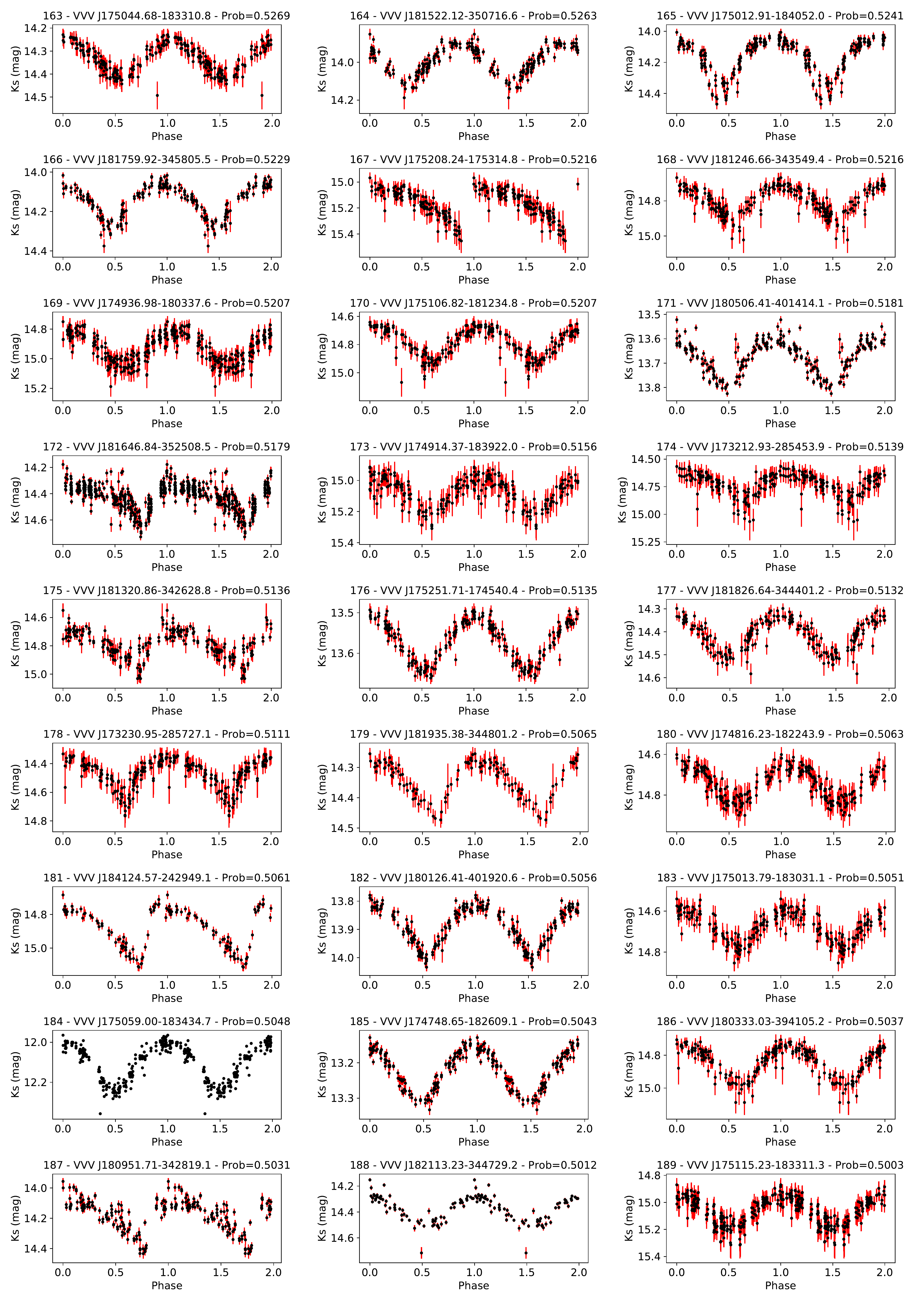}
    \caption{
        Folded light curve for the candidates 163 to 189 to RRL sorted by probability of being  an RRL. All details are the same as in Fig.~\ref{fig:cat_lc1}.}
\end{figure*}

\begin{figure*}[h!]
\centering
    \includegraphics[width=.92\textwidth]{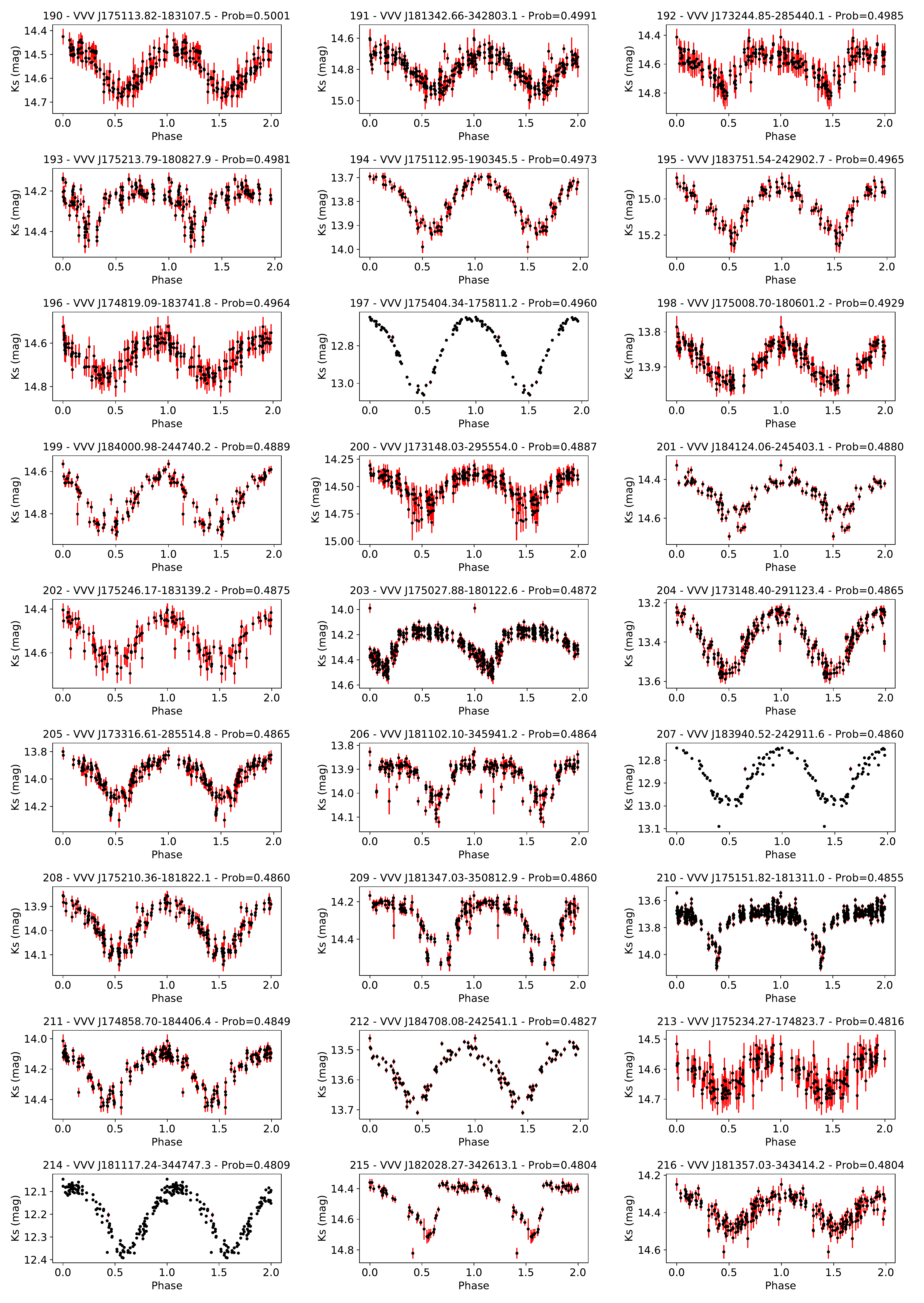}
    \caption{
        Folded light curve for the candidates 190 to 216 to RRL sorted by probability of being \ an RRL. All details are the same as in Fig.~\ref{fig:cat_lc1}.}
\end{figure*}

\begin{figure*}[h!]
\centering
    \includegraphics[width=.92\textwidth]{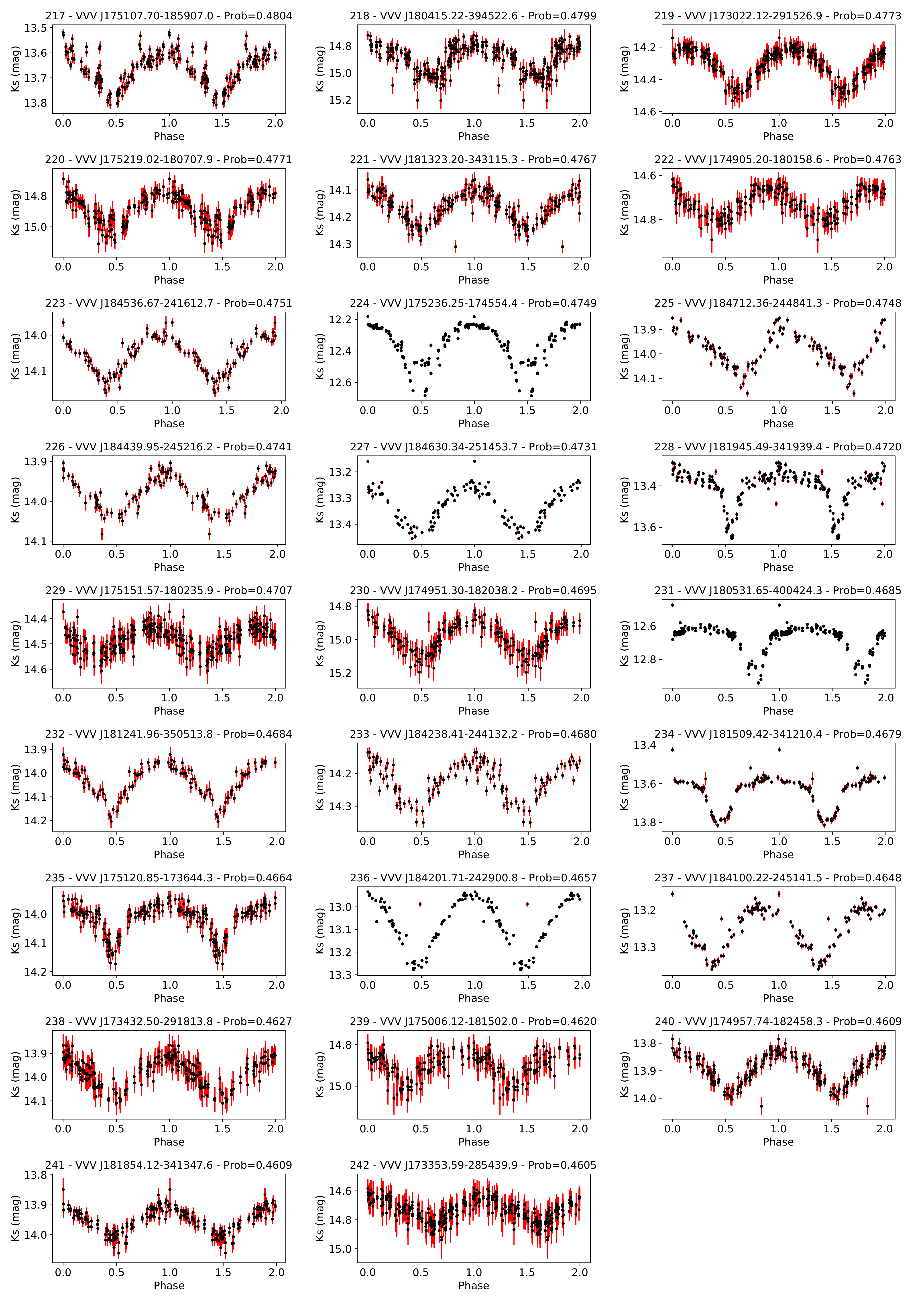}
    \caption{
        Folded light curve for the candidates 217 to 242 to RRL sorted by probability of being  an RRL. All details are the same as in Fig.~\ref{fig:cat_lc1}.}
\end{figure*}

\end{document}